\documentclass[twocolumn,a4paper,reprint,amssymb,aps,prd,groupedaddress,superscriptaddress,nofootinbib]{revtex4-2}
\pdfoutput=1  

\usepackage{booktabs}
\usepackage{multirow}
\usepackage[colorlinks = true,
            linkcolor = blue,
            urlcolor  = blue,
            citecolor = blue,
            anchorcolor = blue]{hyperref}

\usepackage{dcolumn}
\usepackage{bm}
\usepackage{amsmath}
\usepackage{mathtools}
\usepackage{amsfonts}
\usepackage{pbox}
\usepackage{color}
\usepackage[dvipsnames]{xcolor}
\usepackage{tabularx}
\usepackage{subfigure}
\usepackage{graphicx}
\usepackage{psfrag}
\usepackage{lipsum}
\usepackage{enumitem}
\usepackage{changes}
\usepackage{array}
\usepackage{hyperref}
\usepackage{CJK}
\usepackage[english]{babel}
\usepackage{newtxtext}
\usepackage{threeparttable,booktabs}

\usepackage{blindtext}
\usepackage[utf8]{inputenc}
\usepackage{graphicx}
\usepackage{placeins}
\usepackage{siunitx}
\usepackage{amssymb}
\usepackage{mathtools}
\usepackage{physics}
\usepackage{tikz}
\usepackage[titletoc]{appendix}
\usepackage{listings}
\usepackage{float}
\usepackage{soul}
\usepackage{array}
\providecommand{\U}[1]{\protect\rule{.1in}{.1in}}
\newcommand{\be}{\begin{equation}}
\newcommand{\ee}{\end{equation}}

\newcommand{\mincir}{\raise
-3.truept\hbox{\rlap{\hbox{$\sim$}}\raise4.truept\hbox{$<$}\ }}
\newcommand{\magcir}{\raise
-3.truept\hbox{\rlap{\hbox{$\sim$}}\raise4.truept\hbox{$>$}\ }}

\usepackage{color}
\definecolor{darkgreen}{rgb}{0., 0.65, 0.1}

\newcommand{\colvec}[2][.8]{%
  \scalebox{#1}{%
    \renewcommand{\arraystretch}{.8}%
    $\begin{bmatrix}#2\end{bmatrix}$
  }
}

\begin{document}

\title{Testing spatial curvature and anisotropic expansion on top of the $\Lambda$CDM model}

\author{\"{O}zg\"{u}r Akarsu}
\email{akarsuo@itu.edu.tr}
\affiliation{Department of Physics, Istanbul Technical University, Maslak 34469 Istanbul, Turkey}

\author{Eleonora Di Valentino}
\email{e.divalentino@sheffield.ac.uk}
\affiliation{School of Mathematics and Statistics, University of Sheffield, Hounsfield Road, Sheffield S3 7RH, United Kingdom}

\author{Suresh Kumar}
\email{suresh.math@igu.ac.in}
\affiliation{Department of Mathematics, Indira Gandhi University, Meerpur, Haryana 122502, India}

\author{Maya \"{O}zyi\u{g}it}
\email{ozyigita@itu.edu.tr}
\affiliation{Department of Physics, Istanbul Technical University, Maslak 34469 Istanbul, Turkey}

 \author{Shivani Sharma}
\email{shivani@ddu.du.ac.in}
\affiliation{Deen Dayal Upadhyaya College, University of Delhi, Dwarka, New Delhi 110078, India}

\begin{abstract}
We explore the possible advantages of extending the standard $\Lambda$CDM model by more realistic backgrounds compared to its spatially flat Robertson--Walker (RW) spacetime assumption, while preserving the underpinning physics; in particular, by simultaneously allowing non-zero spatial curvature and anisotropic expansion on top of $\Lambda$CDM, viz., the An-$o\Lambda$CDM model. This is to test whether the latest observational data still support spatial flatness and/or isotropic expansion in this case, and, if not, to explore the roles of spatial curvature and expansion anisotropy (due to its stiff fluid-like behavior) in addressing some of the current cosmological tensions associated with $\Lambda$CDM. We first present the theoretical background and explicit mathematical construction of An-$o\Lambda$CDM; in the simplest manner, combining the simplest anisotropic generalizations of the RW spacetime, viz., the Bianchi type I, V, and IX spacetimes, in one Friedmann equation. Then we constrain the parameters of this model and its particular cases, namely, An-$\Lambda$CDM, $o\Lambda$CDM, and $\Lambda$CDM, by using the data sets from different observational probes, viz., Planck cosmic microwave background (CMB) with or without lensing (Lens), baryonic acoustic oscillations (BAO), type Ia Supernovae (SnIa) Pantheon, and cosmic chronometers (CC) data, and discuss the results in detail. Ultimately, we conclude that, within the setup under consideration, (i) the observational data confirm the spatial flatness and isotropic expansion assumptions of $\Lambda$CDM, though a very small amount of present-day expansion anisotropy ($\Omega_{\sigma0}$) cannot be excluded, e.g., $\Omega_{\sigma0}\lesssim10^{-18}$ (95\% C.L.) for An-$\Lambda$CDM from CMB+Lens data, (ii) the introduction of spatial curvature or anisotropic expansion, or both, on top $\Lambda$CDM does not offer a possible relaxation to the $H_0$ tension, and (iii) the introduction of anisotropic expansion neither affects the closed space prediction from the CMB data nor does it improve the drastically reduced value of $H_0$ led by the closed space. We discuss why it is important and indispensable to maintain the geometric generalization work program, especially in models that offer solutions to cosmological tensions, even though our findings do not appear to favor the geometric generalizations of $\Lambda$CDM considered in this work.
\end{abstract}

\maketitle 

\section{Introduction}
\label{sec:intro} 

In the past decades, the scenario assumed as the standard model in cosmology is the so called Lambda cold dark matter ($\Lambda$CDM). This scenario fits a wide range of observational data on different scales and epochs of the universe~\cite{Riess:1998cb,SupernovaCosmologyProject:1998vns,Planck:2018vyg,Alam:2020sor,DES:2021wwk}. However, some discrepancies and inconsistencies in the estimated values of some important cosmological parameters have emerged in the last decade~\cite{DiValentino:2020vvd,DiValentino:2020srs,DiValentino:2020zio,DiValentino:2021izs,Perivolaropoulos:2021jda,Abdalla:2022yfr}. These tensions, even if present with different statistical significance, and possibly due in part to systematic errors in the experiments, demand an explanation. Among all the possibilities, the necessity of physics beyond the well established fundamental theories that underpin, and even extend $\Lambda$CDM, is making its own way. Alternatively, it would be interesting to leave the physics completely untouched, and proceed by considering more realistic spacetimes compared to the spatially flat Friedmann--Lema\^{i}tre--Robertson--Walker (FLRW) spacetime on which $\Lambda$CDM is established.

Here we explore the possible advantages of a pure geometric generalization of the standard $\Lambda$CDM model, in particular, allowing non-flat and anisotropic space. The standard $\Lambda$CDM model, relying on canonical inflationary paradigm~\cite{Starobinsky:1980te,Guth:1980zm,Linde:1981mu,Albrecht:1982wi}, assumes a flat and, in accordance with the cosmological principle, maximally symmetric (homogeneous and isotropic) space, i.e., the spatially flat Robertson--Walker (RW) spacetime, for describing the geometry of the universe on the largest scales, and general relativity (GR) with a positive cosmological constant for describing the dynamics of the universe depending on its material content. It is important to investigate whether these geometric assumptions would still be favored by the latest observational data when the spatially flat RW spacetime is generalized to a more realistic one. Sticking to the cosmological principle, the first step towards a more realistic cosmological model is to allow spatial curvature on top of $\Lambda$CDM. It is known that the Planck cosmic microwave background (CMB) data alone favors positive (closed space) spatial curvature, which mimics a negative energy density that varies as $\propto s^{-2}$, where $s$ is the average expansion scale factor~\cite{Planck:2018vyg}. Moreover, spatially closed models are in agreement with not only the low CMB anisotropy quadrupole of Planck, but also the WMAP CMB data~\cite{Linde:2003hc,Efstathiou:2003hk}. However, the drastic exacerbation of the $H_0$ tension in this case, and the favoring of spatial flatness with extremely high precision by the Planck CMB data in combination with the baryon acoustic oscillations (BAO) data, and the astrophysical data such as type Ia Supernovae (SNIa) and cosmic chronometers (CC), have stimulated the debate on the spatial flatness assumption~\cite{Planck:2018vyg,Handley:2019tkm,DiValentino:2019qzk,DiValentino:2020hov,Vagnozzi:2020zrh,Efstathiou:2020wem,Vagnozzi:2020dfn,Acquaviva:2021gty}; particularly, given that it is in line with the canonical inflationary paradigm---yet, note that inflation is not necessarily in conflict with a non-flat space~\cite{Linde:1995xm,Gratton:2001gw,Linde:2003hc,WMAP:2003ggs}, but spatially closed inflationary models are significantly fine-tuned~\cite{Linde:2003hc}. See also, e.g., Refs.~\cite{Park:2018tgj,Benisty:2020otr,Cao:2020evz,Ryan:2021loh,Khadka:2021ukv,Zuckerman:2021kgm,Cao:2021irf,Cao:2022ugh,Dhawan:2021mel,Chen:2016eyp,Park:2018bwy,Ryan:2018aif,Park:2018fxx,Yang:2021hxg,Cao:2021ldv,Gonzalez:2021ojp,Farrugia:2021zwx,Nilsson:2021ute,Bargiacchi:2021hdp,Yang:2022kho}, which investigate spatial flatness/non-flatness in the RW framework using different observational data sets and analysis methods and/or taking into account extensions (modified gravity, dark energy, etc.) of the standard $\Lambda$CDM model. On the other hand, the canonical inflationary paradigm---assumes a single canonical scalar field, with suitable self-interaction potential, which is minimally coupled to gravity described by GR---makes sense if the observable universe exhibits isotropic expansion (ignoring the tilted models and the possibility of being broken by, for instance, anisotropic dark energy in the late universe). Indeed, the mathematically tractable next step to generalize $\Lambda$CDM geometrically is to allow anisotropic expansion, which then leads to the generalized Friedmann equation bringing in average Hubble parameter along with the shear scalar term (quantifying the anisotropic expansion) and the spatial curvature term that deviates from $\propto s^{-2}$ behavior if the anisotropic expansion is accompanied by an anisotropic spatial curvature~\cite{Collins,Collins:1972tf,Ellis:1998ct,GEllisBook}. In the simplest anisotropic generalizations, viz., when the spatial curvature itself is isotropic or has negligibly small anisotropy, the spatial curvature mimics a source with a positive (open space) or negative (closed space) energy density that varies as $\propto s^{-2}$ and the shear scalar mimics a stiff fluid~\cite{Zeldovich:1961sbr,Barrow1978} (or a canonical scalar field with a negligible potential energy compared to its kinetic energy) with a positive energy density varying as $\propto s^{-6}$, which thus dilutes faster than any other physical source (for which the stiff fluid sets the causality limit~\cite{GEllisBook}) as the universe expands. The stiff fluid-like behavior of the shear scalar is typical for general relativistic anisotropic universes with isotropic spatial curvature filled only with isotropic perfect fluids with no peculiar velocities~\cite{GEllisBook}. Hence, irrespective of the inflationary paradigm, it is not expected there to be an anisotropic expansion at measurable levels in the observable universe. And, given that inflation (canonical) isotropizes the universe very efficiently, leaving almost no anisotropy even in the very early universe (cosmic no-hair theorem~\cite{Wald:1983ky,Starobinsky:1982mr}), any detection of a non-zero shear scalar today would have far reaching consequences on the standard cosmology.\footnote{\label{footnote1} In our discussion here, we ignore the possibility of inflationary models with anisotropic hair, as it requires a departure from the canonical inflationary paradigm, which is part of the standard cosmological model intended to be investigated with its pure geometric generalization in this study. For example, non-scalar fields, in particular gauge fields, are ubiquitous in high-energy models of particle physics related to inflationary energy scales, and they can turn on in the background during inflation (or become relevant at the level of cosmic perturbations), and then lead to the violation of cosmic no-hair theorem~\cite{Watanabe:2009ct,Maleknejad:2012as,Maleknejad:2012fw}. And, departures from GR, e.g., quadratic curvature corrections, scalar-tensor theories of gravity such as Brans-Dicke theory with $\omega_{\rm BD}\sim-1$ (limit of the theory relevant to the string theories and $d$-brane constructions) can generate (or retain) anisotropic expansion~\cite{Barrow:2005qv,Barrow:2009gx,Akarsu:2019pvi,Mimoso:1995ge}. Also, there are tilted spatially homogeneous cosmological models (with the fluid flow lines not orthogonal to the surfaces of spatial homogeneity; the components of the fluid's peculiar velocity enter as further variables) for which the tilt does not vanish at late times (once the inflation ends, anisotropic modes again occur); to an observer moving with the fluid, such models will not seem to isotropize, yet may appear isotropic in another frame at late times~\cite{King:1972td,Goliath:1998na,Turner:1991dn,1984MNRAS.206..377E,Tsagas:2009nh} (see also Refs.~\cite{Ellis:1998ct,GEllisBook}).}

For instance, it is possible to generate anisotropic expansion well after decoupling, which also was suggested to address the low CMB anisotropy quadrupole~\cite{Bennett11,Schwarz:2015cma,Akrami:2019bkn}, though we ignore this possibility in this work, as it demands modifications beyond pure geometry, namely, the introduction of anisotropic stresses, effective well after decoupling, from some actual sources such as vector fields or extended theories of gravity such as Brans--Dicke theory (see Refs.~\cite{Campanelli:2006vb,Campanelli:2007qn,Koivisto:2007bp,Rodrigues:2007ny,Koivisto:2008xf,Campanelli:2009tk,Koivisto:2005mm,BeltranAlmeida:2019fou,Akarsu:2020pka,Akarsu:2020vii,Orjuela-Quintana:2020klr,Battye:2006mb,Koivisto:2008ig,Cooray:2008qn,Akarsu:2013dva,Chang:2013xwa,Koivisto:2014gia,Heisenberg:2016wtr,Yang:2018ubt,Pimentel89,Madsen88,Faraoni:2018qdr,Akarsu:2019pvi,Mimoso:1995ge,Mota:2007sz,appleby10,Appleby:2012as,Amendola:2013qna,Amendola:2016saw}). In tilted universes also, it is possible to generate/retain anisotropic expansion at late times (see footnote~\ref{footnote1}), which was suggested to address the discrepancies between the dipole amplitudes and directions from the different cosmological observations (e.g., between the radio and CMB dipoles) and even the $H_0$ and $S_8$ tensions, though we confine ourselves with orthogonal universes in this work, as tilted ones may go as far as to stop looking for corrections on top of $\Lambda$CDM and look for corrections on top of the Einstein--de Sitter universe; tilted observers (typical observers in a galaxy like the Milky Way) within the bulk flow can be misled into concluding that the universe is accelerating---it has been suggested that the indications for cosmic acceleration found in the SnIa data disappear, if a bulk flow induced anisotropy is allowed in the SnIa data, and thus, a bulk flow dipole aligned with the local bulk flow is identified while any monopole (which can be attributed to $\Lambda$) is consistent with zero~\cite{2011MNRAS.414..264C,Wilczynska:2020rxx,Migkas:2020fza,Chang:2017bbi,Secrest:2020has,Siewert:2020krp,Migkas:2021zdo,Rahman:2021mti,Heinesen:2021azp,Park:2017xbl,Park:2019emi,Macpherson:2021gbh,Yeung:2022smn,Krishnan:2021dyb,Krishnan:2021jmh,Luongo:2021nqh,Cea:2022mtf,Camarena:2022iae,Tsagas:2011wq,Tsagas:2015mua,Colin:2019opb,Asvesta:2022fts,Lin:2015rza,Li:2015uda} (see also Refs.~\cite{Perivolaropoulos:2021jda,Abdalla:2022yfr} for recent reviews and Refs.~\cite{Bennett11,Schwarz:2015cma,Akrami:2019bkn} for hints of unexpected features in CMB and other independent cosmological data types). For such reasons, the interest in anisotropic cosmologies has never ceased and, moreover, has recently begun to rise again.

Besides these more fundamental aspects we discussed above with regard to considering the pure geometric generalization of $\Lambda$CDM by introducing spatial curvature and/or anisotropic expansion, it is tempting to explore whether these geometric generalizations by alone could be beneficial in addressing the tensions related to the $\Lambda$CDM model. Indeed, it was recently reported in Ref.~\cite{Akarsu:2019pwn} that the non-tilted Bianchi Type-I spacetime extension of $\Lambda$CDM, which simply allows different scale factors in three orthogonal directions on top of the spatially flat RW spacetime, relaxes the $H_0$ tension; it predicts $H_0\sim70.0$ km s${}^{-1}$ Mpc${}^{-1}$, which agrees, within $2\sigma$, with both the $\Lambda$CDM Planck 2018 prediction $H_0=67.27 \pm 0.60$ km s${}^{-1}$ Mpc${}^{-1}$~\cite{Planck:2018vyg} and the Hubble Space Telescope (HST) and the SH0ES team local measurements, e.g., $H_0=73.04 \pm 1.04$ km s${}^{-1}$ Mpc${}^{-1}$~\cite{Riess:2021jrx} (see also Refs.~\cite{Freedman:2019jwv,Riess:2020fzl}). This observation naturally makes us wondering whether the shear scalar, which is non-negative by definition, can compensate the negative energy density-like contribution of the positive (closed space) spatial curvature to the Friedmann equation, and then avoid the drastically reduced $H_0$ value accompanying the closed space prediction of the CMB with or without lensing (Lens) data. Thus, here we will study the extension of $\Lambda$CDM allowing both non-zero spatial curvature and anisotropic expansion for mainly two reasons: (i) To test whether the latest observational data still support spatial flatness and/or isotropic expansion in this case, and, if not, (ii) to explore the roles of spatial curvature and expansion anisotropy (via its stiff fluid-like contribution to the Friedmann equation) in addressing some of the current cosmological tensions associated with the $\Lambda$CDM model.

It is also pertinent to mention that the model-independent upper bounds on the present-day expansion anisotropy in terms of its corresponding present-day density parameter ($\Omega_{\sigma 0}$) are of the order of $\mathcal{O}(10^{-3})$, e.g., from type Ia Supernovae~\cite{Campanelli:2010zx,Wang:2017ezt,Jimenez:2014jma,Soltis:2019ryf,Zhao:2019azy,Hu:2020mzd,Kalus:2012zu}. This is consistent with the model-dependent constraints, $\Omega_{\sigma 0}\lesssim10^{-3}$, obtained from the $H(z)$ and/or SNIa data (relevant to $z\lesssim 2.4$) by considering the stiff fluid-like behavior of expansion anisotropy on top of $\Lambda$CDM~\cite{Akarsu:2019pwn,Amirhashchi:2018nxl}. This amount of expansion anisotropy today, within the simplest anisotropic, i.e., the Bianchi type I spacetime, generalization of $\Lambda$CDM, implies the domination of expansion anisotropy at $z\sim 10$ and hence the spoilt of the successful description of the earlier ($z\gtrsim 10$) universe. Indeed, the model-dependent upper bounds are usually much tighter; spanning a range from $\Omega_{\sigma 0}\lesssim10^{-11}$ to $10^{-23}$~\cite{Ellis:1998ct,GEllisBook}. When the stiff fluid-like behavior of expansion anisotropy is considered on top of $\Lambda$CDM, the constraint $\Omega_{\sigma 0}\lesssim 10^{-3}$ from the combined $H(z)$ and SnIa Pantheon data set (relevant to $z\lesssim 2.4$) is tightened to $\Omega_{\sigma 0}\lesssim 10^{-15}$ when the combined CMB+BAO data set (relevant to $z\sim1100$) is also included, and it is further constrained to $\Omega_{\sigma 0}\lesssim 10^{-23}$ not to spoil the successes of the standard Big Bang Nucleosynthesis (BBN) (relevant to $z\sim10^9$)~\cite{Akarsu:2019pwn}. We have also the typical upper bound $\Omega_{\sigma 0} \lesssim 10^{-20}$ derived from the observed CMB quadrupole temperature fluctuation $\Delta T/T \sim 10^{-5}$, which provides an upper bound at the same order of magnitude on the expansion anisotropy at the recombination era ($\sqrt{\Omega_{\sigma}
^{\rm rec}} \sim 10^{-5}$ at $z_{\rm rec}\sim10^3$)~\cite{Martinez95,Bunn:1996ut,Kogut:1997az,Barrow:1997sy,Saadeh:2016sak}. The upper bounds obtained from CMB reaches the level $\Omega_{\sigma 0}\lesssim10^{-22}$ from the rigorous analyses of the full Planck CMB temperature and polarization data~\cite{Pontzen16,Saadeh:2016bmp,Saadeh:2016sak}. The BBN light element abundances also provide tight upper bounds, but at different levels depending on the presence of spatial curvature anisotropy in addition to expansion anisotropy; namely, $\Omega_{\sigma 0}\lesssim10^{-22}$ and $\Omega_{\sigma 0}\lesssim10^{-21}$ for the Bianchi type I and type V spacetimes (the anisotropic generalizations of the spatially flat and open RW spacetimes, respectively, and yield isotropic spatial curvature), and these relax to $\Omega_{\sigma 0}\lesssim10^{-13}$ for the Bianchi type IX and type VII$_0$ spacetimes (the anisotropic generalizations of the spatially closed and open RW spacetimes, respectively, and yield anisotropic spatial curvature)~\cite{Barrow:1976rda,Campanelli:2011aa,Barrow:1997sy}.

The paper is structured as follows: In Section~\ref{S2}, we describe the basic equations of the model; in Section~\ref{data}, we present the data/methodology used; in Section~\ref{results}, we discuss the results; and in Section~\ref{conclusions}, we summarize our findings.

\section{Basic Equations and the Model}\label{S2} 

We study a two parameter extension of the $\Lambda$CDM model that allows both non-zero spatial curvature and anisotropic expansion, both which are geometric corrections on top of the base $\Lambda$CDM model. The fundamental idea here is to extend $\Lambda$CDM just by relaxing its spatially flat RW metric assumption to a more realistic one. Accordingly, we consider the generalized Friedmann equation which, compared to the usual one, includes two new parameters, viz., the spatial curvature (pertaining to deviation from flat space) and the shear scalar (pertaining to deviation from isotropic expansion);
\begin{equation}\label{GFREqn}
    3H^2+\frac{1}{2}\, {^{3}R}-\sigma^2=8\pi G \rho,
\end{equation}
where $H$ is the average Hubble parameter defined as $H\equiv\frac{1}{3}\Theta$ ($\Theta={\rm D}^{\mu}u_{\mu}$ is the volume expansion rate with ${\rm D}^{\mu}$ being the fully orthogonally projected covariant derivative and $u_{\mu}$ is the four-velocity that satisfies $u_\mu u^{\mu}=-1$ and $\nabla_{\nu}u^{\mu}u_{\mu}=0$), $^{3}R$ is the 3-Ricci scalar (the Ricci scalar of the spatial section of the spacetime metric), $\sigma^2$ is the shear scalar quantifying the anisotropic expansion ($\sigma^2=\frac{1}{2}\sigma_{\alpha\beta} \sigma^{\alpha\beta}$, where $\sigma_{\alpha\beta}=\frac{1}{2}(u_{\mu;\nu}+u_{\nu;\mu}) h^{\mu}_{\:\alpha} h^{\nu}_{\:\beta}-\frac{1}{3} u^{\mu}_{\:;\mu} h_{\alpha\beta}$ is the shear tensor with $h_{\mu\nu}=g_{\mu\nu}+u_{\mu}u_{\nu}$ being the projection tensor into the instantaneous rest frame of comoving observers), $G$ is the Newton's gravitational constant, and $\rho$ is the energy density of the total physical ingredient of the universe (see Refs.~\cite{Collins,Collins:1972tf,Ellis:1998ct,GEllisBook}). In this work, we limit our investigations to the orthogonal (non-tilted) models---the fluid flow lines are orthogonal to the surfaces of spatial homogeneity---, as the tilted ones---the fluid flow lines are not orthogonal to the surfaces of spatial homogeneity---are significantly more complicated and bringing in the components of the fluid’s peculiar velocity as additional variables on top of the the fluid’s energy density and pressure; the spatial curvature and shear scalar already introduce two new free parameters to be constrained on top of the $\Lambda$CDM model (see Refs.~\cite{Ellis:1998ct,GEllisBook}).

We use the Einstein field equations (EFE) of GR;
\begin{equation}\label{fieldeqn}   
G_{\mu\nu}\equiv R_{\mu\nu}-\frac{1}{2}g_{\mu\nu} R= 8\pi GT_{\mu\nu},
\end{equation}    
where $G_{\mu\nu}$ is the Einstein tensor, $R_{\mu\nu}$ is the Ricci tensor, $R$ is the Ricci scalar, and $g_{\mu\nu}$ is the metric tensor. We suppose all types of matter distributions (viz., the usual cosmological sources such as radiation, baryons, etc.) are perfect fluids with no peculiar velocities, described by the energy-momentum tensor (EMT) of the form $T^{\nu}_{\,\,\mu} = \text{diag} [-\rho, p, p, p]$, 
where $\rho$ and $p$ are the energy density and pressure, respectively. EFE~\eqref{fieldeqn} satisfy $T^{\mu\nu}_{\;\;\;;\nu}=0$ (the conservation equation for the total EMT representing all sources in the universe) via $G^{\mu\nu}_{\;\;\;;\nu}=0$ as a consequence of the twice-contracted Bianchi identity. This leads to the continuity equation for the total EMT: $\dot{\rho}+3H(\rho+p)=0$, where the dot represents the derivative with respect to the proper time $t$. We consider the usual cosmological sources, namely, pressureless fluid (CDM, baryons) described by the equation of state (EoS) $p_{\rm m}/\rho_{\rm m}=0$, radiation (photons $\gamma$, neutrinos $\nu$) described by the EoS $p_{\rm r}/\rho_{\rm r}=\frac{1}{3}$, and DE mimicked by a cosmological constant (viz., $\rho_{\Lambda}=\frac{\Lambda}{8\pi G}=\rm const$) described by the EoS $p_{\Lambda}/\rho_{\Lambda}=-1$.
We suppose these sources interact only gravitationally, so that the continuity equation is satisfied separately by each source, and this leads to
\begin{equation}\label{sources}
\rho\equiv\rho_\text{r}+\rho_\text{m}+\rho_{\Lambda}= \rho_{\text{r}0}s^{-4}+\rho_{\text{m}0} s^{-3}+\rho_{\Lambda}
\end{equation} 
where $s\equiv v^{1/3}$ is the mean scale factor with $v$ being the volume scale factor. The present-day value of $s$ is set as $s_0=1$. Here and onward, a subscript 0 attached to any quantity implies its value in the present-day universe. Note that the present-day photon energy density $\rho_{\text{r}0}$ is extremely well constrained by the absolute CMB monopole temperature measured by FIRAS $T_0 =2.7255 \pm 0.0006 \,{\rm K}$~\cite{Fixsen09}, and thereby its present-day density parameter given by $\Omega_{\rm r0}\equiv \rho_{\rm r0}/3H_0^2=2.469\times10^{-5}h^{-2}(1+0.2271N_{\rm eff})$---where $h=H_0/100\, {\rm km\,s}^{-1}{\rm Mpc}^{-1}$ is the dimensionless reduced Hubble constant and $N_{\rm eff}=3.046$ is the standard number of effective neutrino species with minimum allowed mass $m_{\nu}=0.06$~eV---is not subject to our observational analysis.

To fully determine the modified Friedmann equation~\eqref{GFREqn}, viz., $H(s)$, however, it is necessary to determine the evolution of the two remaining parameters, namely, the functions ${^{3}R(s)}$ and $\sigma^2(s)$. In what follows, we first present explicit constructions of different cases extending $\Lambda$CDM and then write a single $H(s)$ function including all these cases to be constrained with the observational data. To do so, we start with the spatially maximally symmetric metric, the RW  metric, for the purpose of comparison with the three spatially homogeneous but not necessarily isotropic metrics, namely, the Bianchi type I, Bianchi type V, and Bianchi type IX metrics which are the simplest anisotropic generalizations of the spatially flat, open, and closed RW metrics, respectively.

\subsection{Spatially homogeneous and isotropic universe}

We start with the RW metric; respecting the cosmological principle, it yields maximally symmetric spatial section (not necessarily flat), and it can be written in Cartesian coordinates as follows:
\begin{equation}\label{RW}   
    {\rm d}s^2=-\text{d}t^2 + s^2\frac{{\rm d}x^2 + {\rm d}y^2 + {\rm d}z^2}{\left[1+\frac{\kappa}{4}(x^2+y^2+z^2)\right]^2},
\end{equation}
where $\kappa<0$, $\kappa=0$, and $\kappa>0$ correspond to spatially open, flat, and closed universes, respectively. EFE~\eqref{fieldeqn} for the RW metric~\eqref{RW} lead to the following set of differential equations, viz., the usual Friedmann equations;
\begin{align}
    3\frac{{\dot s}^2}{s^2}+3\frac{\kappa}{s^2}=8\pi G \rho,\\
    -2\frac{\ddot s}{s}-\frac{{\dot s}^2}{s^2}-\frac{\kappa}{s^2}=8\pi G p.
\end{align}
The shear scalar and 3-Ricci scalar for the RW metric~\eqref{RW} read
\begin{equation}
    \sigma^2=0\quad\textnormal{and}\quad {^{3}R}={^{3}R}_0s^{-2},
\end{equation}
where ${^{3}R}_0=6\kappa$.

In what follows, we will refer this well known generalization of the standard $\Lambda$CDM model, which simply allows non-zero spatial curvature on top it, to as the $o\Lambda$CDM model, where "$o$" stands for "spatial curvature".

\subsection{Spatially homogeneous, flat and anisotropic universe}     
  
We continue with the Bianchi type I metric, which simply allows different scale factors in three orthogonal directions on top of the spatially flat ($\kappa=0$) RW metric, while preserving isotropic spatial curvature;
\begin{equation}\label{BI}
    \text{d}s^2=-\text{d}t^2 + a^2\text{d}x^2 + b^2\text{d}y^2 + c^2\text{d}z^2,
\end{equation} 
where $\{a,b,c\}=\{a(t),b(t),c(t)\}$ are the directional scale factors along the principal axes $\{x,y,z\}$ and are functions of $t$ only. The corresponding average expansion scale factor is $s=(abc)^{\frac{1}{3}}$ and from which the average Hubble parameter is defined as  $H=\frac{\dot{s}}{s}=\frac{1}{3}\left(H_x + H_y+H_z\right)$, where $H_x =\frac{\dot{a}}{a}$, $H_y =\frac{\dot{b}}{b}$, and $H_z =\frac{\dot{c}}{c}$ are the directional Hubble parameters defined along the $x$-, $y$-, and $z$-axes, respectively.

EFE~\eqref{fieldeqn} for the Bianchi type I metric~\eqref{BI} lead to the following set of differential equations:
\begin{align}
    \frac{\dot{a}}{a}\frac{\dot{b}}{b}+ \frac{\dot{b}}{b}\frac{\dot{c}}{c}+ \frac{\dot{a}}{a}\frac{\dot{c}}{c}&=8\pi G \rho \label{BI.1}, \\ 
    -\frac{\Ddot{b}}{b}-\frac{\Ddot{c}}{c}- \frac{\dot{b}}{b}\frac{\dot{c}}{c}&=8\pi G p \label{BI.2}, \\ 
      -\frac{\Ddot{a}}{a}-\frac{\Ddot{c}}{c}- \frac{\dot{a}}{a}\frac{\dot{c}}{c}&=8\pi G p \label{BI.3} , \\
        -\frac{\Ddot{a}}{a}-\frac{\Ddot{b}}{b}- \frac{\dot{a}}{a}\frac{\dot{b}}{b}&=8\pi G p \label{BI.4} . 
\end{align}
The shear scalar, in terms of the directional Hubble parameters, reads
\begin{equation}\label{BIshear} 
\sigma^2 = \frac{(H_x -H_y)^2 + (H_y - H_z)^2 + (H_z -H_x)^2}{6}.
\end{equation} 
We can rewrite equations~\eqref{BI.2}-\eqref{BI.4} as follows; 
\begin{align} 
     \dot{H}_x-\dot{H}_y+3H(H_x-H_y)=0,\label{hubble1} \\
      \dot{H}_y-\dot{H}_z+3H(H_y-H_z)=0,\label{hubble2} \\  
       \dot{H}_z-\dot{H}_x+3H(H_z-H_x)=0\label{hubble3},    
\end{align} 
and then, using these along with the time derivative of $\sigma^2$ given in Eq.~\eqref{BIshear}, we reach the shear propagation equation
\begin{equation}\label{BIshearp}     
    \dot{\sigma}+3H\sigma=0. 
\end{equation}     
Its integration yields
\begin{equation}\label{BIsigma2}  
\sigma^2 = \sigma_0^2 s^{-6}.
\end{equation}
The 3-Ricci scalar for the Bianchi type I metric~\eqref{BI} is null;
\begin{equation}
^{3}R=0,
\end{equation}
which corresponds to a flat space, likewise the spatially flat ($\kappa=0$) RW metric.

In what follows, we refer to the Bianchi type I extension of the standard $\Lambda$CDM model as An-$\Lambda$CDM model, where "An" stands for "anisotropic".

\subsection{Spatially homogeneous, open and anisotropic universe}  

We next continue with the Bianchi type V metric, which simply allows different scale factors in three orthogonal directions on top of the spatially open ($\kappa<0$) RW metric, while preserving isotropic spatial curvature;
\begin{equation}\label{BV}     
\begin{aligned} 
\text{d}s^2=-\text{d}t^2 + a^2e^{-2mz}\text{d}x^2 + b^2e^{-2mz}\text{d}y^2+c^2\text{d}z^2,            
\end{aligned}      
\end{equation}
where $m\neq0$ is a non-zero real constant. It can be observed that Bianchi type V metric~\eqref{BV} reduces to the Bianchi type I metric~\eqref{BI} when $m=0$.               

EFE~\eqref{fieldeqn} for the Bianchi type V metric~\eqref{BV} lead to the following set of differential equations:     
\begin{align}  
    \frac{\dot{a}}{a}\frac{\dot{b}}{b}+ \frac{\dot{a}}{a}\frac{\dot{c}}{c}+ \frac{\dot{b}}{b}\frac{\dot{c}}{c}-3\frac{m^2}{c^2}  &= 8\pi G \rho ,\label{BV.1} \\   
    -\frac{\Ddot{b}}{b}-\frac{\Ddot{c}}{c}-\frac{\dot{b}}{b}\frac{\dot{c}}{c}+\frac{m^2}{c^2} &= 8\pi G p  ,\label{BV.2} \\  
      -\frac{\Ddot{a}}{a}-\frac{\Ddot{c}}{c}-\frac{\dot{a}}{a}\frac{\dot{c}}{c}+\frac{m^2}{c^2}&= 8\pi G p ,\label{BV.3} \\   
       -\frac{\Ddot{a}}{a}-\frac{\Ddot{b}}{b}-\frac{\dot{a}}{a}\frac{\dot{b}}{b}+\frac{m^2}{c^2}&= 8\pi G p ,\label{BV.4} \\
       m\left(2\frac{\dot{c}}{c}-\frac{\dot{a}}{a}-\frac{\dot{b}}{b}\right)&=0. \label{BV.5}
\end{align}
In Eq.~\eqref{BV.5}, we see the component $G^0_{\;\;3}=8\pi G\, T^0_{\;\;3}$, where we take $T^0_{\;\;3}=0$ as we consider only perfect fluid distributions; i.e., $T^{0i}$ the energy flux density (which equals the momentum density $T^{i0}$) is not allowed. As a result of this, solving Eq.~\eqref{BV.5}, it turns out that
\begin{align}\label{BVc}      
     c=c_1 (ab)^\frac{1}{2},        
\end{align}  
where $c_1>0$ is the integration constant. As we set $s_0=1$, the present-day values of the directional scale factors, i.e., $a_0>0$, $b_0>0$, and $c_0>0$, must satisfy $a_0b_0c_0=1$, and therefore $c_1=(a_0   b_0)^{-\frac{3}{2}}$. Accordingly, we re-express the average expansion scale factor as $s=(a_0 b_0)^{-\frac{1}{2}}(ab)^\frac{1}{2}$, which in turn implies $c=a_0b_0 s$. Using Eq.~\eqref{BVc}, the set of field equations~\eqref{BV.1}-\eqref{BV.4} reduces to
\begin{align}
2\frac{\dot{a}}{a}\frac{\dot{b}}{b}+\frac{1}{2}\frac{\dot{a}^2}{a^2}+\frac{1}{2}\frac{\dot{b}^2}{b^2}-3\frac{m^2}{ab}&= 8\pi G \rho ,\label{BV.21} \\ 
-\frac{3}{2}\frac{\Ddot{b}}{b}-\frac{1}{2}\frac{\Ddot{a}}{a}+\frac{1}{4}\frac{\dot{a}^2}{a^2}-\frac{1}{4}\frac{\dot{b}^2}{b^2}-\frac{\dot{a}}{a}\frac{\dot{b}}{b}+\frac{m^2}{ab} &= 8\pi G p \label{BV.22},\\
-\frac{3}{2}\frac{\Ddot{a}}{a}-\frac{1}{2}\frac{\Ddot{b}}{b}-\frac{1}{4}\frac{\dot{a}^2}{a^2}+\frac{1}{4}\frac{\dot{b}^2}{b^2}-\frac{\dot{a}}{a}\frac{\dot{b}}{b}+\frac{m^2}{ab} &= 8\pi G p \label{BV.23}, \\
 -\frac{\Ddot{a}}{a}-\frac{\Ddot{b}}{b}-\frac{\dot{a}}{a}\frac{\dot{b}}{b}+\frac{m^2}{ab}&= 8\pi G p \label{BV.24}.
\end{align}    
Also, the average Hubble parameter reduces to
\begin{equation}\label{BVhubble}
    H=\frac{1}{2}(H_x+H_y),
\end{equation}
and the shear scalar~\eqref{BIshear} reduces to
\begin{equation}\label{BVshear2}       
\sigma^2 = \frac{1}{4}(H_x-H_y)^2. 
\end{equation}      
Subtracting Eq.~\eqref{BV.24} from Eq.~\eqref{BV.22} [or Eq.~\eqref{BV.23}] we obtain
\begin{equation}\label{BVsubtract}
    2\frac{\Ddot{a}}{a}-2\frac{\Ddot{b}}{b}+\frac{\dot{a}^2}{a^2}-\frac{\dot{b}^2}{b^2}=0,
\end{equation}  
which, after using $H$~\eqref{BVhubble} and  $\sigma^2$~\eqref{BVshear2}, leads to the shear propagation equation in the form  
\begin{equation}\label{BVshearp}
\dot{\sigma}+3H\sigma=0.    
\end{equation} 
From its integration, we obtain
 \begin{equation}\label{BVsigma2}  
\begin{aligned}
     \sigma^2=\sigma_0^2s^{-6},  
\end{aligned}
\end{equation}
as in the case of the Bianchi type I metric, see Eq.~\eqref{BIsigma2}. The 3-Ricci scalar for the Bianchi type V metric~\eqref{BV} reads
\begin{equation}
{^{3}R}={^{3}R}_0s^{-2},
\end{equation}
which is negative definite, as here ${^{3}R}_0=-6c_0m^2<0$, corresponding to an open space; e.g., to compare with the spatially open ($\kappa<0$) RW metric, when we set $a=b=c$ (so that $a_0=b_0=c_0=1$ as $s_0=1$), we find $-m^2=\kappa<0$.
             
\subsection{Spatially homogeneous, closed and anisotropic spacetime} 
Finally, we consider the Bianchi type IX metric, which allows a different scale factor along only one [as, for simplicity sake, we consider the locally rotationally symmetric (LRS) case] of the principal axes on top the spatially closed RW metric, described by
\begin{equation}\label{BIX}
\begin{aligned}  
\text{d}s^2=&\,-\text{d}t^2 + a^2\text{d}x^2 + b^2\text{d}y^2\\&
+\left(b^2\sin^2{2ny}+a^2\cos^2{2ny}\right)\text{d}z^2\\&
-2a^2\cos{2ny}\,\text{d}x\text{d}z,                
\end{aligned}        
\end{equation}  
where $n\neq0$ is a non-zero real constant. It reduces to the LRS Bianchi type I metric (i.e., the Bianchi type I metric~\eqref{BI} with $c=b$, leading to $H_z=H_y$) for $n=0$. We note that, unlike the Bianchi type I and type V metrics, this metric yields anisotropic spatial curvature, which would alter the evolution $\sigma^2\propto s^{-6}$ of the shear scalar. The corresponding average expansion scale factor and average Hubble parameter respectively read as  $s=(ab^2)^{\frac{1}{3}}$ and  $H=\frac{1}{3}\left(H_x + 2H_y\right)$.    

EFE~\eqref{fieldeqn} for the Bianchi type IX metric~\eqref{BIX} lead to the following set of differential equations:
\begin{align}    
     2\frac{\dot{a}}{a}\frac{\Dot{b}}{b}+\frac{\Dot{b}^2}{b^2}+n^2\left(\frac{4}{b^2}-\frac{a^2}{b^4}\right) &= 8\pi G \rho , \label{BIX.1} \\     
      -2\frac{\Ddot{b}}{b}-\frac{\Dot{b}^2}{b^2}-n^2\left(\frac{4}{b^2}-3\frac{a^2}{b^4}\right) &= 8\pi G p , \label{BIX.2} \\   
       -\frac{\Ddot{a}}{a}-\frac{\Ddot{b}}{b}-\frac{\dot{a}}{a}\frac{\Dot{b}}{b}-n^2\frac{a^2}{b^4} &= 8\pi G p.      \label{BIX.3}
\end{align}

The shear scalar, in terms of the directional Hubble parameters, reads
\begin{equation}\label{BIXshear}
 \sigma^2=\frac{1}{3}(H_x-H_y)^2.      
\end{equation}    
Subtracting Eq.~\eqref{BIX.3} from Eq.~\eqref{BIX.2}, we obtain    
\begin{equation} \label{BIXsubstract}
 \frac{\Ddot{b}}{b}+ \frac{\Dot{b^2}}{b^2} - \frac{\Ddot{a}}{a}-\frac{\dot{a}}{a}\frac{\Dot{b}}{b}+4n^2\left(\frac{1}{b^2}-\frac{ a^2}{b^4}\right)=0.              
\end{equation} 
Rewriting the first four terms here in terms of $H$ and  $\sigma^2$ given in Eqn.~\eqref{BIXshear}, we obtain the shear propagation equation as follows:
\begin{equation}\label{BIXshearp}
\dot{\sigma}+3H\sigma=\frac{4n^2}{\sqrt{3}}  \left(\frac{1}{b^2}-\frac{a^2}{b^4}\right).   
\end{equation}  

 The 3-Ricci scalar for the Bianchi type IX metric~\eqref{BIX} reads
\begin{equation}\label{riccisBIX}
   {^{3}R}= 2n^2\left(\frac{4}{b^2}-\frac{a^2}{b^4}\right),
\end{equation}
with the present day value, ${^{3}R}_0=2n^2\left(\frac{4}{b^2_0}-\frac{a^2_0}{b^4_0}\right)$.

We now suppose that the universe, all the way to the  recombination era, exhibits similar expansion histories along all the principal axes, i.e.,
\begin{equation} \label{assumption}  
     a\simeq b \simeq s,
\end{equation}
so that $a_0\simeq b_0\simeq s_0$, satisfying $s^3_0=a_0b_0^2=1$. This is in line with the observed CMB quadrupole temperature fluctuation ($\Delta T/T \sim 10^{-5}$) setting an upper limit at the same order of magnitude on the expansion anisotropy at the recombination era ($\sqrt{\Omega_{\sigma}
^{\rm rec}} \sim 10^{-5}$ at $z_{\rm rec}\sim10^3$), which transforms to the typical upper limit $\Omega_{\sigma 0} \sim 10^{-20}$ derived from $\sigma^2\propto s^{-6}$ relation~\cite{Martinez95,Bunn:1996ut,Kogut:1997az,Barrow:1997sy,Saadeh:2016sak}.
Accordingly, the shear propagation equation~\eqref{BIXshearp} can approximately be written as follows:
\begin{equation}\label{BIXshearp}
\dot{\sigma}+3H\sigma\simeq0,
\end{equation}
from which the shear scalar approximately reads  
\begin{equation}
    \sigma^2\simeq\sigma^2_0s^{-6}.
\end{equation}
Similarly, the 3-Ricci scalar~\eqref{riccisBIX} can approximately be written as follows:
\begin{equation}
   {^{3}R}\simeq{^{3}R}_0 s^{-2},
\end{equation}
which is positive definite, as here ${^{3}R}_0\simeq 6 n^2>0$, corresponding to a closed space; e.g., to compare with the spatially closed ($\kappa>0$) RW metric, when we set $a=b=s$ (so that $a_0=b_0=1$ as $s_0=1$) we find $n^2=\kappa>0$. Thus, in the case, for small anisotropies, the modified Friedmann equation~\eqref{GFREqn} can approximately be written by using $\sigma^2=\sigma^2_0 s^{-6}$, likewise in the cases of the Bianchi type I and type V metrics, and  ${^{3}R}={^{3}R}_0 s^{-2}>0$, likewise in the case of the spatially closed ($\kappa>0$) RW metric.

\subsection{Spatially homogeneous, but not necessarily flat and isotropic, spacetime extension of the $\Lambda$CDM model}
\label{sec:thefinalmodel}

The generalized Friedmann equation for the spatially homogeneous, but not necessarily flat and isotropic, spacetime metric and the non-tilted perfect fluids that we will confront with the observational data, can be written as follows:
\begin{equation}
    3H^2=8\pi G (\rho+\rho_{\kappa}+\rho_{\sigma}),
\end{equation}
where $\rho$ is the energy density of the total physical ingredient of the universe given in Eq.~\eqref{sources}, while $\rho_{\kappa}$ and $\rho_{\sigma}$ are the effective energy densities corresponding to spatial curvature (viz., the 3-Ricci scalar) and expansion anisotropy (viz., the shear scalar) through the following definitions, correspondingly,
\begin{equation}
    \rho_{\kappa}=\frac{-\frac{1}{2}\, {^{3}R}}{8\pi G} \quad\textnormal{and}\quad \rho_{\sigma}=\frac{\sigma^2}{8\pi G}.
\end{equation}
Consequently, using $\rho$ from Eq.~\eqref{sources} along with the 3-Ricci scalars and the shear scalars obtained above, we reach the following single generalized Friedmann equation, describing the spatially homogeneous, but not necessarily flat and isotropic, and non-tilted extension of the $\Lambda$CDM model, to be confronted with the observational data:
\begin{equation}\label{themodel} 
    \frac{H^2(s)}{H_0^2}  =\Omega_{\sigma0}s^{-6}+\Omega_{\text{r}0}s^{-4}+\Omega_{\text{m}0}s^{-3}+ \Omega_{\kappa0}s^{-2}+\Omega_{\Lambda0},   
\end{equation}  
where the present-day density parameters satisfy $\Omega_{\sigma0} + \Omega_{\text{r}0}+ \Omega_{\text{m}0}+\Omega_{\kappa0}+\Omega_{\Lambda0}=1$ and among which $\Omega_{\kappa0}$ and $\Omega_{\sigma0}$ are geometric terms; namely, $\Omega_{\kappa0}$ stands for the spatial curvature, which can be $\Omega_{\kappa0}=0$, $\Omega_{\kappa0}>0$, and $\Omega_{\kappa0}<0$, corresponding to the spatially flat, open, and closed universes, respectively, and $\Omega_{\sigma0}\geq0$ stands for expansion anisotropy (viz., the shear scalar), which is non-negative definite.

In what follows, we call this model, described by Eq.~\eqref{themodel}, An-$o\Lambda$CDM model. We note that the An-$o\Lambda$CDM model is \textit{mathematically} equivalent to adding stiff/Zeldovich fluid~\cite{Zeldovich:1961sbr,Barrow1978} and a fluid of disordered cosmic strings (CS)~\cite{Vilenkin:2000jqa} on top of the $\Lambda$CDM model, though these two models are \textit{physically} different. Namely, the new terms $\Omega_{\sigma 0}s^{-6}$ and $\Omega_{\kappa0}s^{-2}$ on top of the $\Lambda$CDM model contribute to the Friedmann equation like sources called stiff fluid (described by the EoS of the form $p_{\rm s}/\rho_{\rm s}=1$) and string gas (viz., CS, described by the EoS of the form $p_{\rm cs}/\rho_{\rm cs}=-1/3$), respectively, but in the An-$o\Lambda$CDM model, these terms are geometric in origin and arise from the anisotropic expansion and spatial curvature, correspondingly. A few direct consequences of being physically different are that these two models behave differently at the perturbative level; unlike $\Lambda$CDM+stiff fluid, the Hubble constant in An-$o\Lambda$CDM depends on the direction of view in the sky; unlike $\Lambda$CDM+cosmic strings, the interrelations of cosmological distance measures in An-$o\Lambda$CDM are different from those in $\Lambda$CDM (e.g., the comoving angular diameter distance is no longer proportional to the line-of-sight comoving distance for non-flat space). In what follows in this section, elaborating some aspects of these points little bit more, we will lay out the approach we take in this work.

In the case of the An-$o\Lambda$CDM model, the $H(s)$ in~\eqref{themodel} corresponds to the average expansion rate and the expansion rates along the different principal axes need not necessarily be the same. Yet, we can define an average redshift as $z=-1+\frac{1}{s}$ from the average expansion scale factor $s$, and then rewrite this equation in terms of $z$ for observational analysis purposes at the background level. The observational analysis relying on this approach, which we will follow in this work, would then reflect \textit{only} the stiff fluid-like effect of the expansion anisotropy on the average expansion rate, $H(s)$, of the An-$o\Lambda$CDM model. This in turn implies that the constraints, that we will obtain in this work, can also be adopted for the cosmological models that contain a stiff ``fluid'' on top of the $\Lambda$CDM and $o\Lambda$CDM models.\footnote{\label{fn:2} Cosmologies with a stiff fluid era stemming from different underpinning theories are ubiquitous, but differ in details (in particular, their perturbations and hence their imprints on the CMB anisotropies are different~\cite{Hu:1998kj,Battefeld:2004jh}), see Ref.~\cite{Chavanis:2014lra} for a general discussion (theoretical) on the $\Lambda$CDM+stiff fluid cosmology, and Refs.~\cite{Zeldovich:1961sbr,Barrow1978,Copeland:1997et,Tamanini:2014mpa,Odintsov:2017cfr,Poulin:2018cxd,Smith:2019ihp,Kamionkowski:2022pkx,Jiang:2022uyg,Mavromatos:2020kzj,Banks:2001px,Banks:2003ta,Banks:2004cw,Mimoso:1994wn,Chavanis:2014hba,Andersson:2000cv,Lankinen:2016ile,Poulin:2018cxd,Smith:2019ihp,Kamionkowski:2022pkx,Easson:2021amu,Benisty:2021sul,Akarsu:2020vii,Akarsu:2018zxl,Dutta:2017fjw,Benisty:2021cin,Copeland:1994vi,Lidsey:1999mc,Maartens:2003tw} for some particular models of this type that include sources (actual or effective) that resemble stiff fluid (permanently or temporarily).}

For the observational analysis of the models, we will use SNIa and cosmic chronometers (CC) data relevant to the late universe (viz., $z\lesssim2.4$), and BAO and CMB data relevant to the universe since the recombination (viz., the drag redshift $z_{\rm d}\sim1060$ and last scattering redshift $z_*\sim 1090$ involved in BAO and CMB data analyses, respectively). In order to quantify the amount/effects of expansion anisotropy within the framework of the considered theoretical model described by Eq.~\eqref{themodel} using the observational data, we have two main options: (i) We can perform a full likelihood treatment of the data; so that, using the late time observational data, which incorporates the location of objects (e.g., SNIa) in the sky, and/or full CMB data, which requires involving perturbations in our investigations (see Ref.~\cite{Pereira:2007yy} for the theory of cosmological perturbations in an anisotropic universe and Refs.~\cite{Saadeh:2016sak,Pontzen16,Saadeh:2016bmp} for the constraints on anisotropic expansion from the full Planck CMB temperature and polarization spectra measurements) can in principle be able to distinguish between anisotropic expansion and actual stiff fluid. (ii) We can explore the observational constraints on the model at the background level. We prefer to proceed with the second approach for the reasons explained in detail below.

A full likelihood treatment of the late time observational data, which incorporates the location of objects (e.g., SNIa) in the sky, can in principle be able to distinguish between anisotropic expansion and stiff fluid, as these likelihoods use the differing $z$ and $H(z)$ values along different axes of line of sight and incorporate these in an object-by-object differentiation of the cosmological distances entering the likelihoods. Such an approach would inevitably be necessary, e.g., in tilted models, since in this case determining the dipole amplitudes and directions from different cosmological observations would have far-reaching consequences that may be beyond seeking corrections on top of $\Lambda$CDM, as discussed also in the introduction (Sec.~\ref{sec:intro}). However, in our case, assuming non-tilted perfect fluids as in the standard cosmological model, this would work most efficiently if the shear scalar was not predicted to resemble a stiff fluid ($\sigma^2\propto s^{-6}$) by GR, but a source that exhibits a much flatter average expansion scale factor dependence. For example, suppose the shear scalar resembles dust ($\sigma^2\propto s^{-3}$) via an anisotropic dark energy or a modified theory of gravity (see Refs.~\cite{Campanelli:2006vb,Campanelli:2007qn,Koivisto:2007bp,Rodrigues:2007ny,Koivisto:2008xf,Campanelli:2009tk,Koivisto:2005mm,BeltranAlmeida:2019fou,Akarsu:2020pka,Akarsu:2020vii,Orjuela-Quintana:2020klr,Battye:2006mb,Koivisto:2008ig,Chang:2013xwa,Cooray:2008qn,Akarsu:2013dva,Koivisto:2014gia,Heisenberg:2016wtr,Yang:2018ubt,Pimentel89,Madsen88,Faraoni:2018qdr,Akarsu:2019pvi,Mimoso:1995ge,Mota:2007sz,appleby10,Appleby:2012as,Amendola:2013qna,Amendola:2016saw}). In this case, unless a full likelihood treatment of the observational data is performed, the constraints on the present-day density parameters of a dust-like shear scalar and actual dust components would degenerate. On the other hand, in the case anisotropic expansion is allowed on top of $\Lambda$CDM or $o\Lambda$CDM, the stiff fluid approach is not less robust in determining constraints on the $\Omega_{\sigma0}$ parameter, although it cannot map the state of expansion anisotropy in the sky. In particular, when using the data relevant to relatively late universe (e.g., SNIa), the model independent constraints---which naturally consider full likelihoods---on $\Omega_{\sigma 0}$ and the ones from the stiff fluid like behaviour of the shear scalar only provide us with upper bounds, and these are at almost the same order of magnitude in the two different approaches, viz., $\Omega_{\sigma 0}\lesssim 10^{-3}$~\cite{Campanelli:2010zx,Wang:2017ezt,Jimenez:2014jma,Soltis:2019ryf,Zhao:2019azy,Hu:2020mzd,Kalus:2012zu,Akarsu:2019pwn,Amirhashchi:2018nxl}. Also, in the stiff fluid approach on top of $\Lambda$CDM, the upper bounds obtained by using the compressed BAO and/or CMB data are already much stronger, $\Omega_{\sigma0}\lesssim 10^{-15}$; leaving anisotropic expansion on top of $\Lambda$CDM only as a correction (see~\cite{Akarsu:2019pwn} and references therein), at least all the way the recombination. That is, even if the compressed BAO and/or CMB data are used instead of full data, it does not matter for the determination of $\Omega_{\sigma0}$ whether the full likelihoods of the data relevant to late universe (e.g., SNIa, CC) are used or not.

It is also worth noting that introducing isotropic (or almost isotropic) spatial curvature, viz., the $\Omega_{\kappa0}s^{-2}$ term, on top of $\Lambda$CDM is similar to introducing a source described by the EoS of the form $p_{\rm cs}/\rho_{\rm cs}=-1/3$, such as cosmic strings (which can also have negative mass density~\cite{Cramer:1994qj}). Accordingly, the An-$o\Lambda$CDM model, the pure geometric generalization of $\Lambda$CDM discussed here, resembles a cosmological model with cosmic strings (CS) and a stiff fluid with a positive energy density on top of $\Lambda$CDM. However, strictly speaking, this should only be considered true at the background level; a spatial curvature term in the background and a source with $w=-1/3$, and similarly, the shear scalar term and a source with $w=1$, give identical contributions to $H(s)$, the volumetric expansion rate of the Universe, but are not similar in their contribution to CMB anisotropies~\cite{Hu:1998kj,Battefeld:2004jh,Pereira:2007yy}. Not to mention that the perturbations, hence the CMB anisotropies, will differ when anisotropic expansion is allowed on top of the RW spacetime, moreover they also differ depending on the theoretical models (see footnote \ref{fn:2}) that introduce a stiff fluid like contribution on top of $\Lambda$CDM~\cite{Hu:1998kj,Battefeld:2004jh,Pereira:2007yy}.  It should be noted here that despite the correspondence between spatial curvature and cosmic strings mentioned above, we will not follow this correspondence in our analysis; because, they differ not only perturbatively, but also due to the fact that the interrelations of cosmological distance measures differ in spatially flat models and non-flat models, see, e.g., Sec.~\ref{subsec:BAO} where the various distance measures used in our analysis are given.

For these reasons, exploring the observational constraints on An-$o\Lambda$CDM at the background level would provide us the opportunity to interpret these results also as constraints on a large family of cosmological models containing stiff fluid on top of $\Lambda$CDM and $o\Lambda$CDM, rather than a particular model (each of which may be a separate research topic); compromising the details of the underpinning physics/theoretical model (see footnote \ref{fn:2}) giving rise to stiff fluid type contributions in the Friedmann equation. These models can, of course, be further constrained when, for instance, the full CMB data are used and they can then be distinguished from each other based on their perturbation differences. Lastly, the strongest upper bounds on anisotropic expansion are obtained from the CMB quadrupole ($l=2$ corresponds to the angular scale $\theta=\pi/2$ on the sky) temperature fluctuation ($\Delta T_{\rm Planck}\approx10\,\mu\,\rm K$~\cite{Ade:2013kta}, relevant to perturbations) and/or BBN, viz., $\Omega_{\sigma0}\lesssim 10^{-21}-10^{-23}$  (see~\cite{Akarsu:2019pwn} and references therein). Compromising these in our analysis, implying relatively weaker constraints, may be seen as a drawback, but this is preferred on purpose in the current work; in this case, we would be able to assess whether the presence of anisotropy, as a correction, yet at a level beyond the one imposed by the full CMB and/or BBN data, on top of $\Lambda$CDM and $o\Lambda$CDM could have consequences on the $H_0$ tension~\cite{DiValentino:2020zio} and/or the spatial curvature ($\Omega_{\kappa0}$) tension~\cite{DiValentino:2020srs}, which, in fact, is the main motivation of the current work, rather than finding new improved constraints on the expansion anisotropy and spatial curvature, when introduced on top of the standard cosmological model.
 
\section{Data and Likelihoods}
\label{data}

In this section, we briefly describe the used data sets, and the adopted methodology to perform the observational analyses.

\subsection{Cosmic Chronometers}

We consider the compilation of 31 cosmic chronometer (CC) measurements of $H(z)$ lying in the redshift range $0.07\leq z \leq 1.965$, as shown in Table~I of~\cite{Vagnozzi:2020dfn}, and the references therein. The principle, underlying these measurements, was first proposed in~\cite{Jimenez:2001gg}, relating the Hubble parameter $H(z)$, redshift $z$, and cosmic time $t$ as
\begin{equation}
    H(z)= -\frac{1}{1+z}\frac{\text{d}z}{\text{d}t}.
\end{equation}

The chi-squared function for these measurements, denoted by $\chi^2_{\rm CC}$, is
\begin{equation}
\chi^2_{\rm CC} = \sum_{i=1}^{31} \frac{[H^{\text{obs}}(z_i)-H^{\text{th}}(z_i)]^2}{\sigma^2_{H^{\text{obs}}(z_i)}},
\end{equation}
where $H^{\text{obs}}(z_i)$ is the observed value of the Hubble parameter with the standard deviation $\sigma^2_{H^{\text{obs}}(z_i)}$ as given in the aforementioned table, and $H^\text{th}(z_i)$ is the theoretical value obtained from the cosmological model under consideration.

\subsection{SnIa (Pantheon 2018)}

The Pantheon 2018 (Pan) sample comprises a large type Ia supernovae sample of 1048 measurements from five subsamples PS1, SDSS, SNLS, low-$z$, and HST spanning in the redshift range $0.01<z<2.3$~\cite{Pan-STARRS1:2017jku,Akarsu:2019pwn,Sharma:2020wio}.

The chi-squared function of the Pan data is given by
\begin{equation}
\chi^2_{\rm Pan}={\Delta \mu}\,{ C}_{\rm Pan}^{-1}\,{\Delta \mu}^{T}\,,
\end{equation}
where ${\Delta\mu}=\mu^{\rm obs}_i-\mu^{\rm th}$. The observed distance modulus ($\mu^{\rm obs}_{i}$)~\cite{SDSS:2014iwm} is evaluated as 
\begin{equation}
\mu^{\rm obs}_{i} = \mu_{B,i}+\mathcal{M},
\end{equation}
where $\mu_{B,i}$ is the observed peak magnitude 
at maximum in the rest frame of the $B$ band for 
redshift $z_{i}$, while the quantity $\mathcal{M}$ is the nuisance parameter. 
On the other hand, we evaluate the theoretical distance modulus as
 \begin{equation}
 \mu^{\text{th}}=5\log_{10}D_L +\mathcal{M},
\end{equation}
where
\begin{equation}
D_{L}= (1+z_{\rm hel})\int_0^{z_{\rm cmb}}\frac{H_0 dz}{H(z)},
\end{equation}
with $z_{\rm hel}$ and $z_{\rm cmb}$ being the heliocentric and CMB rest frame redshifts, respectively.

The covariance matrix is measured as~\cite{Conley11},
\begin{equation}
{ C}_{\rm Pan}= C_{\text{sys}}+D_{\text{stat}},
\end{equation}
where ${C}_{\text{sys}}$ is the systematic covariance matrix. Further, ${D}_{\text{stat}}$ is the diagonal covariance matrix of the statistical uncertainty, calculated as
\begin{equation}
{D}_{\text{stat}, ii} = \sigma^{2}_{\mu_{B,i}}.
\end{equation}
The detailed description and the systematic covariance matrix together with $\mu_{B,i}$, $\sigma^{2}_{\mu_{B,i}}$, $z_{\text{cmb}}$, and $z_{\text{hel}}$ for the $i{\text{th}}$ SnIa are mentioned in~\cite{Pan-STARRS1:2017jku}.

\subsection{BAO}
\label{subsec:BAO}

The completed experiments of the Sloan Digital Sky Survey (SDSS) recently presented the BAO measurements using galaxies, Lyman-$\alpha$ (Ly$\alpha$), and quasars~\cite{eBOSS:2020yzd}. These experiments include the compilation of data from SDSS, BOSS, SDSS-II, and eBOSS, offering independent BAO measurements of Hubble distances and angular-diameter distances relative to the sound horizon from eight different samples as shown in Table~\ref{tab:BAO_measurements}.

\begin{table*}[hbt!]
\caption{\rm Clustering measurements for the BAO samples.}
\scalebox{0.9}{ 
\centering

\resizebox{\textwidth}{!}{%
\begin{tabular}{|l|c|c|c|c|c|c|c|c|}
\hline 

Parameter & MGS & BOSS Galaxy & BOSS Galaxy & eBOSS LRG & eBOSS ELG & eBOSS Quasar & Ly$\alpha$-Ly$\alpha$ & Ly$\alpha$-Quasar \\
\hline

\hline
$z_{\rm eff}$ & 0.15 & 0.38 & 0.51 & 0.70 & 0.85 & 1.48 & 2.33 & 2.33 \\
\hline
$D_V(z)/r_{\rm d}$ & $4.47 \pm 0.17$ & --- & --- & --- & $18.33_{-0.62}^{+0.57}$ & --- & --- & ---  \\
$D_M(z)/r_{\rm d}$ & --- & $10.23 \pm 0.17$ & $13.36 \pm 0.21$ & $17.86 \pm 0.33$ & --- & $30.69 \pm 0.80$ & $37.6 \pm 1.9$ & $37.3 \pm 1.7$ \\
$D_H(z)/r_{\rm d}$ & --- & $25.00 \pm 0.76$ & $22.33 \pm 0.58$ & $19.33 \pm 0.53$ & --- &  $13.26 \pm 0.55$ & $8.93 \pm 0.28$ & $9.08 \pm 0.34$ \\
\hline

\end{tabular} 
}
}  

\label{tab:BAO_measurements}
\end{table*}

The comoving size of the sound horizon ($r_{\rm s}$) at the drag redshift ($z_{\rm d}$), i.e., $r_{\rm d}$, is calculated as:
\begin{equation}
\label{comoving size}
r_{\rm d}=r_{\rm s}(z_{\rm d})=\int_{z_{\rm d}}^\infty \frac{c_{\rm s}\text{d}z}{H(z)}. 
\end{equation}

We use the best fit values of $z_{\rm d}$ for the spatially flat and non-flat cases as obtained in~\cite{Params_table_2018} for the $\Lambda$CDM and $o\Lambda$CDM models, respectively, for the CMB and CMB+Lens cases. For CMB, we use $z_{\rm d}=1059.971$ in $\Lambda$CDM and An-$\Lambda$CDM models, and $z_{\rm d}=1060.390$ in $o\Lambda$CDM and An-$o\Lambda$CDM models. For CMB+Lens, we consider $z_{\rm d}=1059.971$ in $\Lambda$CDM and An-$\Lambda$CDM models, and $z_{\rm d}=1060.123$ in $o\Lambda$CDM and An-$o\Lambda$CDM models. Further, the sound speed of the baryon--photon fluid reads $c_{\rm s}=\frac{c}{\sqrt{3(1+\mathcal{R})}}$, where $\mathcal{R}=\frac{3\Omega_{\rm b0}}{4\Omega_{\rm \gamma 0}(1+z)}$ with $\Omega_{\rm b0}=0.022h^{-2}$ being the present-day physical density of baryons~\cite{Cooke:2016rky} and $\Omega_{\gamma 0}=2.469\times 10^{-5}h^{-2}$ being the present-day physical density of photons ~\cite{Mangano:2005cc,deSalas:2016ztq,Akita:2020szl,Froustey:2020mcq,Bennett:2020zkv}.

BAO measurements directly constrain the quantities $D_H(z)/r_{\rm d}$ and $D_{M}(z)/r_{\rm d}$. The Hubble distance at redshift $z$ is measured as
\begin{equation}
 D_H(z) = \frac{c}{H(z)}.
\end{equation}
The comoving angular diameter distance $D_{M}(z)$ is calculated as
\begin{equation}
    D_{M}(z)=\frac{c}{H_0}S_k\bigg(\frac{D_{ C}(z)}{c/H_0}\bigg),
\end{equation}
where the comoving distance of the line-of-sight is
\begin{equation}
  D_C(z) = {c \over H_0}\int_0^z \text{d}z' {H_0 \over H(z')},
\label{eqn:dcomove}
\end{equation}
and
\begin{equation}
S_{\kappa} (x) =
\begin{cases}
  \sin(\sqrt{-\Omega_{\kappa}} x)/\sqrt{-\Omega_{\kappa}} & \textnormal{for}\quad \Omega_{\kappa}<0, \\
  x & \textnormal{for}\quad \Omega_{\kappa}=0, \\
  \sinh(\sqrt{\Omega_{\kappa}}x)/\sqrt{\Omega_{\kappa}} & \textnormal{for}\quad \Omega_{\kappa}>0. \\
\end{cases}
\end{equation}

Further, these measurements are summarized by a 
quantity $D_{V}(z)/r_{\rm d}$, with the spherically averaged distance $D_V(z)$ as
\begin{equation}
 D_V(z) \equiv \left[z D^2_M(z) D_H(z)\right]^{1/3}.
\end{equation}

We now define the chi-squared function corresponding to each distance considered in Table~\ref{tab:BAO_measurements}. We denote $d_1$ for $D_{V}(z)/r_{\rm d}$, $d_2$ for $D_{M}(z)/r_{\rm d}$, and $d_3$ for $D_{H}(z)/r_{\rm d}$, respectively. The chi-squared function for each measurement is considered as
\begin{eqnarray}
\chi^2_{\rm B_1} =&\displaystyle\sum_{i=1}^{2}\left(\frac{d_1^{\text{obs}}(z_i)-d_1^{\text{th}}(z_i)}{\sigma_{d_1^{\text{obs}}(z_i)}}\right)^2,\nonumber \\
\chi^2_{\rm B_2}=&\displaystyle\sum_{j=1}^{6}\left(\frac{d_2^{\text{obs}}(z_j)-d_2^{\text{th}}(z_j)}{\sigma_{d_2^{\text{obs}}(z_j)}}\right)^2,\nonumber \\
\chi^2_{\rm B_3}=&\displaystyle\sum_{j=1}^{6}\left(\frac{d_3^{\text{obs}}(z_j)-d_3^{\text{th}}(z_j)}{\sigma_{d_3^{\text{obs}}(z_j)}}\right)^2.
\end{eqnarray}
    
Here, $d^{\rm obs}$ is the observed distance value as given in Table~\ref{tab:BAO_measurements}, while $d^{\rm th}$ is the theoretical value calculated for the models under consideration. In the case of $D_{V}(z)/r_{\rm d}$, $z_i$ ($i=1,2$) are respectively the effective redshifts  for the two measurements of samples, MGS and eBOSS ELG. While, in the case of $D_{M}(z)/r_{\rm d}$ and $D_{H}(z)/r_{\rm d}$, $z_j$ ($j=1,2,3,4,5,6$) are respectively the effective redshifts for the six measurements: BOSS Galaxy, BOSS Galaxy, eBOSS Galaxy, eBOSS LRG, eBOSS Quasar, Ly$\alpha$-Ly$\alpha$, and Ly$\alpha$-Quasar.

Thus, the total chi-squared function for the BAO measurements, denoted by $\chi^2_{\rm BAO}$, reads
\begin{equation}
\chi^2_{\rm BAO}=\chi^2_{\rm B_1} +\chi^2_{\rm B_2}+\chi^2_{\rm B_3}.
\end{equation}

\subsection{CMB}
\label{subsec:CMB}

In~\cite{Aubourg:2014yra}, various cosmological models are constrained using the compressed CMB likelihood information. In this approach, CMB measurements are compressed to the variables that govern the expansion history of the universe. The compressed form helps in simplifying the computations, and allows to fit complex models. Here, we follow a similar approach and use the compressed likelihood information of the CMB data from Planck 2018 chains~\cite{Params_table_2018} as described in the following.

In the case of CMB data, we fix the photon-decoupling surface $z_*=1089.920$ for the spatially flat models ($\Lambda$CDM and An-$\Lambda$CDM), and the chi-squared function denoted by $\chi^2_{\rm C_1}$, reads as
\begin{equation}
\label{eq:chi2C1}
\chi^2_{\rm C_1}=\Delta p_i {C_1}^{-1}(\Delta p_i)^T,
\hskip .3cm
\Delta p_i= p_i^{\rm th} - p_i^{\rm mean},
\end{equation}
where $i=1,2,3$ with $p_1=\omega_{\rm b}$, $p_2= \omega_{\rm cdm}$, and $p_3=100\theta_{\rm MC}$. $p_i^{\rm mean}$ are the mean values of the parameters from Planck 2018 chains, and $C_1$ is the covariance matrix for $(\omega_{\rm b}, \omega_{\rm cdm}, 100\theta_{\rm MC})$, given by
\begin{equation}\nonumber
\left[
\begin{array}{ccc}   
2.18\times 10^{-8} & -1.16\times 10^{-7} & 1.60\times 10^{-8} \\   
-1.16\times 10^{-7} & 1.85\times 10^{-6} & -1.41\times 10^{-7} \\    
1.60\times 10^{-8} & -1.41\times 10^{-7} & 9.62\times 10^{-8} \\   
\end{array}
\right ].
\label{eq:C1}
\end{equation}

Further, we fix $z_*=1089.411$ for the spatially non-flat models ($o\Lambda$CDM and An-$o\Lambda$CDM), and the chi-squared function denoted by $\chi^2_{\rm C_2}$, reads as
\begin{equation}
\label{eq:chi2C2}
\chi^2_{\rm C_2}=\Delta q_i {C_2}^{-1}(\Delta q_i)^T,
\hskip .3cm
\Delta q_i= q_i^{\rm th} - q_i^{\rm mean},
\end{equation}
where $i=1,2,3,4$ with $q_1=\omega_{\rm b}$, $q_2= \omega_{\rm cdm}$, $q_3=100\theta_{\rm MC}$ and $q_4=\Omega_{\kappa}$. Here $q_i^{\rm mean}$ are the mean values of the parameters from Planck 2018 chains, and $C_2$ is the covariance matrix for $(\omega_{\rm b}, \omega_{\rm cdm}, 100\theta_{\rm MC}, \Omega_{\kappa})$, given by
\begin{equation}\nonumber
\colvec[.9]{2.88\times 10^{-8} & -1.68\times 10^{-7} & 2.53\times 10^{-8} & -1.30\times 10^{-6} \\
-1.68\times 10^{-7} & 2.18\times 10^{-6} & -2.18\times 10^{-7} & 1.04\times 10^{-5} \\    
2.53\times 10^{-8} & -2.18\times 10^{-7} & 1.05\times 10^{-7}& -1.60\times 10^{-6} \\   
-1.30\times 10^{-6} & 1.04\times 10^{-5} & -1.60\times 10^{-6} & 2.94\times 10^{-4}}.   
\end{equation}

Similarly, we evaluate the chi-squared distribution for spatially flat and non-flat models in the case of CMB+Lens data. In other words, we consider the photon-decoupling surface $z_*=1089.914$ for the spatially flat models ($\Lambda$CDM and An-$\Lambda$CDM), and $z_*=1089.606$ for the spatially non-flat models ($o\Lambda$CDM and An-$o\Lambda$CDM).

In order to constrain the parameters of the models under consideration with different combinations of the aforementioned data sets, we use a Python interface for \texttt{Multinest}~\cite{Feroz:2007kg,eroz:2008xx,Feroz:2013hea}, namely, the \texttt{Pymultinest} code~\cite{Buchner:2014nha}. We further analyze our models with the Bayesian inference tool, nested sampling~\cite{Skilling} that helps in model comparison by calculating Bayesian evidence.
We define the total multivariate joint Gaussian likelihood function as
\begin{equation}\label{27}
     \mathcal{L}_{\rm tot} \propto \text{exp}\left(\frac{-\chi^2_{\rm tot}}{2}\right),
\end{equation}
where the total chi-square function of all the data sets reads
\begin{equation}
\chi^2_{\rm tot} =  \chi^2_{\rm CC}+ +\chi^2_{\rm Pan} + \chi^2_{\rm BAO} + \chi^2_{\rm CMB}.
\end{equation}

In our study, we choose uniform prior distribution for all the model parameters, viz., $35< H_0~[{\rm km ~s{}^{-1} Mpc{}^{-1}}]< 85$, $0.1< \Omega_{\rm m0} < 0.7$, $-40< \log_{10}(\Omega_{\sigma 0})<0$ [except the analysis for the data sets CC, Pan, and CC+Pan, for which we use $-4< \log_{10}(\Omega_{\sigma 0})<0$], and $-0.3<\Omega_{\kappa 0}<0.3$. \footnote{In the CMB and BAO data likelihoods, we have used some fixed high redshifts values, viz., the drag redshift $z_{\rm d}$ and the last-scattering redshift $z_*$, and therefore a small amount of anisotropy is expected in our results, as a correction on the top of standard $\Lambda$CDM. The redshift $z\sim1100$ of recombination where the Universe is supposed to be matter dominated, is physically closely related to $z_*$ and $z_{\rm d}$. Therefore, using $\Omega_{\rm m}(z\sim1100)\approx 1$ and say, $\Omega_{\sigma}(z\sim1100)\lesssim 10^{-2}$ into $\frac{\Omega_{\sigma}}{\Omega_{\rm m}}=\frac{\Omega_{\sigma 0}}{\Omega_{\rm m0}}(1+z)^{3}$, the upper bound for $\frac{\Omega_{\sigma 0}}{\Omega_{\rm m0}}$ is found to be $\sim 10^{-11}$. The test runs of the code suggested that the prior range $-40< \log_{10}(\Omega_{\sigma0})< 0$ is good enough to extract the information about $\Omega_{\sigma 0}$.}\\

Finally, note that we aim in this study to constrain the allowed amount of expansion anisotropy, quantified by its stiff fluid-like contribution, $\rho_{\sigma}\propto (1+z)^6$, to the Friedmann equation~\eqref{themodel}, from the observational data on top of the $\Lambda$CDM and $o\Lambda$CDM models, and assess its effects (see section \ref{sec:thefinalmodel} for more discussion). Notice that $\Omega_{\sigma0}s^{-6}$ is the fastest growing term with decreasing $s$, i.e., increasing $z$, in the Friedmann equation. Therefore, we expect to typically achieve constraints getting tighter on $\Omega_{\sigma 0}$ as we use data from higher redshifts; as is reflected in the sequence of our discussion in the next section. Also, as the deviations from the isotropic models would be pronounced more and more with increasing redshift, for guaranteeing expansion anisotropy to remain as a correction all the way to the largest redshifts relevant to the CMB and BAO data used in our analyses, we fix the drag redshift $z_{\rm d}$ (involved in the BAO data analyses, see section \ref{subsec:BAO}) and the last scattering redshift $z_*$ (involved in the CMB data analyses, see section \ref{subsec:CMB}) for An-$o\Lambda$CDM and An-$\Lambda$CDM by using respectively the values obtained for $o\Lambda$CDM and $\Lambda$CDM in the Planck 2018 release~\cite{Params_table_2018}.

\section{Results and Discussions} \label{results}

In this section, we discuss the observational constraints on the cosmological parameters of the models (viz., An-$o\Lambda$CDM described in~\eqref{themodel} and its particular cases $o\Lambda$CDM, An-$\Lambda$CDM, and $\Lambda$CDM) obtained by using the data sets from the different observational probes (viz., CC, SnIa Pantheon, BAO, and CMB) and their various combinations, and the methods as discussed in Sec.~\ref{data}. In Tables~\ref{tab:R3},~\ref{tab:R1}, and~\ref{tab:R2}, we report the bounds at 68\% and 95\% Confidence Level (C.L.) for $H_0$, $\Omega_{\rm m0}$, $\Omega_{\kappa 0}$, and $\log_{10}(\Omega_{\sigma 0})$, while we show the corresponding one and two-dimensional marginalized posteriors in the figures from Fig.~\ref{fig:NF1} to Fig.~\ref{fig:NF6}.

Table~\ref{tab:R3} displays the constraints on the model parameters from the CC, Pan, and CC+Pan data, which are relatively low redshifts ($z\lesssim 2.4$) data. The accompanying  Fig.~\ref{fig:NF1} compares the constraints of different models for a given data set while  Fig.~\ref{fig:NF2} compares the constraints from different data sets for a given model.  We notice that, regardless of whether or not anisotropic expansion is allowed, the CC data set constrains $H_0$ better than $\Omega_{\kappa0}$, while the Pan data set constrains $\Omega_{\kappa0}$ better than $H_0$. The combination of CC and Pan data provides stronger constraints on the parameters of the models in general, but constraints on the expansion anisotropy, viz., $\Omega_{\sigma 0}$, are not affected significantly, see Fig.~\ref{fig:NF2}. When we use only the CC data or Pantheon 2018 data, or the combination of these two, the upper bounds on the expansion anisotropy are allowed to be large; up to $\log_{10}(\Omega_{\sigma 0})<-1.77$ at 95\% C.L. (An-$o\Lambda$CDM with Pan). The tightest upper bound is $\log_{10}(\Omega_{\sigma 0})<-2.23$ at 95\% C.L. (An-$o\Lambda$CDM with CC) and it is slightly looser than those obtained from the model independent constraints, e.g., from the SnIa data, $\log_{10}(\Omega_{\sigma 0})\lesssim-3$ assuming LRS Bianchi type I background~\cite{Campanelli:2010zx,Wang:2017ezt,Jimenez:2014jma,Soltis:2019ryf,Zhao:2019azy,Hu:2020mzd,Kalus:2012zu}. We notice that the models with spatial curvature are predicted to be well consistent with the spatially flat ($\Omega_{\kappa0}=0$) models in all these analyses. We find no significant difference among the predicted $H_0$ values, yet, it might be interesting to notice that the introduction of expansion anisotropy leads to a slight increase in the predicted mean value of $H_0$ for the CC data and to a slight decrease in it for the Pan data. Also, for these cases, the introduction of spatial curvature does not change the constraints of the parameters, because these models are predicted to be well consistent with a spatially flat universe.

\begin{table}[t] 

\caption{\label{tab:R3}Constraints on the An-$o\Lambda$CDM, $o\Lambda$CDM, An-$\Lambda$CDM, and $\Lambda$CDM model parameters from CC, Pan, and CC+Pan data.
The upper bound of $\log_{10}(\Omega_{\sigma 0})$ is at 95\% C.L. The parameter $H_{\rm 0}$ is measured in units of km s${}^{-1}$ Mpc${}^{-1}$.
}
    \centering
    \resizebox{0.95\columnwidth}{!}
    {%
    \renewcommand{\arraystretch}{1.2}
    \begin{tabular}{|c|c|c|c|c|}
        \hline
        Parameter & CC & Pan & CC+Pan\\
         \hline
         An-$o\Lambda$CDM & & &   \\
        \hline
        $H_0$ &
        $68.2^{+3.1+6.2}_{-3.1-5.8}$ &
        $70^{+8+10}_{-8-10}$ &
        $69.0^{+1.9+3.8}_{-1.9-3.8}$
        \\

        $\Omega_{\rm m 0}$ &
        $0.302^{+0.081+0.15}_{-0.081-0.15}$ &
        $0.286^{+0.065+0.11}_{-0.065-0.12}$ &
        $0.300^{+0.067+0.11}_{-0.058-0.12}$
        \\  
        
        $\Omega_{\kappa 0}$ &
        $0.02^{+0.20+0.26}_{-0.16-0.28}	$ &
        $-0.01^{+0.14+0.26}_{-0.14-0.26}$ &
        $-0.02^{+0.14+0.26}_{-0.14-0.26}$
        \\ 
        
        $\log_{10}(\Omega_{\sigma 0})$ &
        $<-2.23$ &
        $<-1.77$ & 	
        $<-2.22$
        \\
      \hline 
        
\hline
        $o\Lambda$CDM & & &  \\
        \hline
        $H_0$ &
        $67.9^{+3.0+6.2}_{-3.0-5.9}$ &
        $71^{+9+10}_{-9-10}$ &
        $69.1^{+2.0+4.0}_{-2.0-3.9}$
        \\

        $\Omega_{\rm m 0}$ &
        $0.328^{+0.081+0.13}_{-0.072-0.14}$ &
        $0.318^{+0.062+0.10}_{-0.052-0.11}$ &
        $0.321^{+0.062+0.10}_{-0.053-0.11}$ 
        \\  

        $\Omega_{\kappa 0}$ &
        $0.00^{+0.16+0.28}_{-0.16-0.28}$ &
        $-0.05^{+0.12+0.26}_{-0.16-0.24}$ &
        $-0.05^{+0.13+0.27}_{-0.17-0.24}$
        \\ 
        \hline
        
        \hline
        \hline
        
        An-$\Lambda$CDM & & & \\
        \hline
        $H_0$ &
        $68.1^{+3.0+6.1}_{-3.0-5.7}$ &
        $69^{+9+10}_{-10-10}$ &
        $68.8^{+1.8+3.6}_{-1.8-3.6}$
        \\
     
        $\Omega_{\rm m 0}$ &
        $0.311^{+0.064+0.13}_{-0.064-0.13}$ &
        $0.284^{+0.031+0.056}_{-0.025-0.059}$ &
        $0.292^{+0.023+0.045}_{-0.023-0.046}$
        \\

        $\log_{10}(\Omega_{\sigma 0})$ &
        $<-2.22$ & 
        $<-1.80$ & 
        $<-2.19$
        \\
      \hline

\hline
        $\Lambda$CDM & & & \\
        \hline
        $H_0$ &
        $67.8^{+2.9+5.6}_{-2.9-5.6}$ &
        $70^{+9+10}_{-9-10}$ &
        $69.0^{+1.8+3.5}_{-1.8-3.5}$
        \\

        $\Omega_{\rm m 0}$ &
        $0.330^{+0.052+0.12}_{-0.065-0.11}	$ &
        $0.298^{+0.022+0.044}_{-0.022-0.041}$ &
        $0.302^{+0.020+0.041}_{-0.021-0.039}$
        \\  
        \hline
        
    \end{tabular}%
    }
\end{table}

\begin{table}[t] 

\caption{\label{tab:R1}Constraints on the An-$o\Lambda$CDM, $o\Lambda$CDM, An-$\Lambda$CDM, and $\Lambda$CDM model parameters from BAO+Pan, CMB only, and CMB+Pan data.
The upper bound of $\log_{10}(\Omega_{\sigma 0})$ is at 95\% C.L.  The parameter $H_{\rm 0}$ is measured in units of km s${}^{-1}$ Mpc${}^{-1}$. }
    \centering
    \resizebox{0.95\columnwidth}{!}
    {%
    \renewcommand{\arraystretch}{1.2}
    \begin{tabular}{|c|c|c|c|c|}
        \hline
        Parameter & BAO+Pan &CMB & CMB+Pan  \\
         \hline
         An-$o\Lambda$CDM & & &   \\
        \hline
        $H_0$ &
        $67.3^{+2.9+6.6}_{-3.5-6.1}$ &
        $54.2^{+2.9+7}_{-3.5-6}$ &
        $66.7^{+2.2+5.2}_{-2.6-4.6}$ 
        \\

        $\Omega_{\rm m 0}$ &
        $0.294^{+0.027+0.054}_{-0.027-0.049}$ &
        $0.484^{+0.056+0.12}_{-0.056-0.11}$ &
        $0.319^{+0.023+0.045}_{-0.023-0.0446}$ 
        \\  
        
        $\Omega_{\kappa 0}$ &
        $0.013^{+0.074+0.15}_{-0.074-0.15}$ &
        $-0.046^{+0.017+0.029}_{-0.014-0.032}	$ &
        $-0.0032^{+0.0059+0.011}_{-0.0059-0.012}	$ 
        \\ 
        
        $\log_{10}(\Omega_{\sigma 0})$ &
        $<-14.3$ &
        $<-16.3$ & 
        $<-17.1$ 
        \\
      \hline

\hline
        $o\Lambda$CDM & & &   \\
        \hline
        $H_0$ &
        $67.4^{+3.0+7.2}_{-3.5-6.3}$ &
        $54.3^{+2.6+7}_{-3.9-6}$ &
        $66.7^{+2.5+5.1}_{-2.5-4.8}$ 
        \\
        
        $\Omega_{\rm m 0}$ &
        $0.294^{+0.027+0.054}_{-0.027-0.055}$ &
        $0.483^{+0.058+0.11}_{-0.058-0.12}	$ &
        $0.320^{+0.023+0.047}_{-0.025-0.045}$ 
        \\  

        $\Omega_{\kappa 0}$ &
        $0.010^{+0.076+0.15}_{-0.076-0.15}$ &
        $-0.045^{+0.016+0.031}_{-0.016-0.031}$ &
        $-0.0032^{+0.0068+0.011}_{-0.0056-0.013}$ 
        \\ 
        \hline

        \hline
        An-$\Lambda$CDM & & &    \\
        \hline
        $H_0$ &
        $67.7^{+1.2+2.7}_{-1.5-2.6}$ &
        $67.57^{+0.64+1.3}_{-0.64-1.3}$ &
        $67.75^{+0.60+1.2}_{-0.60-1.2}$ 
        \\

        $\Omega_{\rm m 0}$ &
        $0.297^{+0.014+0.029}_{-0.014-0.028}$ &
        $0.3125^{+0.0085+0.017}_{-0.0085-0.017}$ &
        $0.3102^{+0.0078+0.016}_{-0.0078-0.015}$ 
        \\

        $\log_{10}(\Omega_{\sigma 0})$ &
        $<-14.3$ &
        $<-17.9$ & 
        $<-17.2$ 
        \\
      \hline

\hline
        $\Lambda$CDM & & &    \\
        \hline
        $H_0$ &
        $67.6^{+1.3+2.7}_{-1.3-2.5}	$ &
        $67.59^{+0.65+1.3}_{-0.65-1.3}$ &
        $67.75^{+0.64+1.2}_{-0.64-1.2}$ 
        \\

        $\Omega_{\rm m 0}$ &
        $0.297^{+0.014+0.029}_{-0.014-0.027}$ &
        $0.3122^{+0.0087+0.017}_{-0.0087-0.017}$ &
        $0.3100^{+0.0084+0.017}_{-0.0084-0.015}$ 
        \\  
        \hline
        
\hline
    \end{tabular}
    }
\end{table}

\begin{table*}[ht!] 

\caption{\label{tab:R2}Constraints on the An-$o\Lambda$CDM, $o\Lambda$CDM, An-$\Lambda$CDM, and $\Lambda$CDM model parameters from CMB+Lens, CMB+Lens+BAO, CMB+Lens+Pan, CMB+Lens+BAO+Pan, and CMB+Lens+BAO+Pan+CC data. The upper bound of $\log_{10}(\Omega_{\sigma 0})$ is at 95\% C.L.  The parameter $H_{\rm 0}$ is measured in units of km s${}^{-1}$ Mpc${}^{-1}$.
}

\centering
    \resizebox{2\columnwidth}{!}
    {
    \renewcommand{\arraystretch}{1.2}
    \begin{tabular}{|c|c|c|c|c|c|c|c|c|c|}
        \hline
        Parameter & CMB+Lens & CMB+Lens+BAO & CMB+Lens+Pan  & CMB+Lens+BAO+Pan & CMB+Lens+BAO+Pan+CC\\
        \hline
        An-$o\Lambda$CDM & & & & &   \\
        \hline
        $H_0$ &
        $63.8^{+2.0+4.3}_{-2.2-4.2}$&
        $67.71^{+0.66+1.3}_{-0.66-1.3}$ &
        $66.2^{+1.8+4.0}_{-2.0-3.6}$ &
        $67.78^{+0.65+1.3}_{-0.65-1.3}$ & 
        $67.85^{+0.62+1.2}_{-0.62-1.2}$
        \\
        
        $\Omega_{\rm m 0}$ &
        $0.347^{+0.022+0.043}_{-0.022-0.043}$&
        $0.3093^{+0.0063+0.013}_{-0.0063-0.012}	$ &
        $0.323^{+0.017+0.034}_{-0.017-0.035}$ &
        $0.3086^{+0.0061+0.012}_{-0.0061-0.012}	$& 
        $0.3082^{+0.0059+0.012}_{-0.0059-0.011}$
        \\
        
        $\Omega_{\kappa 0}$ & 
        $-0.0107^{+0.0062+0.012}_{-0.0062-0.012}$&
        $-0.0006^{+0.0021+0.0040}_{-0.0021-0.0040}$ &
        $-0.0045^{+0.0050+0.0097}_{-0.0050-0.010}$ &
        $-0.0005^{+0.0020+0.0040}_{-0.0020-0.0040}$ & 
        $-0.0002^{+0.0019+0.0038}_{-0.0019-0.0037}$ \\  
    
        $\log_{10}(\Omega_{\sigma 0})$ &
        $<-17.7$ & 
        $<-17.5$ & 
        $<-17.6$ & 
        $<-17.4$ & 
        $<-17.1$ 
        \\
      \hline 
        
        \hline

        $o\Lambda$CDM & & & & &   \\
        \hline
        $H_0$ &
        $63.9^{+2.1+4.7}_{-2.4-4.2}$ &
        $67.69^{+0.65+1.3}_{-0.65-1.3}$ &
        $66.2^{+1.9+3.6}_{-1.9-3.5}$ &
        $67.80^{+0.66+1.3}_{-0.66-1.2}$ & 
        $67.86^{+0.60+1.2}_{-0.60-1.2}$
        \\

        $\Omega_{\rm m 0}$ &
        $0.346^{+0.023+0.046}_{-0.023-0.045}$ &
        $0.3096^{+0.0062+0.012}_{-0.0062-0.012}$ &
        $0.323^{+0.017+0.034}_{-0.017-0.032}	$ &
        $0.3085^{+0.0062+0.012}_{-0.0062-0.012}	$& 
        $0.3081^{+0.0058+0.011}_{-0.0058-0.011}$
        \\
        
        $\Omega_{\kappa 0}$ & 
        $-0.0103^{+0.0066+0.012}_{-0.0066-0.013}$ &
        $-0.0005^{+0.0020+0.0039}_{-0.0020-0.0039}$ &
        $-0.0043^{+0.0050+0.0093}_{-0.0050-0.0099}	$ &
        $-0.0004^{+0.0021+0.0040}_{-0.0021-0.0042}$ & 
        $-0.0001^{+0.0020+0.0038}_{-0.0020-0.0038}$ \\  
        
      \hline 
        
\hline
\hline
        An-$\Lambda$CDM & & & & &  \\
        \hline
        $H_0$ &
       
        $67.67^{+0.58+1.2}_{-0.58-1.1}$ &
        $67.85^{+0.43+0.83}_{-0.43-0.80}$ &
        $67.78^{+0.55+1.1}_{-0.55-1.1}$ & 
        $67.90^{+0.42+0.83}_{-0.42-0.81}	$ &
        $67.92^{+0.42+0.85}_{-0.42-0.79}$
        \\

        $\Omega_{\rm m 0}$ &
        $0.3110^{+0.0076+0.015}_{-0.0076-0.015}$ &
        $0.3087^{+0.0056+0.010}_{-0.0056-0.011}	$&
        $0.3096^{+0.0072+0.014}_{-0.0072-0.014}$& 
         $0.3080^{+0.0054+0.011}_{-0.0054-0.010}$&
        $0.3078^{+0.0054+0.010}_{-0.0054-0.010}$
        \\

        $\log_{10}(\Omega_{\sigma 0})$ &
        $<-17.9$ & 
        $<-17.8$ & 
        $<-17.7$ & 
         $<-17.1$ & 
         $<-17.4$
        \\
      \hline 
        
 \hline 
        $\Lambda$CDM & & & & &   \\
        \hline
        $H_0$ &
        $67.63^{+0.58+1.1}_{-0.58-1.1}$ &
        $67.84^{+0.43+0.83}_{-0.43-0.84}$ &
        $67.77^{+0.56+1.1}_{-0.56-1.1}$ & 
        $67.89^{+0.43+0.84}_{-0.43-0.83}	$ &
        $67.89^{+0.43+0.82}_{-0.43-0.83}$
        \\

        $\Omega_{\rm m 0}$ &
        $0.3116^{+0.0075+0.015}_{-0.0075-0.015}$ &
        $0.3087^{+0.0055+0.011}_{-0.0055-0.010}$&
        $0.3097^{+0.0073+0.015}_{-0.0073-0.014}$& 
        $0.3081^{+0.0055+0.011}_{-0.0055-0.011}$&
        $0.3081^{+0.0055+0.011}_{-0.0055-0.010}$
        \\

      \hline 
        
\hline

    \end{tabular}%
    }

\end{table*}

Table~\ref{tab:R1} displays the constraints from the BAO data combined with the Pantheon 2018 data and from the CMB data (alone and in combination with the Pantheon 2018 data). We see that including the BAO data from the measurements at redshifts up to $z=2.33$ (see Table~\ref{tab:BAO_measurements})---but carry information from much higher redshifts via the drag redshift $z_{\rm d}\sim1060$ information---in our analyses with low-redshift data, viz., Pan, improves the constraints on the expansion anisotropy term considerably, as expected; see also the accompanying figures, viz., Fig.~\ref{fig:NF3} and Fig.~\ref{fig:NF4}. Accordingly, when the CMB data set is not in the picture, the most stringent upper bound on the expansion anisotropy is obtained by the BAO+Pan data, which is $\log_{10}(\Omega_{\sigma 0})<-14.3$ at 95\% C.L. irrespective of the presence of spatial curvature, as the cases with spatial curvature are predicted to be well consistent with a spatially flat universe. However we notice that, for the models with spatial curvature, the data predict slightly lower mean values for $H_0$ accompanied by a slight tendency towards a spatially open universe, which can be associated with the negative correlation between $H_0$ and $\Omega_{\kappa0}$, see the top-left panel of Fig.~\ref{fig:NF3}. As discussed above, we expect to achieve the tightest constraints on the expansion anisotropy when the CMB information, the data relevant to the highest redshifts ($z\sim1100$) among the data sets we used, is included in our analyses.  We notice that the constraints, both for the $\Lambda$CDM and $o\Lambda$CDM models, are in exquisite agreement with those obtained by the Planck collaboration~\cite{Planck:2018vyg}. We find that the $H_0$ predicted by the $\Lambda$CDM model with the CMB data only is in tension with the local measurements of $H_0$~\cite{Freedman:2019jwv,Riess:2020fzl,Riess:2021jrx}, and this tension reaches a level more than $5\sigma$ with the SH0ES collaboration measurements, where the latest estimation reads $H_0=73.04 \pm 1.04$ km s${}^{-1}$ Mpc${}^{-1}$~\cite{Riess:2021jrx}. Moreover, we notice that this tension is exacerbated by the introduction of spatial curvature, i.e., in the $o\Lambda$CDM model, see Table~\ref{tab:R1} and the top-right panel of Fig.~\ref{fig:NF3}. This gives evidence for a spatially closed universe at more than $3\sigma$ level while lowering the predicted Hubble constant value~\cite{Handley:2019tkm,DiValentino:2019qzk}. The introduction of the expansion anisotropy, i.e., the An-$o\Lambda$CDM and the An-$\Lambda$CDM models, does not modify these findings, as can be seen from the top-right panel of Fig.~\ref{fig:NF3}. This happens because, as can be seen from the same panel, within $2\sigma$, the CMB data set by alone already predicts that the anisotropic models, An-$o\Lambda$CDM and An-$\Lambda$CDM, remain consistent with their isotropic ($\Omega_{\sigma 0}=0$) counterparts, $o\Lambda$CDM and $\Lambda$CDM. However, the introduction of spatial curvature relaxes the upper bound on $\Omega_{\sigma 0}$ obtained for the An-$\Lambda$CDM model; namely, the constraint on the expansion anisotropy relaxes from $\log_{10}(\Omega_{\sigma 0})<-17.9$ (95\% C.L.) of the An-$\Lambda$CDM model to $\log_{10}(\Omega_{\sigma 0})<-16.3$ (95\% C.L.) of the An-$o\Lambda$CDM model (see also the top-right panel of Fig.~\ref{fig:NF3}). The inclusion of the Pantheon 2018 data in our CMB data only analyses, improves the determination of the parameters by breaking their correlations; in particular, in the models with spatial curvature (see top panels in Fig.~\ref{fig:NF4} and also the bottom panel of Fig.~\ref{fig:NF3}). The introduction of the spatial curvature on top of $\Lambda$CDM or An-$\Lambda$CDM in this case (CMB+Pan) does not affect the results as now $\Omega_{\kappa0}$ is well constrained and is in good agreement with a spatially flat ($\Omega_{\kappa0}=0$) universe, see bottom panel of Fig.~\ref{fig:NF3}. Therefore, the upper bound on the expansion anisotropy remains almost unaltered, viz., $\log_{10}(\Omega_{\sigma 0})<-17.1$ for An-$o\Lambda$CDM and $\log_{10}(\Omega_{\sigma 0})<-17.2$ for An-$\Lambda$CDM, both at 95\% C.L.

It is well known that the prediction of spatially closed background by $o\Lambda$CDM with the CMB only data is accompanied by substantially higher lensing amplitudes compared to $\Lambda$CDM, and that combining the CMB data with the lensing reconstruction (which is consistent with spatial flatness) pulls the parameters of $o\Lambda$CDM back into consistency with a spatially flat background within $2\sigma$~\cite{Planck:2018vyg}. Also, whereas the addition of CMB lensing with the CMB data weakly breaks the geometric degeneracy regarding spatial curvature, the spatial flatness can be tested to high accuracy by adding the BAO data. Therefore, in Table~\ref{tab:R2}, we present first the constraints from the combined data sets CMB+Lens and CMB+Lens+BAO, and then from the data combinations obtained by adding astrophysical data sets, namely, Pantheon 2018 and CC; see also the accompanying figures, viz., Fig.~\ref{fig:NF5} and Fig.~\ref{fig:NF6}. We find that the constraints obtained here for the $\Lambda$CDM and $o\Lambda$CDM models are in excellent agreement with those obtained by the Planck collaboration~\cite{Planck:2018vyg}. We observe that the results in Table~\ref{tab:R2} confirm the aforementioned points regarding $\Lambda$CDM and $o\Lambda$CDM, and the introduction of the expansion anisotropy does not change the picture; the inclusion of the Lens data to our CMB analyses (CMB+Lens) pulls the parameters of both the An-$o\Lambda$CDM and $o\Lambda$CDM models back into consistency with their spatially flat counterparts (An-$\Lambda$CDM and $\Lambda$CDM) within $2\sigma$, and that the inclusion of the BAO data further tightens the constraints to bring the said consistency with spatial flatness even below $1\sigma$ (see also the top panels of Fig.~\ref{fig:NF6}). We notice positive correlation between $H_0$ and $\Omega_{\kappa 0}$ for all of the combinations of the data sets (see also Fig.~\ref{fig:NF6}); the mean values of $H_0$ typically tend to smaller values accompanied by the slightly negative mean values of $\Omega_{\kappa 0}$, and this situation is the most pronounced in the cases of CMB+Lens and CMB+Lens+Pan. With the inclusion of the BAO data---i.e., in the cases of CMB+Lens+BAO, CMB+Lens+BAO+Pan, and CMB+Lens+BAO+Pan+CC---$\Omega_{\kappa 0}$ is predicted to be zero with a precision level of $\sim 2\times10^{-3}$. We notice that the introduction of the expansion anisotropy does not change the results (see Fig.~\ref{fig:NF5}); at 95\% C.L., the upper bounds on the present-day expansion anisotropy remain stable for the different combinations of data sets, namely, between $\log_{10}(\Omega_{\sigma 0})<-17.9$ (CMB+Lens) and $\log_{10}(\Omega_{\sigma 0})<-17.1$ (CMB+Lens+BAO+Pan) for the An-$\Lambda$CDM model, and get slightly weakened when non-zero spatial curvature is allowed, between $\log_{10}(\Omega_{\sigma 0})<-17.7$ (CMB+Lens) and $\log_{10}(\Omega_{\sigma 0})<-17.1$ (CMB+Lens+BAO+Pan+CC) for the An-$o\Lambda$CDM model (see Fig.~\ref{fig:NF6}). We observe that the inclusion of the data sets BAO, Pan, CC or their combinations in the analyses with the CMB+Lens data does not improve the constraints on the expansion anisotropy rather it slightly relaxes the upper bounds on it (see Fig.~\ref{fig:NF6}). Also, irrespective of the introduction of the spatial curvature, even the loosest upper bound on expansion anisotropy, $\Omega_{\sigma0}\lesssim 10^{-17}$ at 95\% C.L. (CMB+Lens+BAO+Pan+CC), obtained in our analyses with the CMB+Lens data is about two orders of magnitude tighter than $\Omega_{\sigma0}\lesssim10^{-15}$, obtained in Ref.~\cite{Akarsu:2019pwn} for the An-$\Lambda$CDM model by using the CMB information along with the $H(z)$, BAO, and Pantheon data. With regard to the $H_0$ tension, it was reported in Ref.~\cite{Akarsu:2019pwn} that the predicted mean values of $H_0$ in the case of the An-$\Lambda$CDM model are systematically larger compared to the $\Lambda$CDM model, though not significantly. However, with the tighter constraints on the expansion anisotropy found in the current work, it has almost no effect on the predicted $H_0$ value compared to the $\Lambda$CDM model. Thus, we conclude that the introduction of expansion anisotropy on top of the standard $\Lambda$CDM model, or its extension allowing spatial curvature as well, does not offer a possible relaxation to the so-called $H_0$ tension; at least, through its stiff fluid-like contribution to the Friedmann equation.\footnote{However, it is important to recall here that we have limited our investigations to orthogonal (non-tilted) models; otherwise it is possible to arrive at completely different conclusions, see, e.g., Refs.~\cite{Krishnan:2021dyb,Krishnan:2021jmh,Luongo:2021nqh} as well as Sec.~\ref{sec:intro} for a discussion and further references on this matter.}

Finally, in order to compare the goodness of the statistical fits of the models, we calculate their Bayesian evidences. To compare a model $\mathcal{M}_a$ with a reference model $\mathcal{M}_b$, we compute the ratio of the posterior probabilities of the models, given by
\begin{equation}
\frac{P(\mathcal{M}_a|D)}{P(\mathcal{M}_b|D)} = B_{ab}\frac{P(\mathcal{M}_a)}{P(\mathcal{M}_b)},
\end{equation}
where $B_{ab}$ is the Bayes' factor given by
\begin{equation}
    B_{ab}=\frac{\mathcal{E}(D|\mathcal{M}_a)}{\mathcal{E}(D|\mathcal{M}_b)}\equiv\frac{\mathcal{E}_a}{\mathcal{E}_b}.
\end{equation}
The Bayes' factor provides the strength of evidence according to Jeffreys' scale~\cite{Kass:1995loi}, viz., it is weak or inconclusive for $|\ln B_{ab}|\in[0,1)$, definite or positive for $|\ln B_{ab}|\in[1,3)$, strong for $|\ln B_{ab}|\in[3,5)$, and very strong for $|\ln B_{ab}|>5$. Here, we choose the $o\Lambda$CDM and $\Lambda$CDM models as the reference models for the An-$o\Lambda$CDM and An-$\Lambda$CDM models, respectively. This is because we have used the compressed CMB likelihood information from the Planck chains for the flat and non-flat $\Lambda$CDM models separately in our observational analyses. In Table~\ref{BE}, we show Bayesian evidences as well as chi-squared minimum differences, where
\begin{align}
\nonumber
\Delta \ln B_{o\rm A} &= \ln {\cal{E}}_{ o\Lambda \rm CDM} - \ln {\cal{E}}_{ {\rm An}-o\Lambda \rm CDM},\\
\nonumber
\Delta \chi^2_{o\rm A} &= \chi^2_{{\rm min}, o\Lambda \rm CDM}-\chi^2_{{\rm min},{\rm An}-o\Lambda \rm CDM},\\
\nonumber
\Delta \ln B_{\Lambda \rm A} &= \ln {\cal{E}}_{\Lambda \rm CDM} - \ln {\cal{E}}_{\rm An-\Lambda \rm CDM },\\
\nonumber
\Delta \chi^2_{\Lambda\rm A} &= \chi^2_{{\rm min},\rm \Lambda \rm CDM}-\chi^2_{{\rm min},\rm An-\Lambda \rm CDM}.
\end{align}
In all of the cases, we notice that the strength of evidence is weak or inconclusive. It is also consistent with the chi-squared minimum differences. It implies that the amount of anisotropic expansion obtained in our analyses is very much allowed on top of the $\Lambda$CDM and $o\Lambda$CDM models by the observational data under consideration.

\begin{table}[t]
\small
\caption{Bayesian evidences and chi-squared  minimum differences. Here, $\Delta \ln B_{o\rm A} = \ln {\cal{E}}_{ o\Lambda \rm CDM} - \ln {\cal{E}}_{ {\rm An}-o\Lambda \rm CDM}$, $\Delta \chi^2_{o\rm A} = \chi^2_{{\rm min}, o\Lambda \rm CDM}-\chi^2_{{\rm min},{\rm An}-o\Lambda \rm CDM}$, $\Delta \ln B_{\Lambda \rm A} = \ln {\cal{E}}_{\Lambda \rm CDM} - \ln {\cal{E}}_{\rm An-\Lambda \rm CDM }$, and $\Delta \chi^2_{\Lambda\rm A} = \chi^2_{{\rm min},\rm \Lambda \rm CDM}-\chi^2_{{\rm min},\rm An-\Lambda \rm CDM}$.}
\label{BE}
\renewcommand{\arraystretch}{1}
\resizebox{0.95\columnwidth}{!}
    {
\begin{tabular} { |l| c| c|c|c| }  \hline 

 & $\Delta \ln B_{o\rm A}$
 & $\Delta  \chi^2_{o \rm A}$
 & $\Delta \ln B_{\Lambda \rm A}$
 & $\Delta  \chi^2_{\Lambda \rm A}$
\\

\hline
 CMB &  0.72 & -0.04  &  0.86  & 0.01 \\
  \hline
 CMB+Pan &  0.60 & 0.03  &  0.57  & 0.08\\
  \hline
 CC &    0.86 & -0.08 &  0.78 &  -0.02 \\
  \hline
  Pan &  -0.02 & 0.41 & 0.72 &  0.35 \\
  \hline
  CC+Pan &  0.46 & 0.08 &  0.37 & 0.06  \\
  \hline
  BAO+Pan &  0.17 & -0.09 & 0.33  & 0.00 \\
  \hline
  CMB+Lens &  0.61 & 0.12 & 0.29 & -0.02  \\
  \hline
  CMB+Lens+BAO &  0.57  & 0.13 & 0.54 &  0.05  \\ 
    \hline
  CMB+Lens+Pan &   0.54 & 0.15 &  0.54 & -0.04\\
    \hline
CMB+Lens+BAO+Pan &  0.52 & -0.15 &  0.53 & 0.05 \\
    \hline
CC+CMB+Lens+BAO+Pan &  0.74 &  0.20 & 0.52 & -0.08 \\
    \hline
\end{tabular}}
\end{table}

\section{Conclusions}
\label{conclusions}

\begin{figure*}[hbt!]
\centering
\begin{subfigure}
    \centering 
    \includegraphics[width=8.5cm]{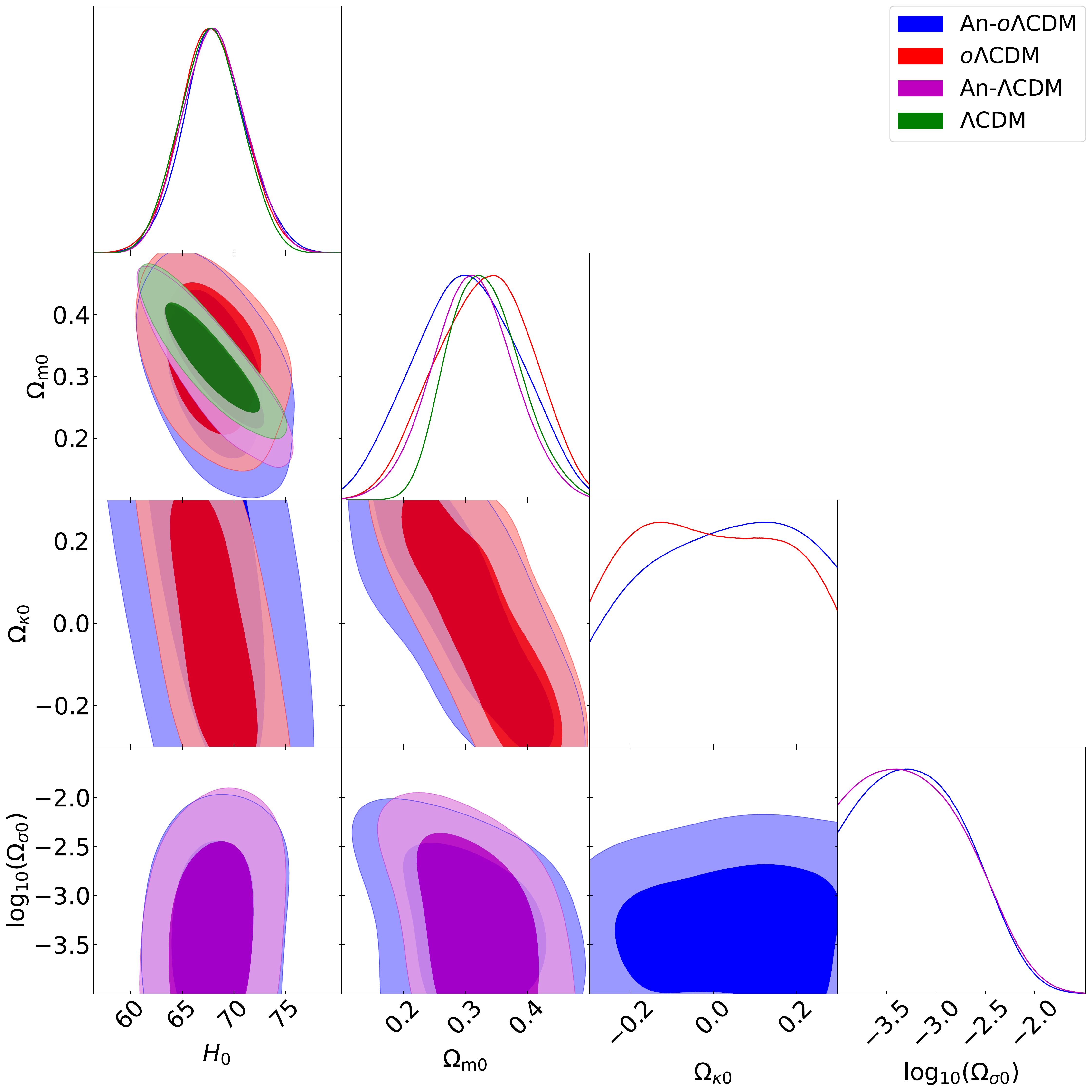}
    \label{fig:sub1}
\end{subfigure}
\begin{subfigure}
    \centering 
    \includegraphics[width=8.5cm]{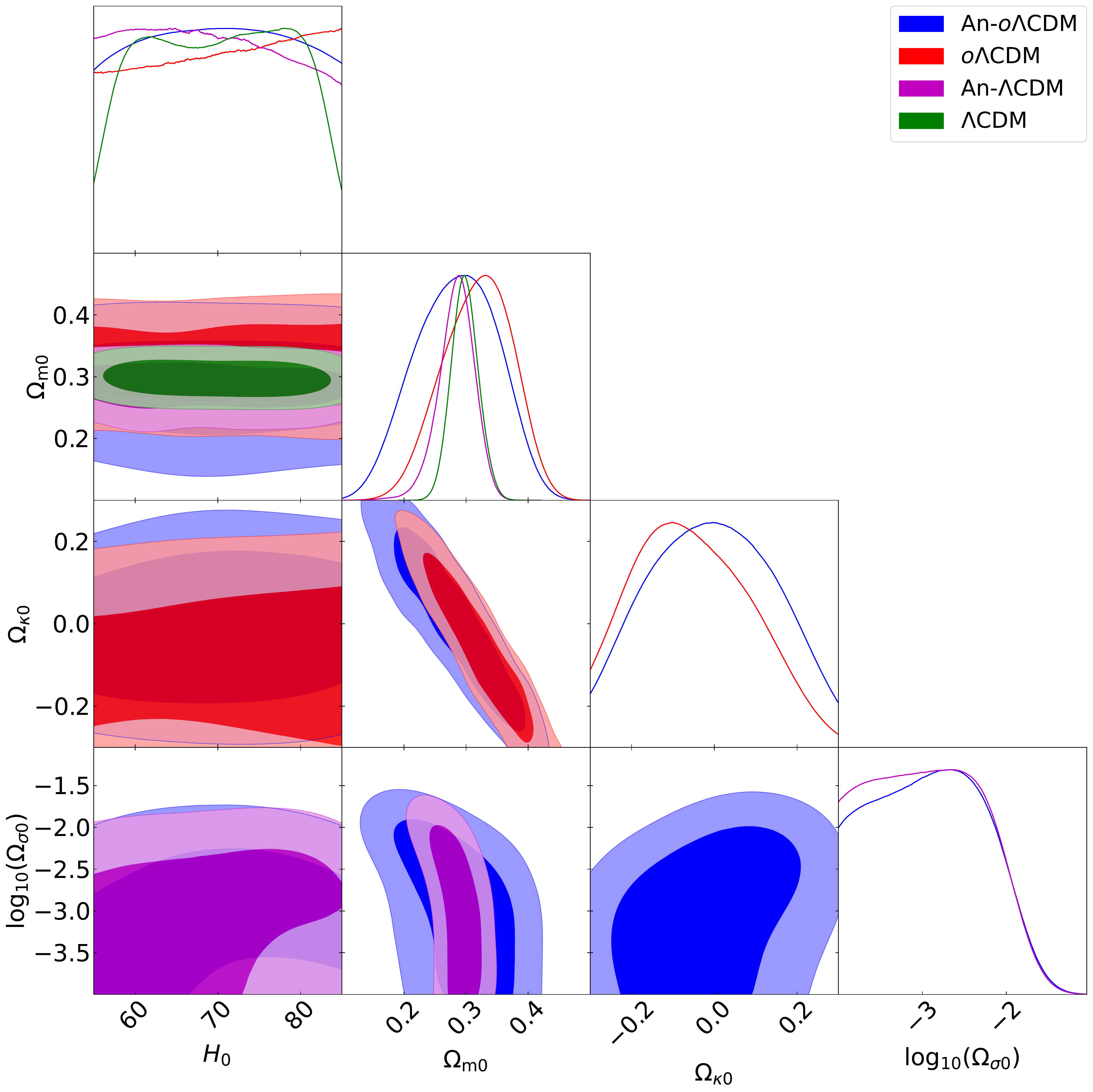}
    \label{fig:sub2}
\end{subfigure}
\begin{subfigure}
    \centering 
    \includegraphics[width=8cm]{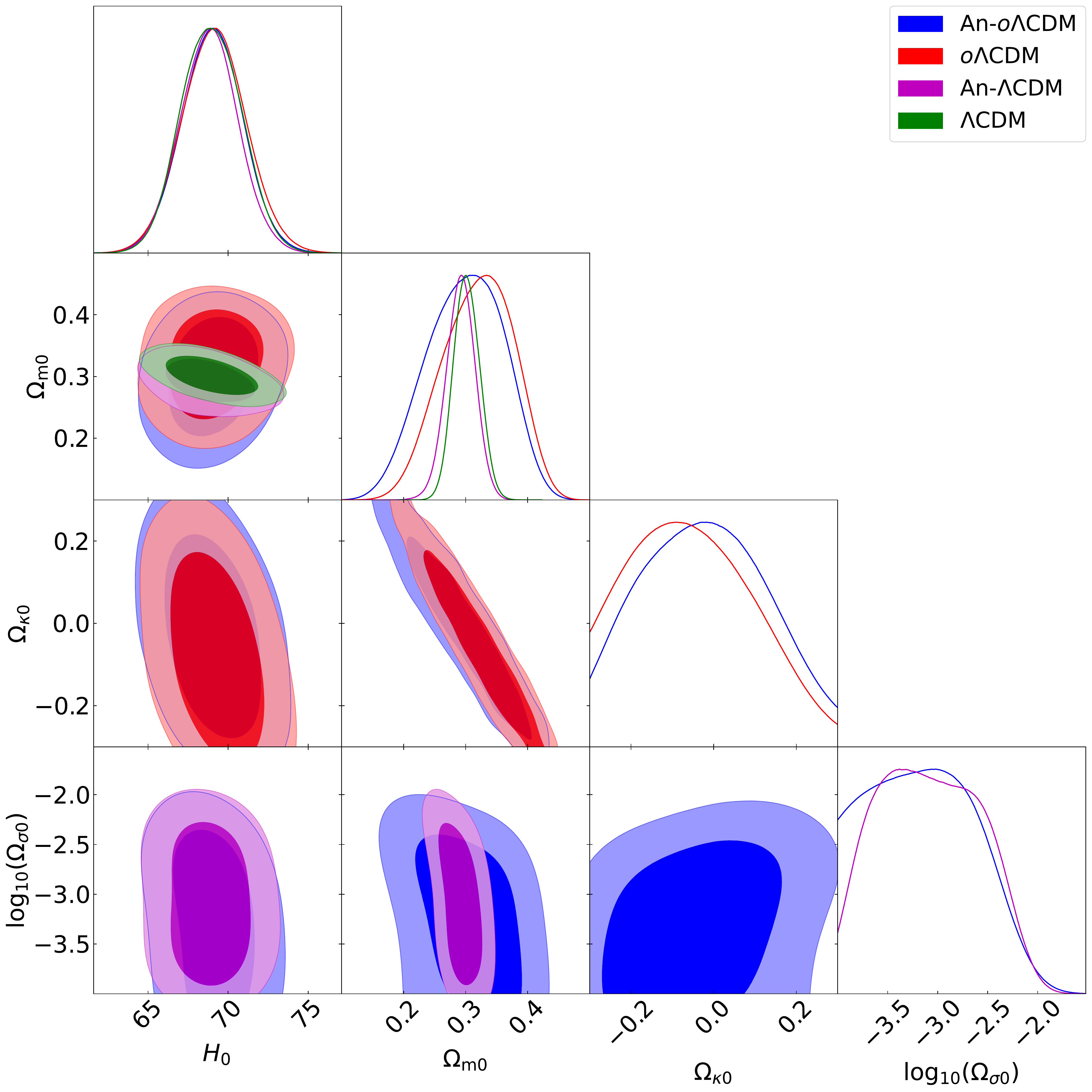}
    \label{fig:sub1}
\end{subfigure}
\caption{One-dimensional and two-dimensional marginalized confidence regions (68\% and 95\%  C.L.) from CC (top-left), Pan (top-right), and CC+Pan (bottom) data for An-$o\Lambda$CDM, $o\Lambda$CDM, An-$\Lambda$CDM, and $\Lambda$CDM model parameters.  The parameter $H_{\rm 0}$ is measured in units of km s${}^{-1}$ Mpc${}^{-1}$.}
\label{fig:NF1}
\end{figure*}

\begin{figure*}[hbt!]
\centering
\begin{subfigure}
    \centering 
    \includegraphics[width=8.5cm]{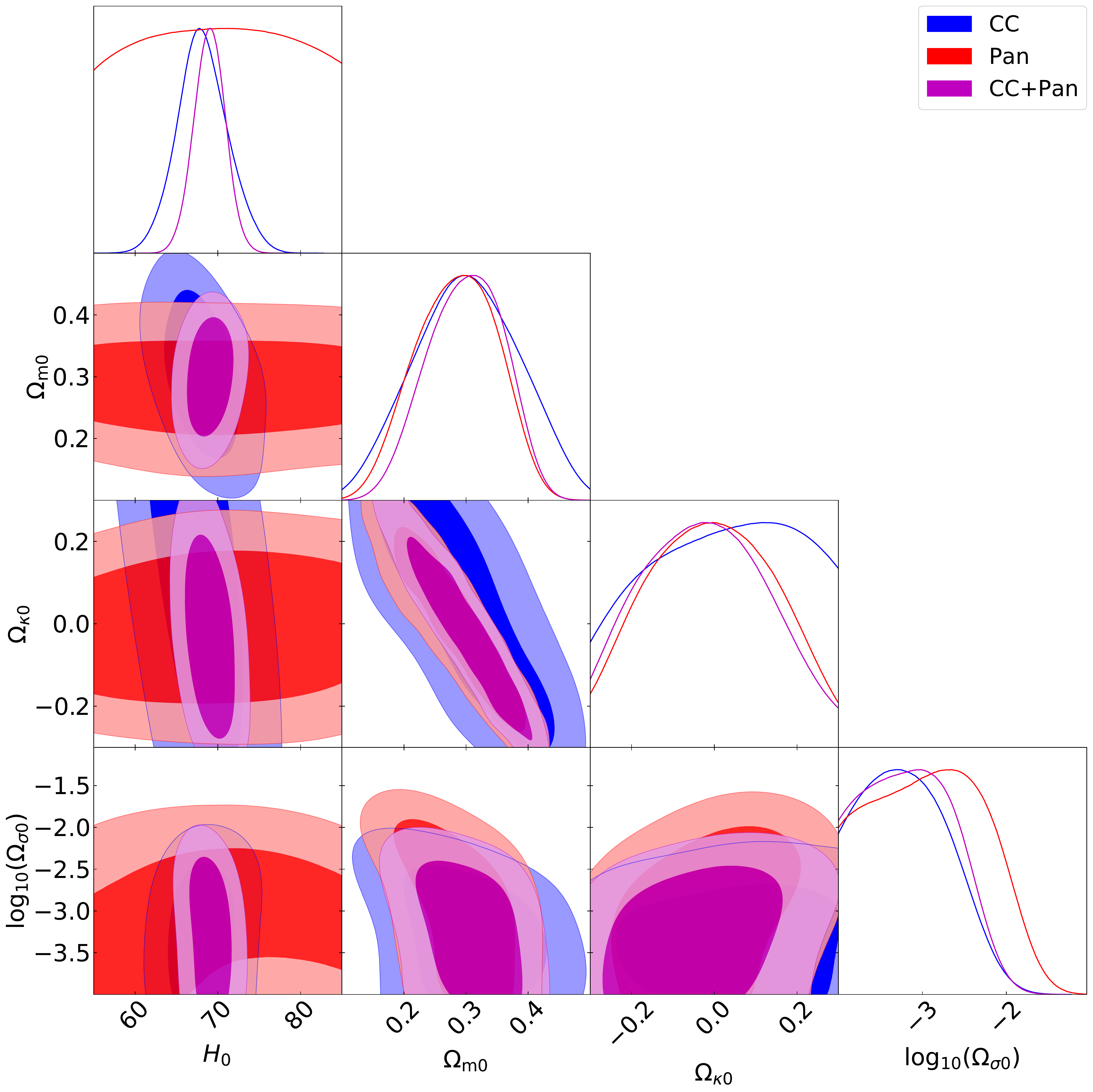}
    \label{fig:sub1}
\end{subfigure}
\begin{subfigure}
    \centering 
    \includegraphics[width=8.5cm]{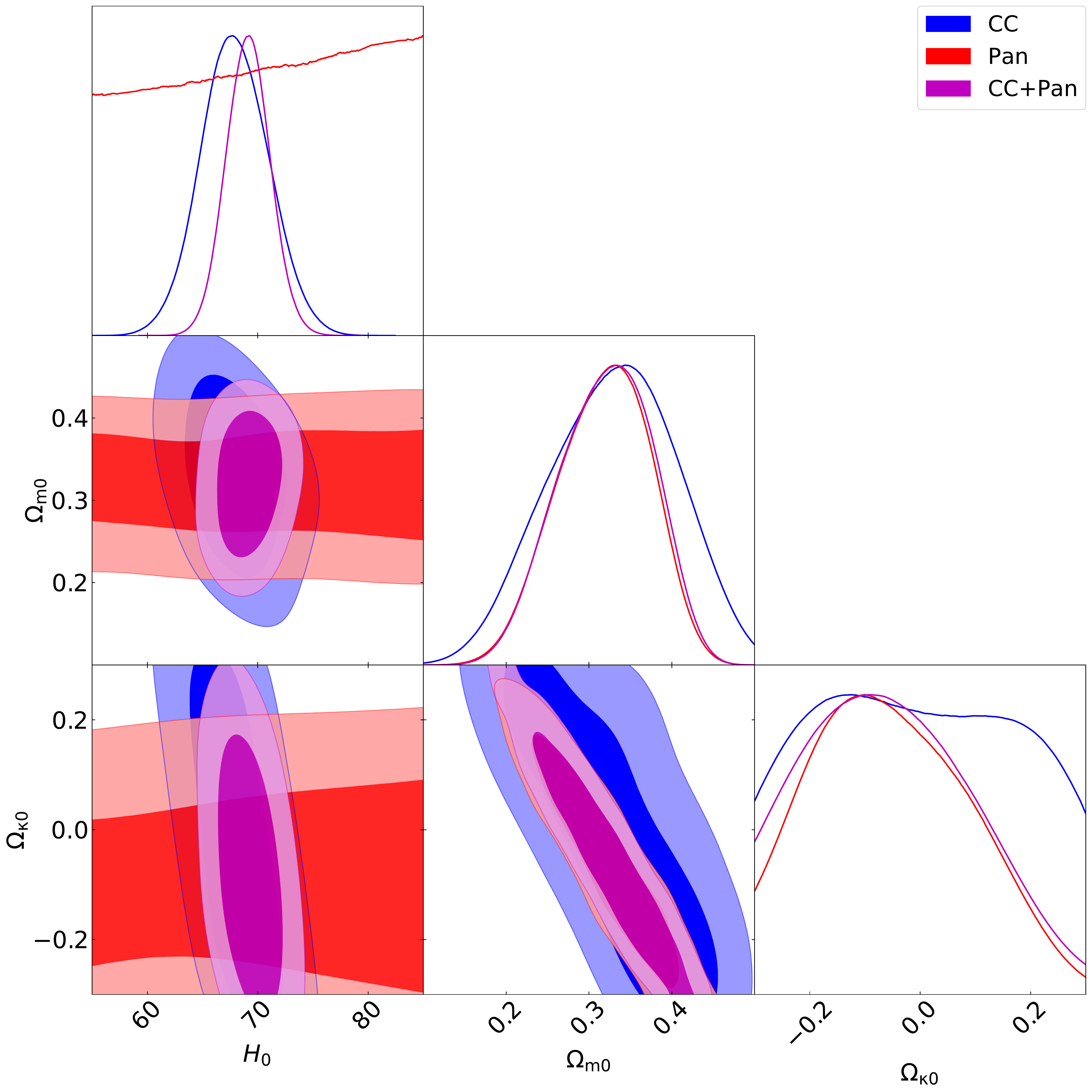}
    \label{fig:sub2}
\end{subfigure}
\begin{subfigure}
    \centering 
    \includegraphics[width=8.5cm]{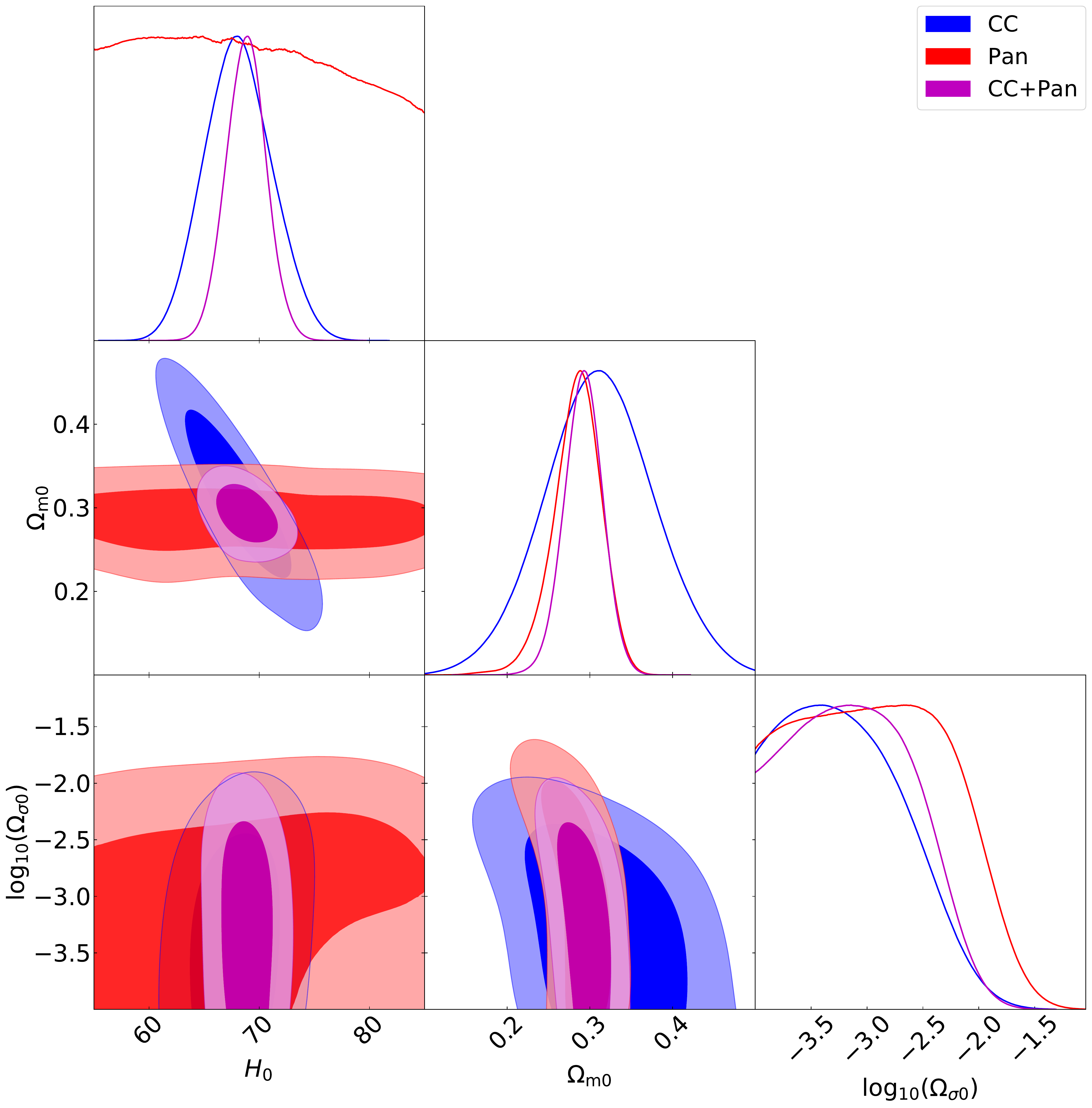}
    \label{fig:sub1}
\end{subfigure}
\begin{subfigure}
    \centering 
    \includegraphics[width=8.5cm]{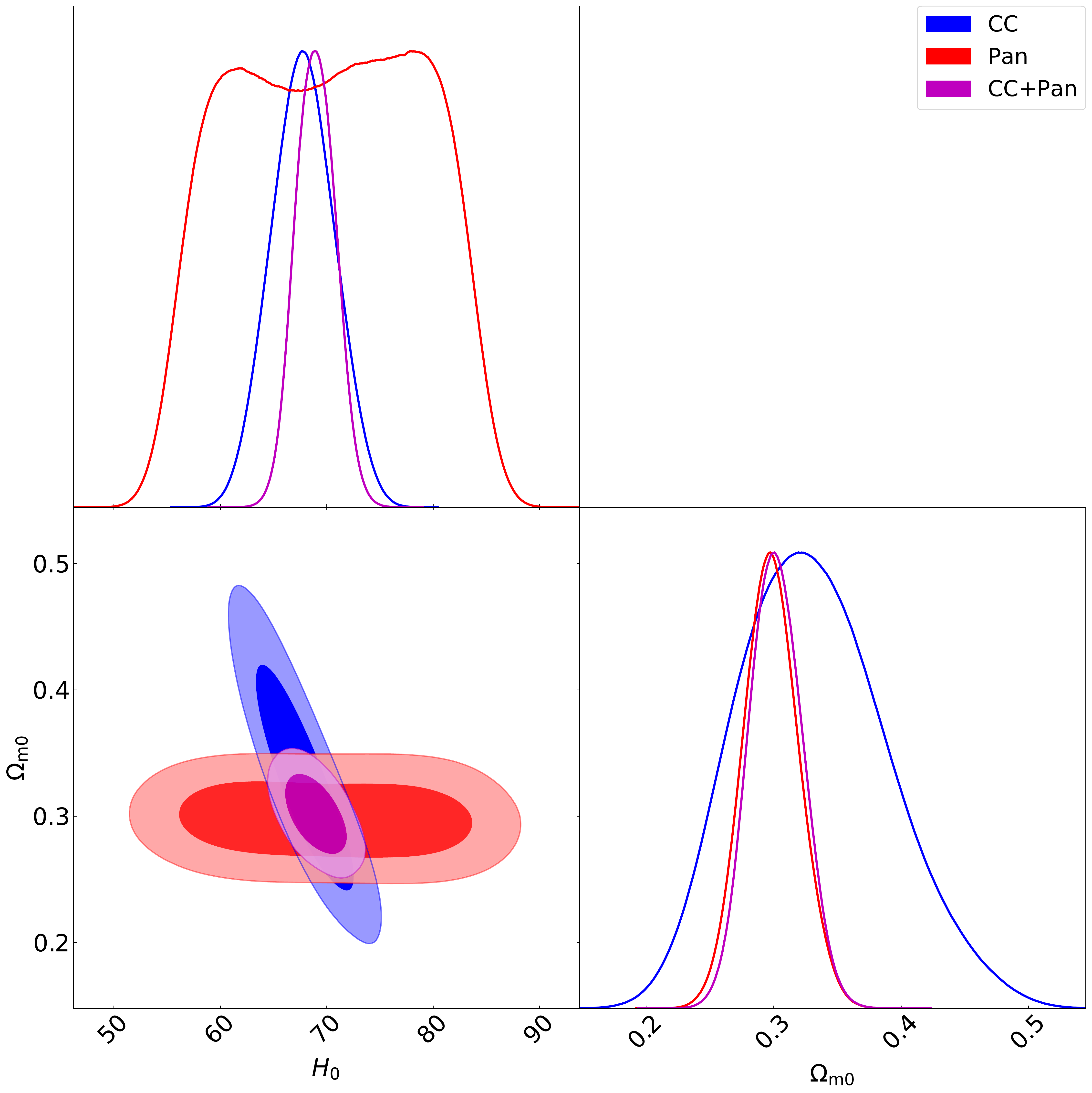}
    \label{fig:sub2}
\end{subfigure}
\caption{One-dimensional and two-dimensional marginalized confidence regions (68\% and 95\%  C.L.) of An-$o\Lambda$CDM (top-left), $o\Lambda$CDM (top-right),  An-$\Lambda$CDM (bottom-left), and $\Lambda$CDM (bottom-right) model parameters from CC, Pan, CC+Pan, and BAO+Pan data combinations. The parameter $H_{\rm 0}$ is measured in units of km s${}^{-1}$ Mpc${}^{-1}$.}
\label{fig:NF2}
\end{figure*}

\begin{figure*}[hbt!]
\centering
\begin{subfigure}
    \centering 
    \includegraphics[width=8cm]{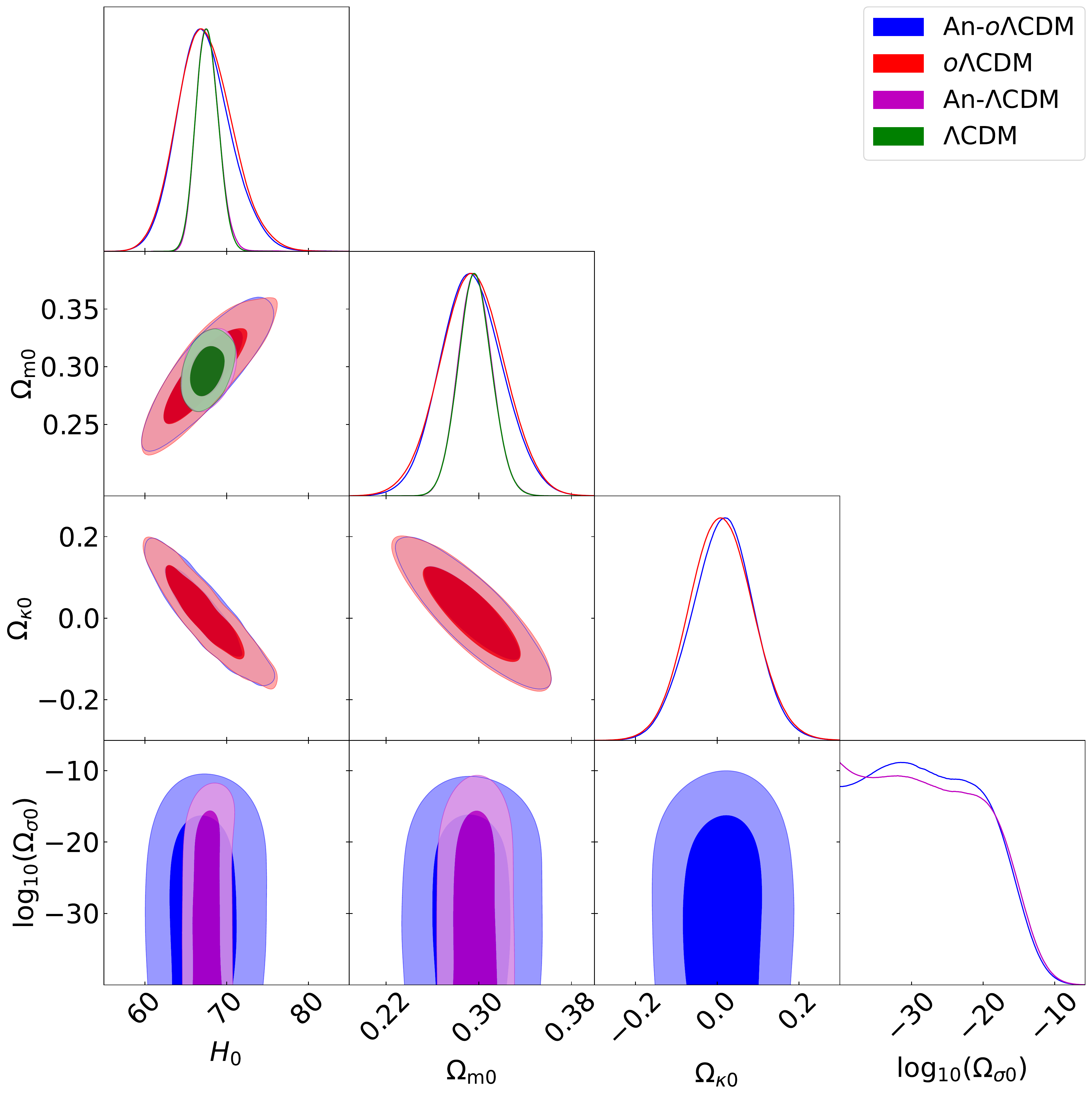}
    \label{fig:sub2}
\end{subfigure}
\begin{subfigure}
    \centering 
    \includegraphics[width=8cm]{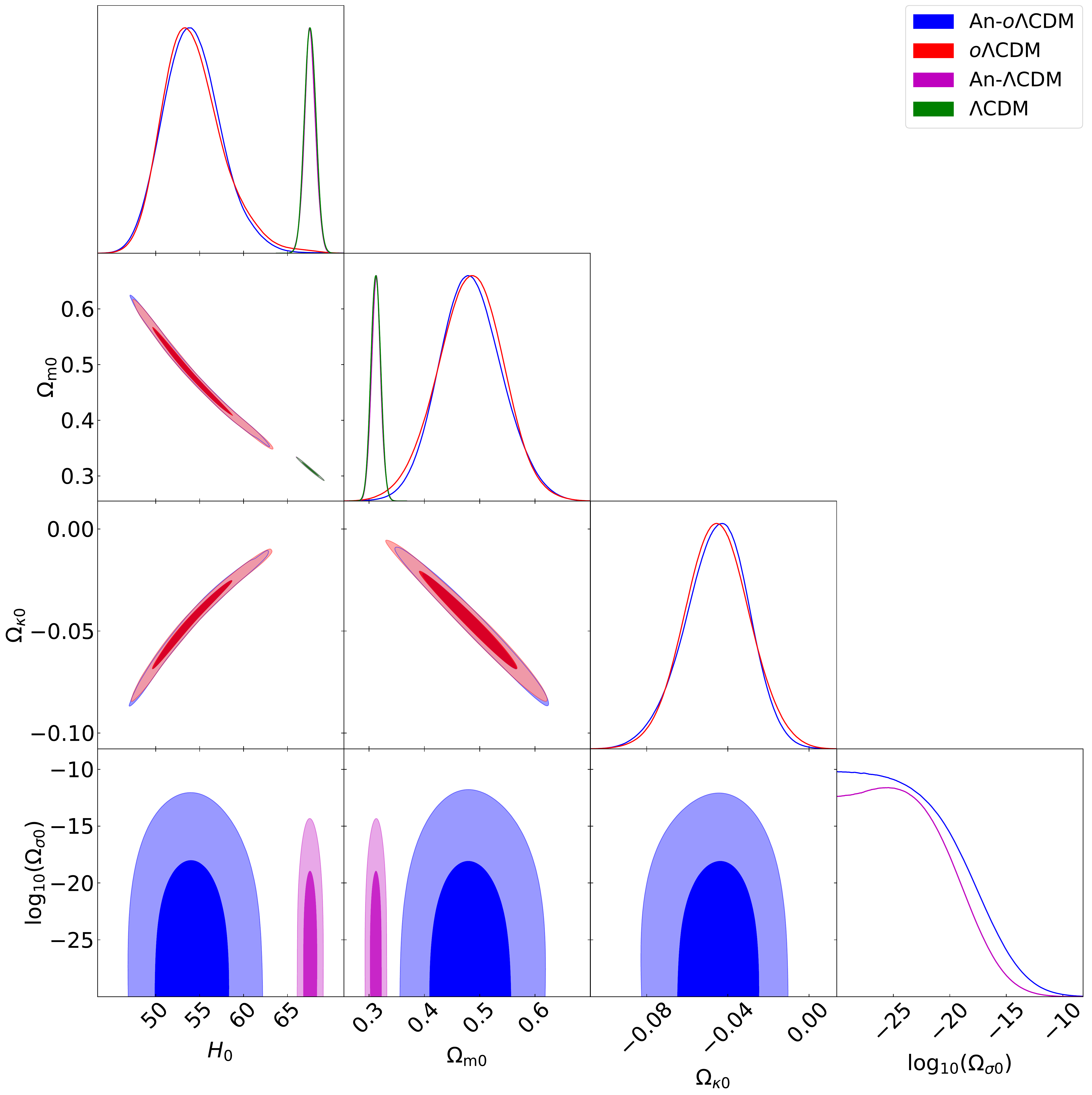}
\end{subfigure}
\begin{subfigure}
    \centering 
    \includegraphics[width=8cm]{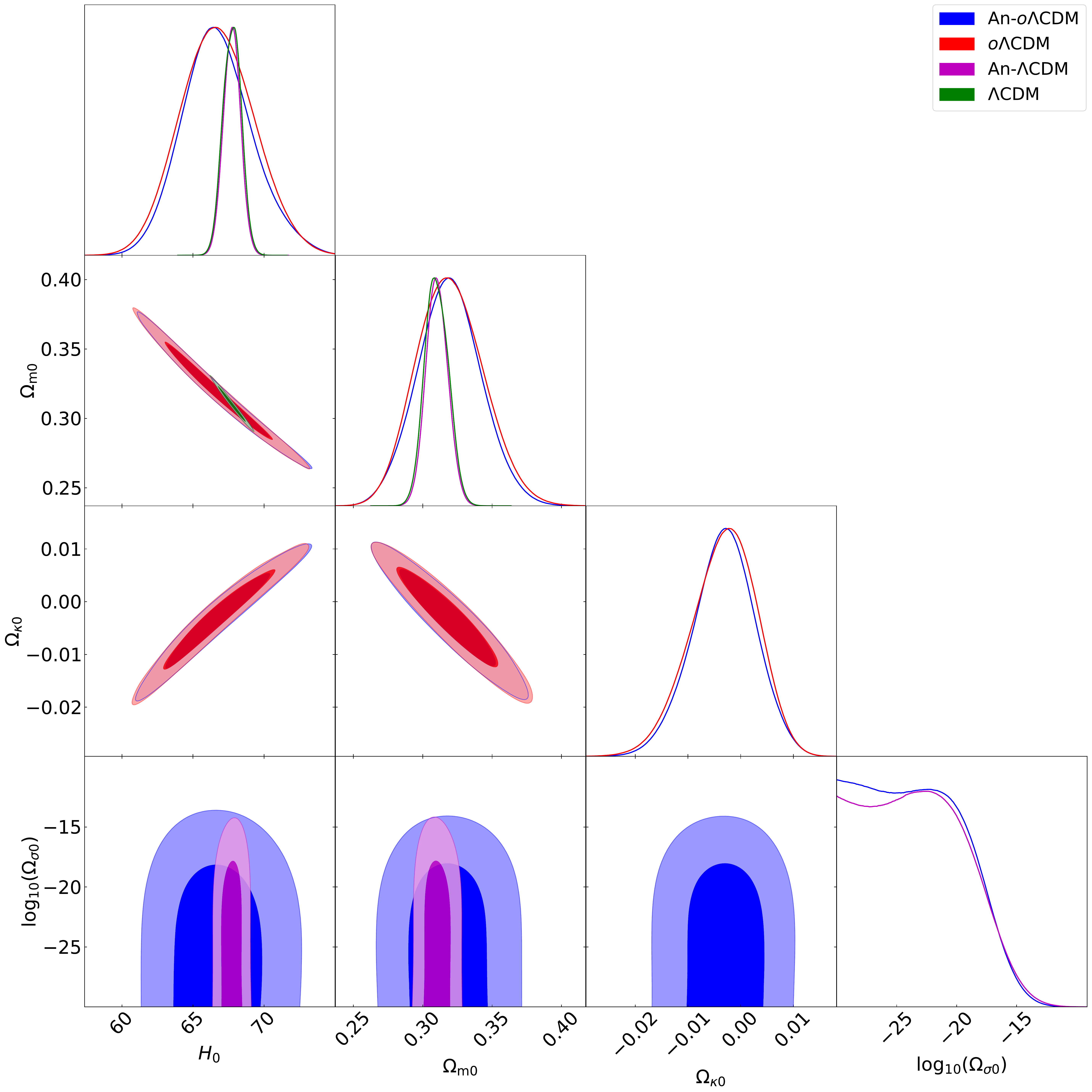}
\end{subfigure}
\caption{One-dimensional and two-dimensional marginalized confidence regions (68\% and 95\%  C.L.) from BAO+Pan (top-left), CMB (top-right) and CMB+Pan (bottom) data for An-$o\Lambda$CDM, $o\Lambda$CDM, An-$\Lambda$CDM, and $\Lambda$CDM model parameters.  The parameter $H_{\rm 0}$ is measured in units of km s${}^{-1}$ Mpc${}^{-1}$.}
\label{fig:NF3}
\end{figure*}

\begin{figure*}[hbt!]
\centering
\begin{subfigure}
    \centering 
    \includegraphics[width=8.5cm]{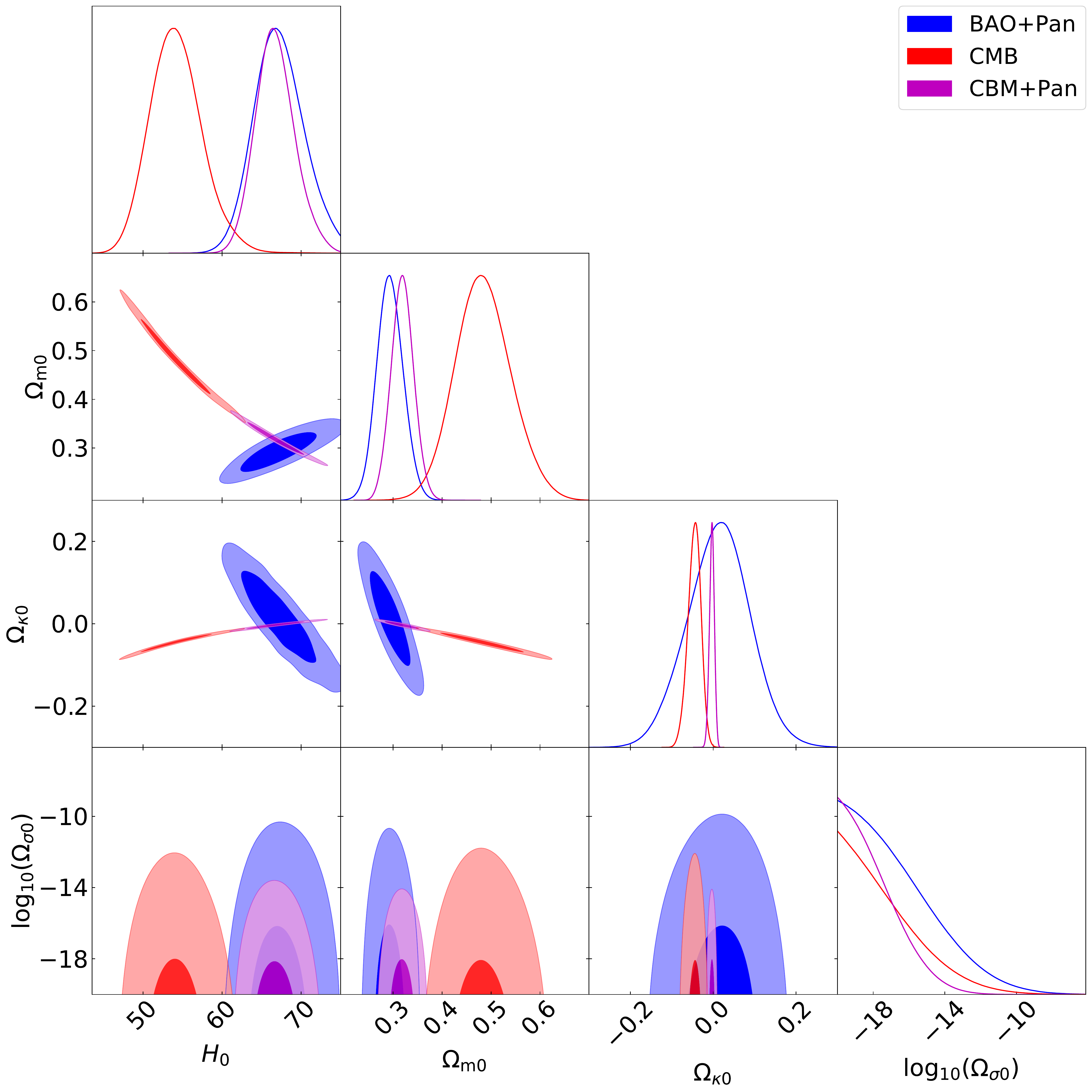}
    \label{fig:sub1}
\end{subfigure}
\begin{subfigure}
    \centering 
    \includegraphics[width=8.5cm]{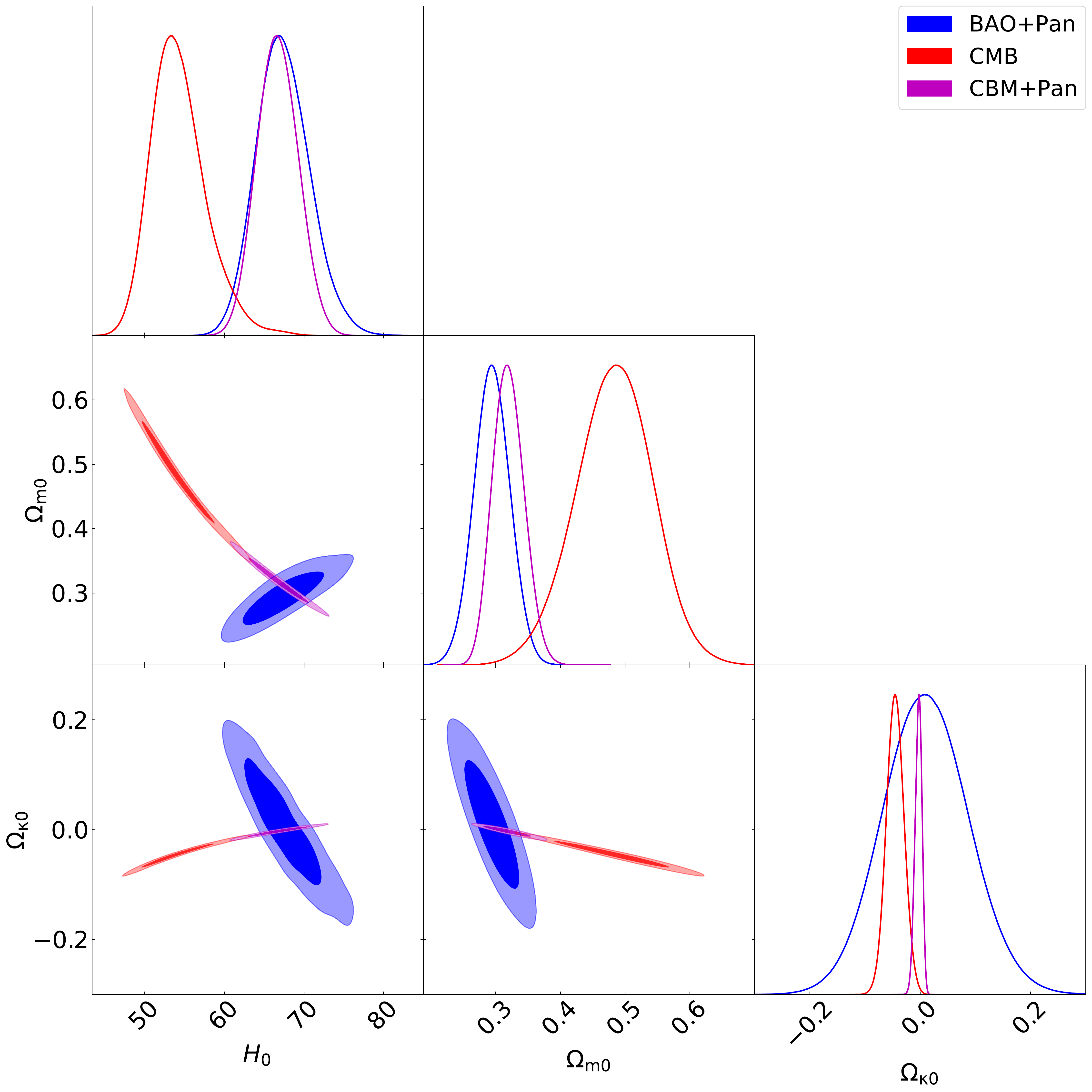}
    \label{fig:sub2}
\end{subfigure}
\begin{subfigure}
    \centering 
    \includegraphics[width=8.5cm]{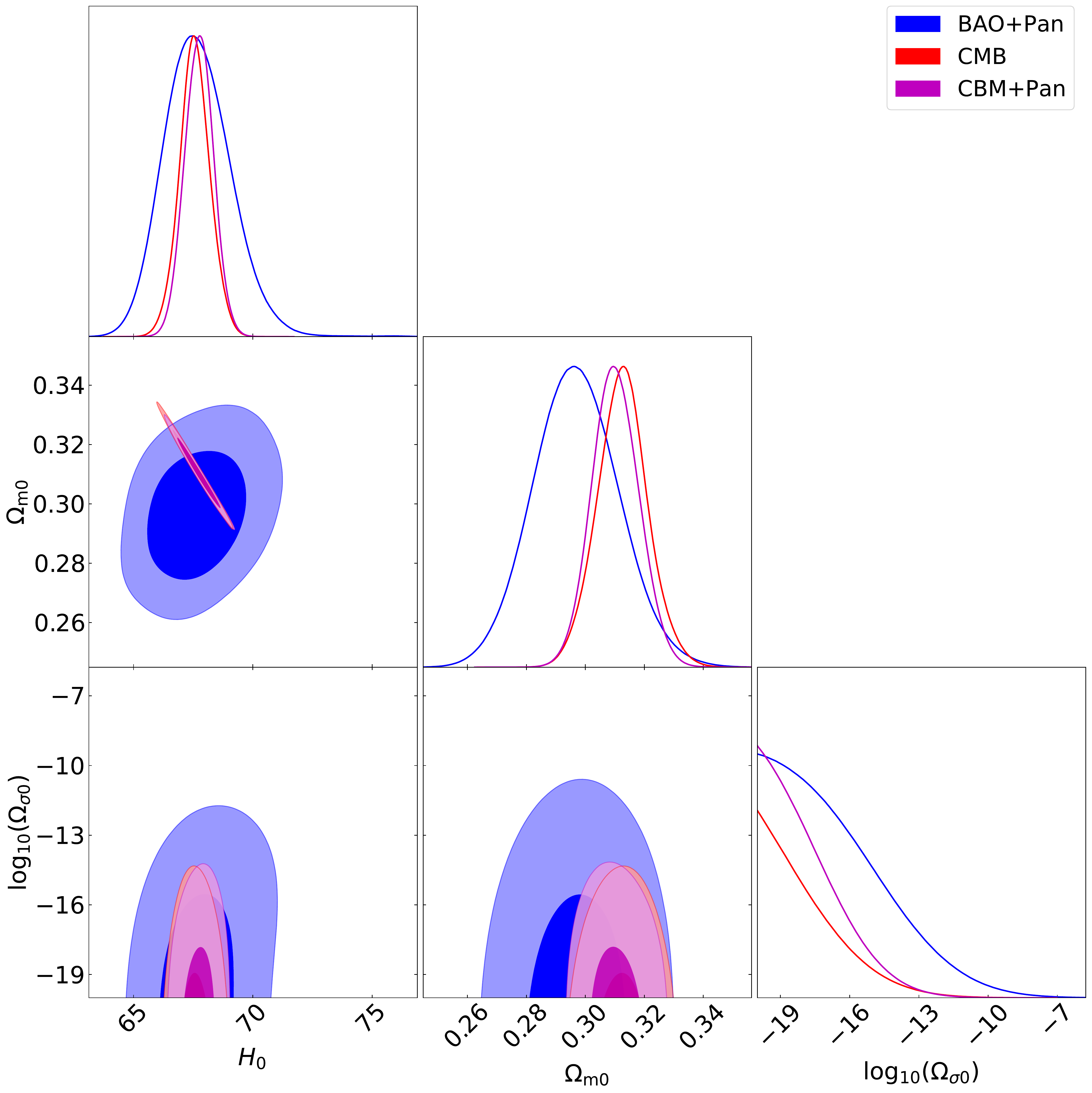}
    \label{fig:sub1}
\end{subfigure}
\begin{subfigure}
    \centering 
    \includegraphics[width=8.5cm]{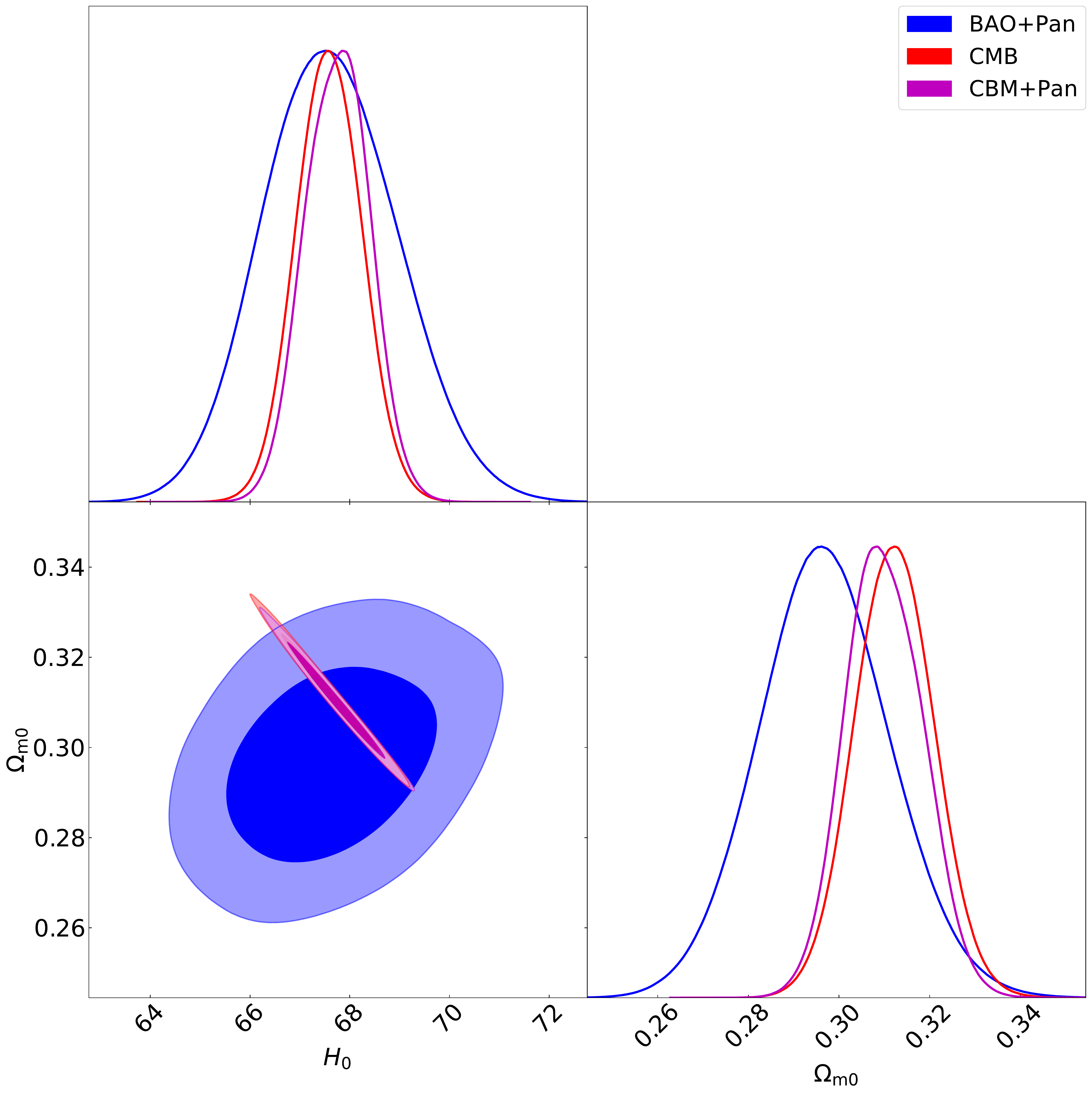}
    \label{fig:sub2}
\end{subfigure}
\caption{One-dimensional and two-dimensional marginalized confidence regions (68\% and 95\%  C.L.) of An-$o\Lambda$CDM (top-left), $o\Lambda$CDM (top-right),  An-$\Lambda$CDM (bottom-left), and $\Lambda$CDM (bottom-right)  model parameters from BAO+Pan, CMB and CMB+Pan data combinations. The parameter $H_{\rm 0}$ is measured in units of km s${}^{-1}$ Mpc${}^{-1}$.}
\label{fig:NF4}
\end{figure*}

\begin{figure*}[b]
\centering
\includegraphics[width=8.5cm]{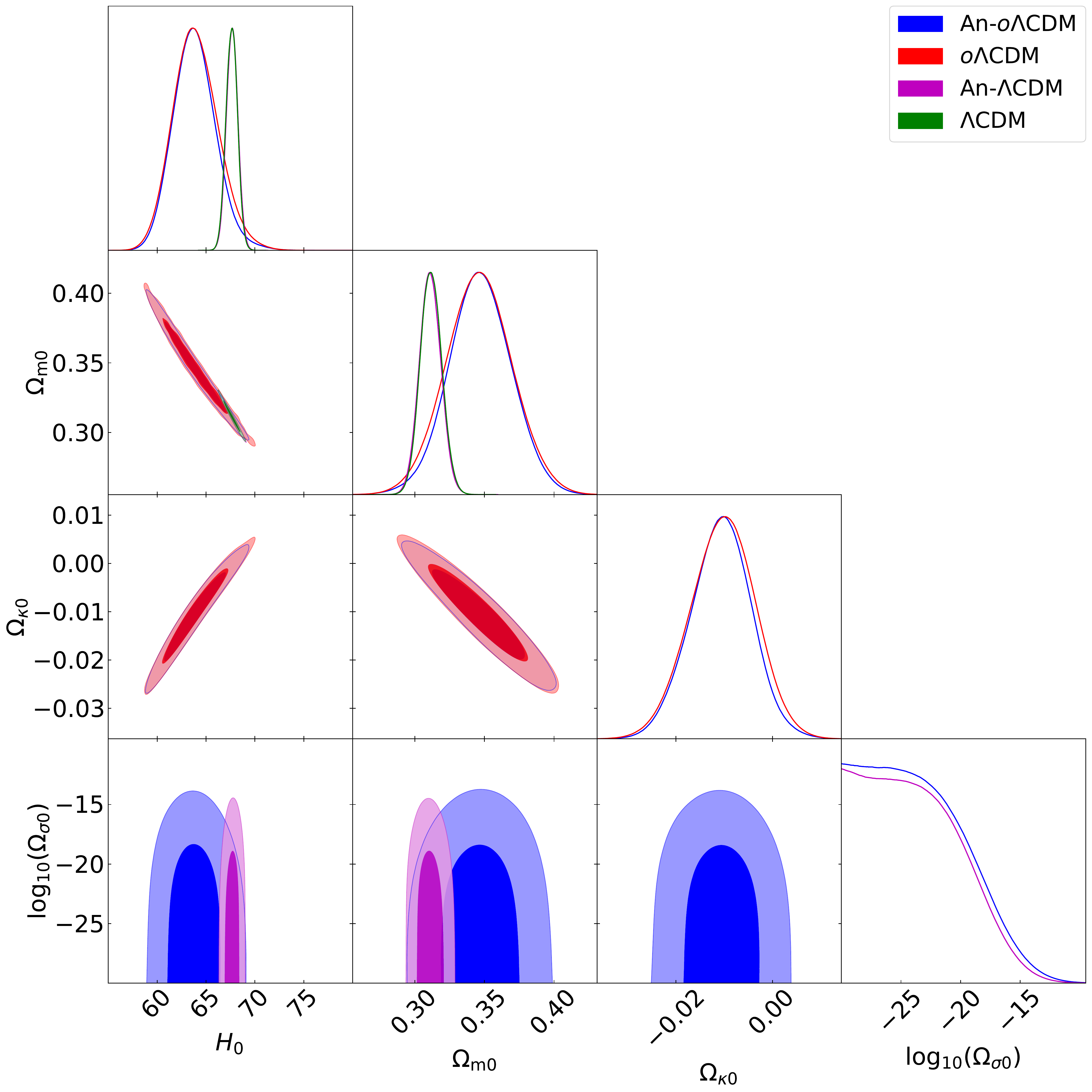}
\includegraphics[width=8.5cm]{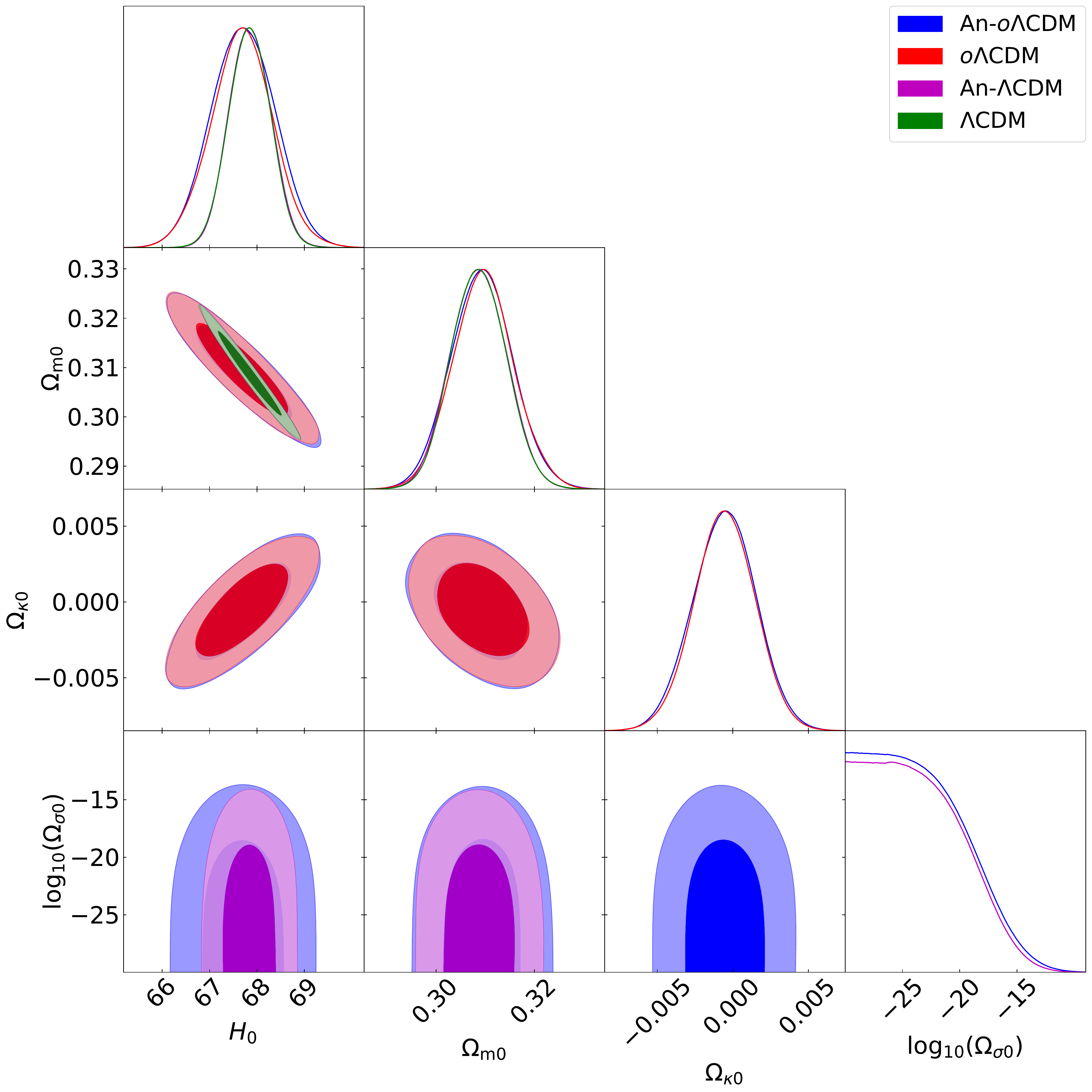}
\includegraphics[width=8.5cm]{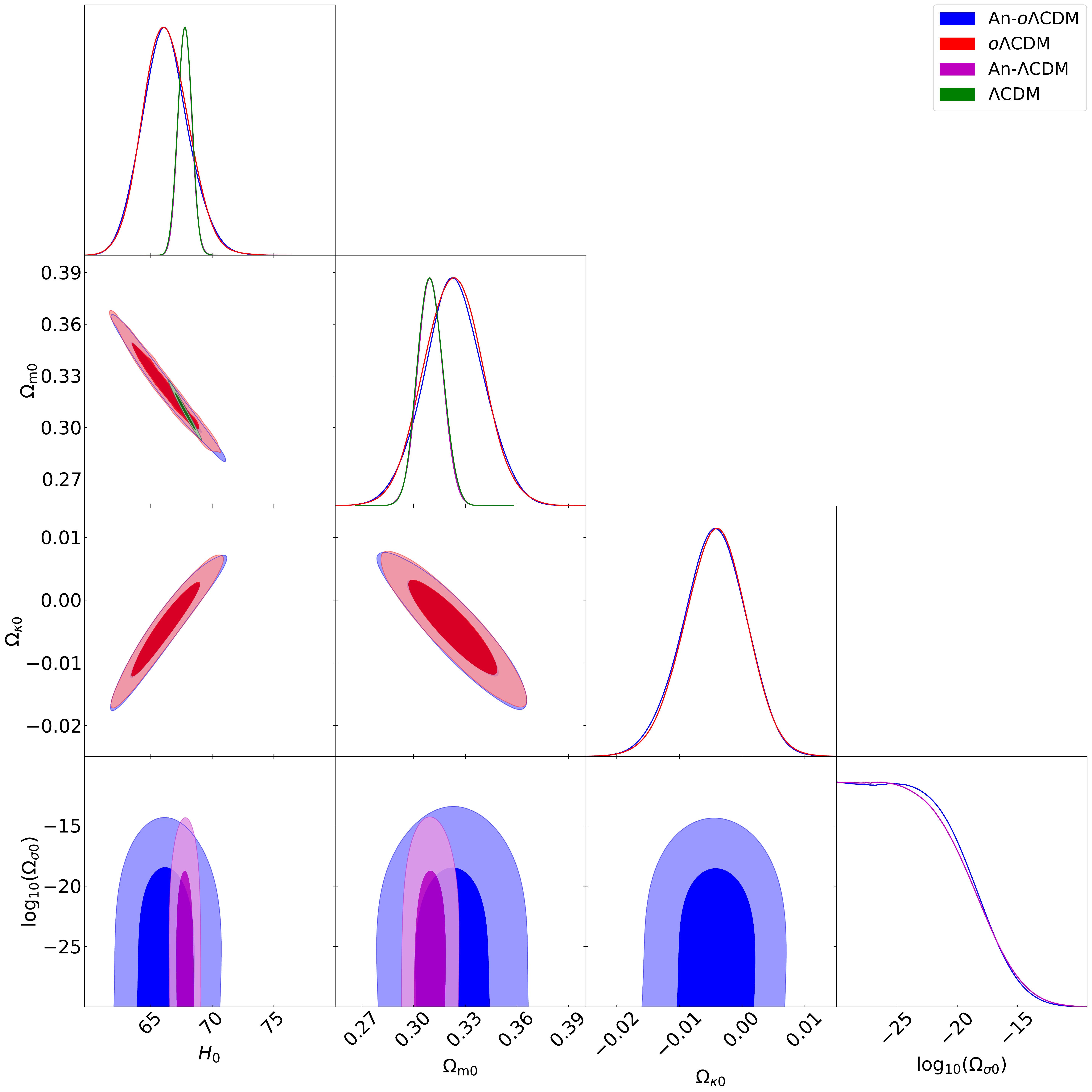}
\includegraphics[width=8.5cm]{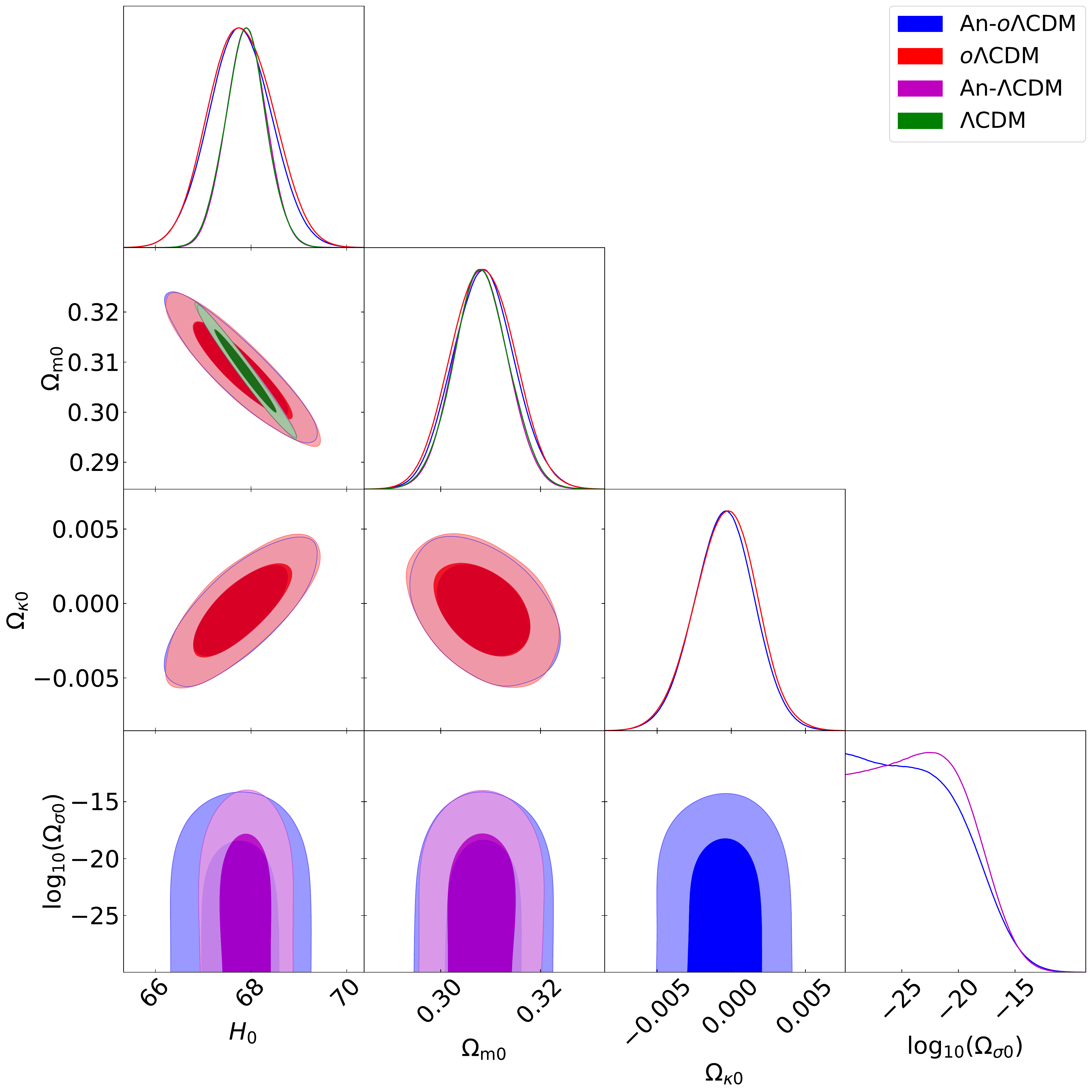}
\includegraphics[width=8.5cm]{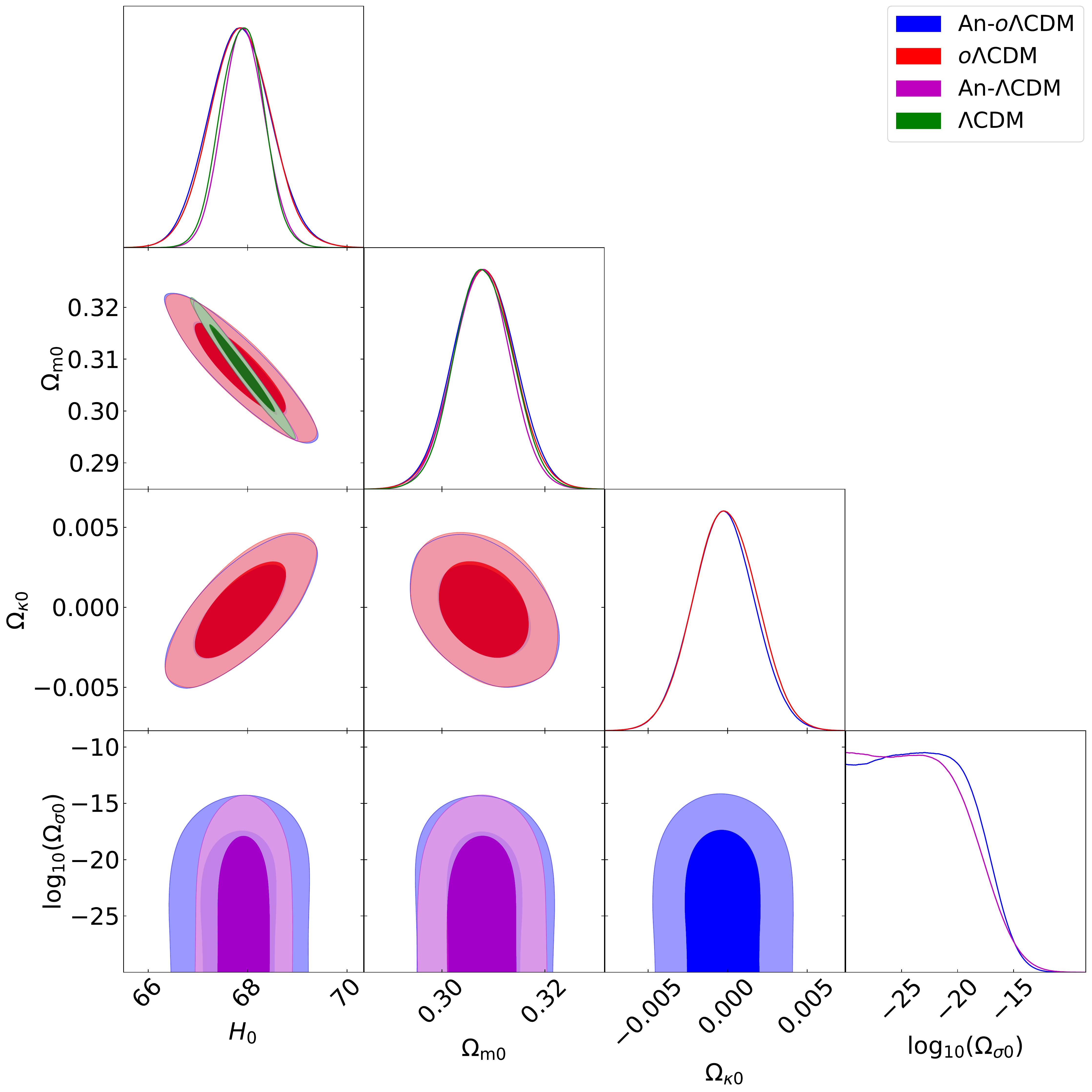}
\caption{One-dimensional and two-dimensional marginalized confidence regions (68\% and 95\%  C.L.) from CMB+Lens (top-left), CMB+Lens+BAO (top-right), CMB+Lens+Pan (middle-left), CMB+Lens+BAO+Pan (middle-right), and CMB+Lens+BAO+Pan+CC (bottom) data for An-$o\Lambda$CDM, $o\Lambda$CDM, An-$\Lambda$CDM, and $\Lambda$CDM model parameters.  The parameter $H_{\rm 0}$ is measured in units of km s${}^{-1}$ Mpc${}^{-1}$.}
\label{fig:NF5}
\end{figure*}

\begin{figure*}[hbt!]
\centering
\begin{subfigure}
    \centering 
    \includegraphics[width=8.5cm]{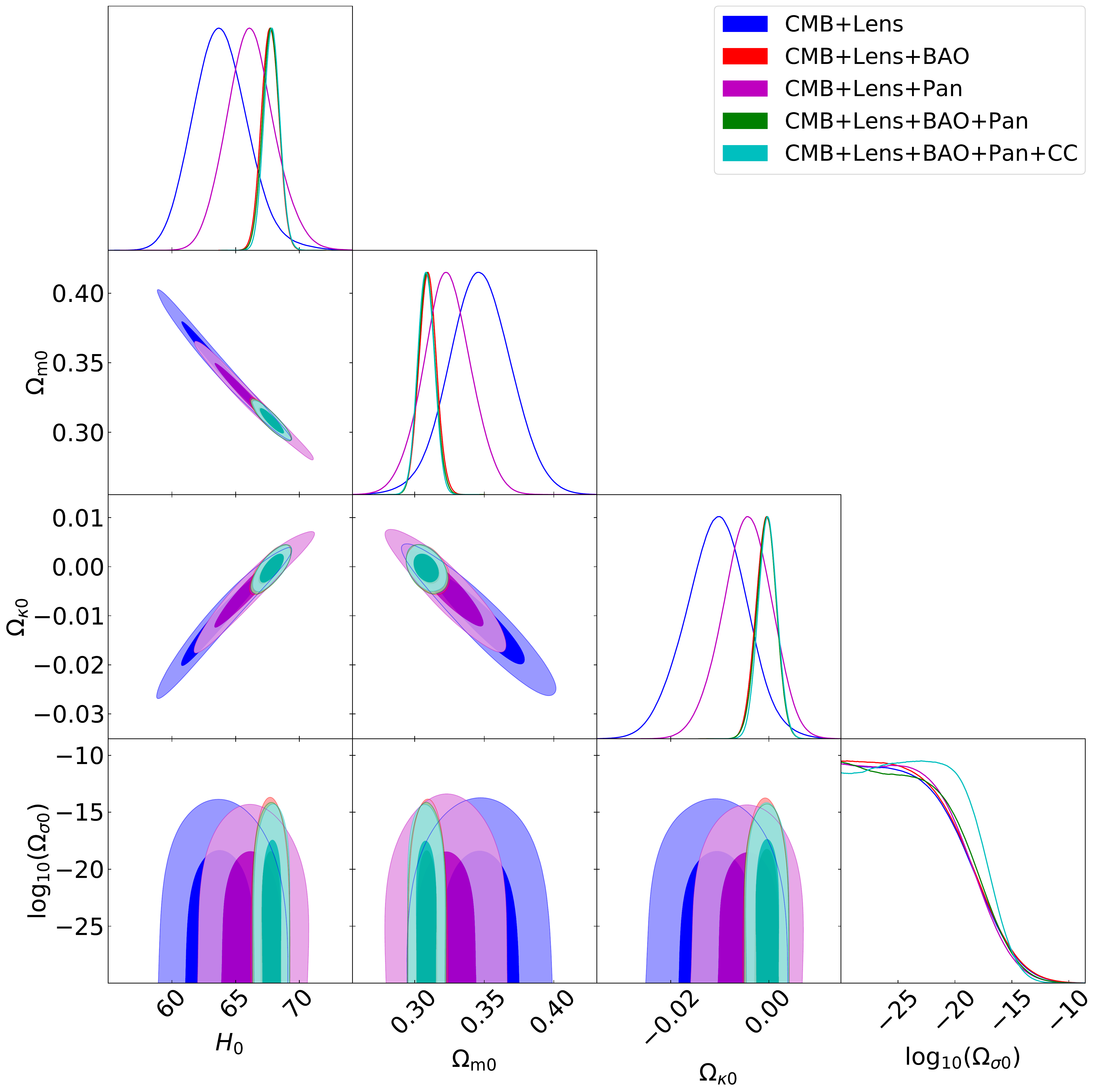}
    \label{fig:sub1}
\end{subfigure}
\begin{subfigure}
    \centering 
    \includegraphics[width=8.5cm]{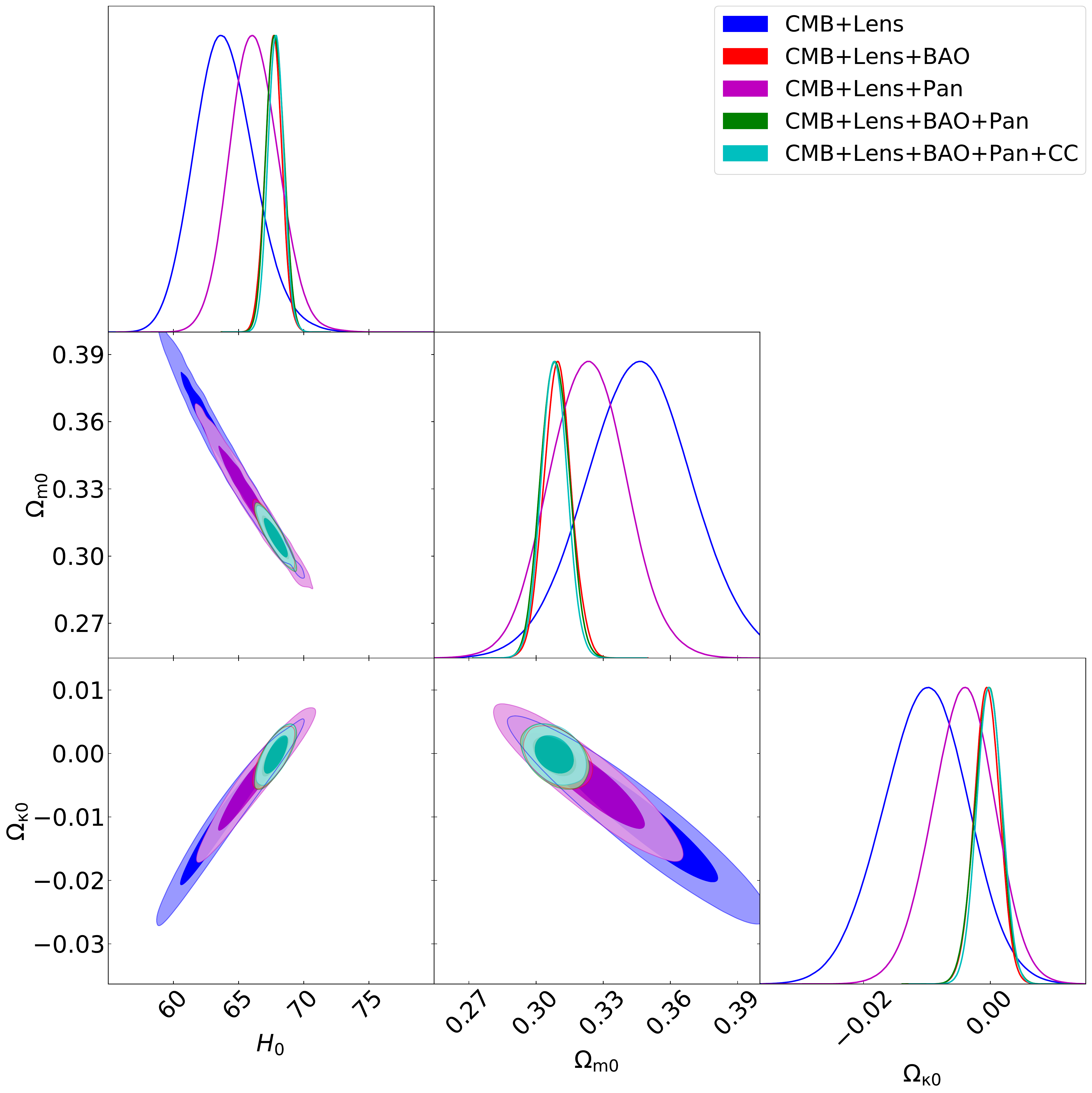}
    \label{fig:sub2}
\end{subfigure}
\begin{subfigure}
    \centering 
    \includegraphics[width=8.5cm]{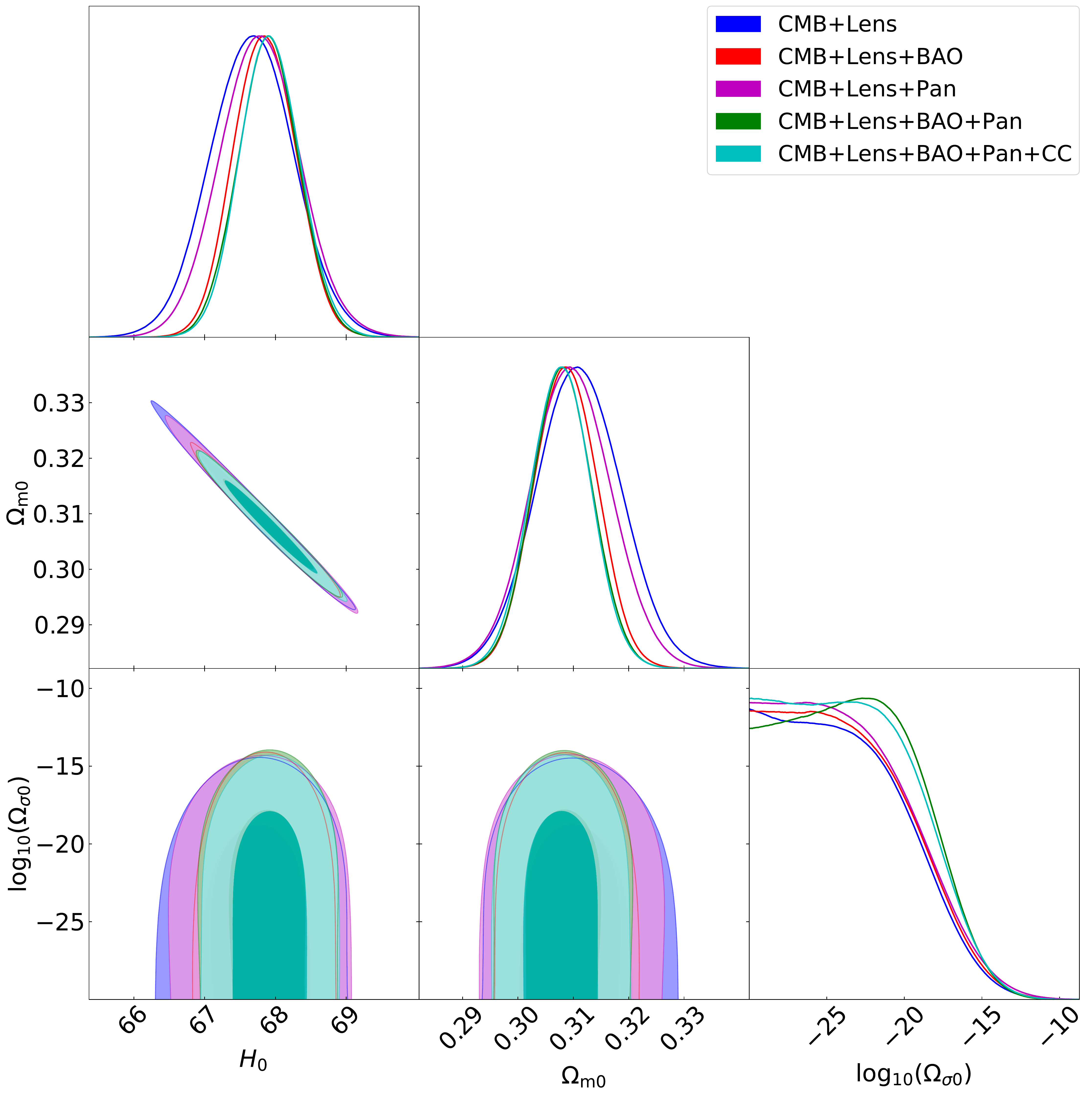}
    \label{fig:sub1}
\end{subfigure}
\begin{subfigure}
    \centering 
    \includegraphics[width=8.5cm]{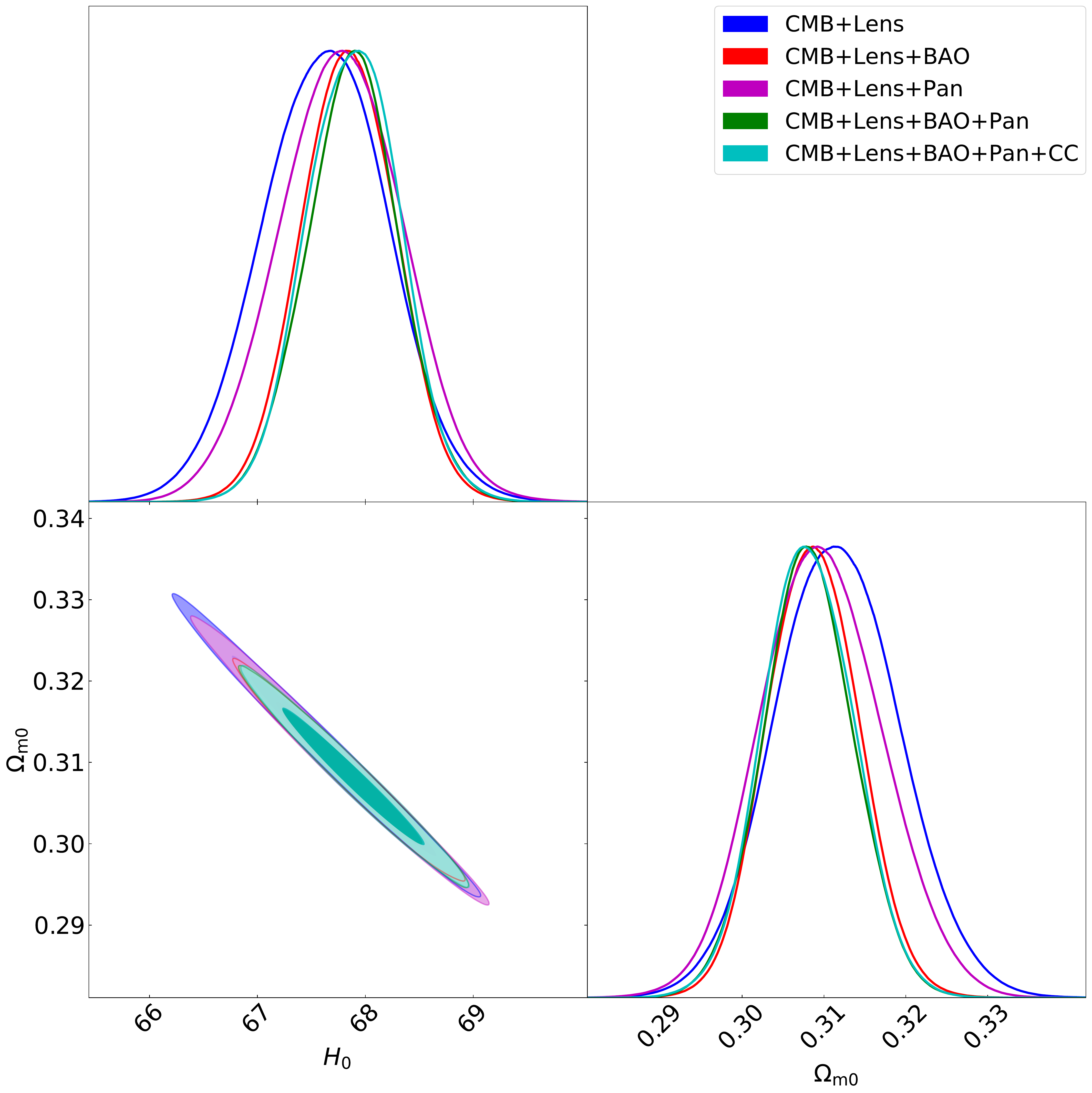}
    \label{fig:sub2}
\end{subfigure}
\caption{One-dimensional and two-dimensional marginalized confidence regions (68\% and 95\%  C.L.) of An-$o\Lambda$CDM (top-left), $o\Lambda$CDM (top-right), An-$\Lambda$CDM (bottom-left) and $\Lambda$CDM (bottom-right) model parameters from CMB+Lens, CMB+Lens+BAO, CMB+Lens+Pan, CMB+Lens+BAO+Pan, and CMB+Lens+BAO+Pan+CC data combinations. The parameter $H_{\rm 0}$ is measured in units of km s${}^{-1}$ Mpc${}^{-1}$.}
\label{fig:NF6}
\end{figure*}

In this work, we have presented an exploration of the possible advantages of the pure geometric extension of the $\Lambda$CDM model, which is strongly motivated by the fact that (i) it implies using more realistic spatial backgrounds than that of the spatially flat and maximally symmetric spacetime, and (ii) it is the only way to generalize the $\Lambda$CDM model without touching the physics that underpins this model. Accordingly, to begin with, we have replaced, in the simplest manner, the spatially flat RW spacetime assumption of the $\Lambda$CDM model with the simplest more realistic background that simultaneously allows non-zero spatial curvature and anisotropic expansion; namely, considered the simplest anisotropic generalizations of the RW spacetime, viz., the Bianchi type I, V, and IX spacetimes (having the simplest homogeneous and flat, open, and closed spatial sections, respectively) combined in one Friedmann equation~\eqref{themodel}. We have then aimed to investigate whether the observational data still support spatial flatness and/or isotropic expansion in this case, and, if not, to explore the roles of spatial curvature and expansion anisotropy (viz., the shear scalar due to its stiff fluid-like behavior) in addressing some of the current cosmological tensions associated with the $\Lambda$CDM model~\cite{DiValentino:2020vvd,DiValentino:2020srs,DiValentino:2020zio,DiValentino:2021izs,Perivolaropoulos:2021jda}. We have presented the theoretical background and explicit mathematical construction of this model, dubbed An-$o\Lambda$CDM, in Sec.~\ref{S2}. We have carried out the analyses of the model, and its particular cases, namely, An-$\Lambda$CDM (allowing anisotropic expansion), $o\Lambda$CDM (allowing non-zero spatial curvature), and $\Lambda$CDM, by using the latest data sets from the different observational probes, viz., CC, SnIa Pantheon 2018, BAO, and Planck CMB(+Lens); see Sec.~\ref{data} for the data sets and the methodology used, and Sec.~\ref{results} for the results (summarized in Tables~\ref{tab:R3},~\ref{tab:R1}, and~\ref{tab:R2} and the figures from Fig.~\ref{fig:NF1} to Fig.~\ref{fig:NF6}) and discussions.

We have found some deviations from spatial flatness and only upper bounds on anisotropic expansion, which also affect the predicted Hubble constant and present-day density parameter of the pressureless fluid, at various significance levels depending on the data set used. However, when the data sets relevant to the low- and high-redshifts are simultaneously employed, all the extended models are predicted to be indistinguishable from the Planck 2018 $\Lambda$CDM model~\cite{Planck:2018vyg}, though we emphasize that when the spatial curvature is free to vary, some tensions appear between the data~\cite{Handley:2019tkm,DiValentino:2019qzk,Vagnozzi:2020zrh}. In particular, with regard to the expansion anisotropy, from the low redshift data, viz., the CC and/or Pan data, we have found $\Omega_{\sigma 0}\lesssim10^{-2}$ (about one order of magnitude looser than the model independent estimations~\cite{Campanelli:2010zx,Wang:2017ezt,Jimenez:2014jma,Soltis:2019ryf,Zhao:2019azy,Hu:2020mzd,Kalus:2012zu}), while it considerably tightens to $\Omega_{\sigma 0}\lesssim10^{-14}$ with the addition of the BAO data. We have obtained the tightest constraints, $\Omega_{\sigma 0}\lesssim10^{-18}$, when the data set relevant to the highest redshifts ($z\sim1100$), viz., the CMB($+$Lens) information is considered in our analyses. Yet, a very small amount of expansion anisotropy in the late universe cannot be excluded; we have not observed a correlation between $\Omega_{\sigma 0}$ and the other cosmological parameters, and therefore the inclusion of expansion anisotropy does not modify their constraints. These conclusions are well supported by the $\chi^2$ and model comparison analysis (see Table~\ref{BE}).

The main lessons that we have learned from this study, which considers the simplest extension of $\Lambda$CDM that simultaneously allows non-zero spatial curvature and anisotropic expansion, are summarized as follows:
\begin{itemize}[noitemsep,topsep=0pt]
\item  The combination at the face value of the current observational data, forgetting about their tension, confirm the spatial flatness and isotropic expansion assumptions of the $\Lambda$CDM model, yet a very small amount of expansion anisotropy in the late universe cannot be excluded.
\item The introduction of the spatial curvature or the expansion anisotropy, or both, in the simplest manner, on top the $\Lambda$CDM model does not offer a possible relaxation to the $H_0$ tension.  
\item The introduction of the anisotropic expansion in the simplest manner neither affects the closed space prediction from the CMB data only analyses, nor does improve the drastically reduced value of $H_0$ which is known to be led by a closed universe (see for example~\cite{Planck:2018vyg,Handley:2019tkm,DiValentino:2019qzk}).
\end{itemize}

We emphasize that our conclusions (with $\Omega_{\sigma0}\lesssim10^{-18}$), which favor the standard $\Lambda$CDM model, are based on our observational analysis of the model at the background level using the compressed CMB data. It is conceivable that these will be more pronounced with the already existing stronger upper bounds on the expansion anisotropy ($\Omega_{\sigma 0}\lesssim10^{-21}$--$10^{-23}$), using the full Planck CMB temperature and polarization spectra measurements and taking into account cosmological perturbations on the anisotropic background~\cite{Pontzen16,Saadeh:2016bmp,Saadeh:2016sak}, and the BBN light element abundances~\cite{Barrow:1976rda,Barrow:1997sy,Campanelli:2011aa,Akarsu:2019pwn}, because of the absence of correlation between $\Omega_{\sigma0}$ and the other cosmological parameters. Therefore, we can conclude that the pure geometric extension of the standard $\Lambda$CDM model is not helpful in dealing with the cosmological tensions, and the solutions should be sought in possible systematic errors or new physics. However, one should be careful while deriving final conclusions from the current work; our findings may also be suggesting further exploration of the role of anisotropic expansion in reaching a successful alternative to the $\Lambda$CDM model.

We have studied an extension of $\Lambda$CDM obtained by considering only the simplest anisotropic generalizations of the RW spacetime, viz., the Bianchi type I, V, and IX spacetimes (among the spatially homogeneous but not necessarily isotropic spacetimes; the Bianchi spacetimes and Kantowski--Sachs spacetime \cite{Ellis:1998ct,GEllisBook}), combined in one Friedmann equation~\eqref{themodel} and then treated the shear scalar as a stiff fluid-like correction to the usual the FLRW model, thereby, ignored the line of sight of the data in the sky in our analyses. On the other hand, it was suggested in \cite{Cea:2022mtf} that, even within the simplest anisotropic extension of $\Lambda$CDM (using LRS Bianchi type I spacetime, which describes spatially flat but ellipsoidal universe), the $H_0$ tension can be addressed when the comoving angular diameter distance is allowed to be direction dependent, and additionally, the $S_8$ tension can be addressed when producing large-scale CMB E-mode correlations via anisotropic expansion is allowed. More importantly, we have limited our investigations to orthogonal (non-tilted) models, yet it is possible to consider tilted models; it is possible to arrive at completely different conclusions through this possibility such as addressing the $H_0$ tension, but, which could go as far as describing the universe we live in by a tilt and anisotropy on top of Einstein--de Sitter (suggesting that the accelerated late universe is a misinterpretation of the data) \cite{2011MNRAS.414..264C,Wilczynska:2020rxx,Migkas:2020fza,Chang:2017bbi,Secrest:2020has,Siewert:2020krp,Migkas:2021zdo,Rahman:2021mti,Heinesen:2021azp,Park:2017xbl,Park:2019emi,Macpherson:2021gbh,Yeung:2022smn,Krishnan:2021dyb,Krishnan:2021jmh,Luongo:2021nqh,Cea:2022mtf,Camarena:2022iae,Tsagas:2011wq,Tsagas:2015mua,Colin:2019opb,Asvesta:2022fts,Lin:2015rza,Li:2015uda} (see also Refs.~\cite{Perivolaropoulos:2021jda,Abdalla:2022yfr} for recent reviews). Having said that, such models bring in additional free parameters to be constrained and require investigation by using the full likelihood of the data, as it is important to determine the amplitudes and directions of the dipole anisotropy in such models. Thus, despite the null results of anisotropy in the simplest pure geometric extension of the $\Lambda$CDM model that simultaneously involves anisotropic expansion and spatial curvature, pure geometric generalizations of $\Lambda$CDM without touching the underpinning physics deserve further investigations.

In our analyses with BAO and/or CMB data, we assumed that the standard cosmological model successfully describes the expansion dynamics of the pre-recombination universe, and accordingly, fixing the drag redshift (involved in analysis with BAO data) and the last scattering surface redshift (involved in analysis with CMB data) to their Planck 2018 $\Lambda$CDM and $o\Lambda$CDM values, we have forced the expansion anisotropy to remain as a correction all the way to the largest redshifts associated with the BAO and CMB data. Therefore, it is clear that our analysis can be extended by relaxing these conditions, say, by allowing the expansion anisotropy to be large enough to affect the average expansion rate of the pre-recombination universe. In this case, one might expect the stiff fluid-like behavior of the shear scalar to decrease sound horizon at drag epoch and cause $H_0$ to increase, as the early dark energy (EDE)~\cite{Poulin:2018cxd,Smith:2019ihp,Kamionkowski:2022pkx} does; namely, EDE is characterized by that its energy dilutes away like radiation or faster (fastest at its stiff fluid-like behavior limit) for $z\lesssim 3000$. However, unlike EDE, which behaves like a cosmological constant for $z\gtrsim3000$, anisotropic expansion would spoil the standard BBN as it would dominate radiation in the BBN era due to the persistent stiff fluid-like behavior of the shear scalar (see Ref.~\cite{Akarsu:2019pwn} and references therein). Nevertheless, it is in principle possible to control the behavior of the shear scalar by introducing anisotropic sources (see, e.g., Refs.~\cite{Barrow:1997sy,Akarsu:2020pka}), which suggests that exploring the possible anisotropic sources that, together with shear scalar, can lead to expansion dynamics of the universe similar to that caused by EDE, is an interesting alternative to scalar field models of EDE (see Ref.~\cite{Akarsu:2020pka}, which suggests that any canonical scalar field can be emulated via anisotropically deformed vacuum energy).

It is worth emphasizing that generalizing the spatially flat and maximally symmetric spacetime assumption of the $\Lambda$CDM model is of a different nature than generalizing its GR and/or $\Lambda$ assumptions; it is to consider a more \textit{realistic} space, rather than a new/more general gravity or DE theory/model. While the pure geometric generalization of $\Lambda$CDM considered in this study does not offer a solution to the $H_0$ tension, the interrelation between such generalizations and cosmological models that propose solutions to some cosmological tensions~\cite{DiValentino:2020vvd,DiValentino:2020srs,DiValentino:2020zio,DiValentino:2021izs,Perivolaropoulos:2021jda,Abdalla:2022yfr}, stands as a new research topic worth exploring---such studies, albeit so far limited to non-flat RW spacetime generalizations, have recently started to appear in the literature~\cite{Acquaviva:2021gty,Yang:2021hxg,Cao:2021ldv,Gonzalez:2021ojp,Farrugia:2021zwx,Nilsson:2021ute,Bargiacchi:2021hdp,Yang:2022kho}. What will be the response of models that propose solutions to cosmological tensions to a spatially more general/realistic spacetime? Will these models continue to be successful? In the case of geometric generalizations of these models, how will observational constraints on spatial non-flatness and anisotropy be affected? These questions can be even more interesting when it comes to some specific models proposed to address some cosmological tensions. For instance, in the case of DE models that attain negative energy densities in the past, such as the graduated dark energy (gDE) and the $\Lambda_{\rm s}$CDM model~\cite{Akarsu:2019hmw,Akarsu:2021fol,Akarsu:2022typ}; since the three-Ricci scalar in a spatially closed universe (favored by the CMB data) is reminiscent of cosmic strings with negative energy density. And, the EDE model~\cite{Poulin:2018cxd,Smith:2019ihp,Kamionkowski:2022pkx}, as, in some cases, it exhibits a stiff fluid-like behavior (for $z\lesssim3000$) reminiscent of the shear scalar (quantifying anisotropic expansion). And also, models that suggest modifications in both early and late universe dynamics~\cite{Vagnozzi:2021tjv}; since, in line with this, while non-zero spatial curvature (dilutes slower than dust, faster than DE) would be most effective after matter-radiation equality before DE domination, anisotropic expansion would be most effective in the early universe (as the shear scalar dilutes faster than dust and radiation). Of course, having an excessive number of free parameters in such studies would be a disadvantage, as it is against Occam's razor. However, there can still be a lot to learn from these studies, which may be crucial given that the $\Lambda$CDM model seems to suffer from multiple discrepancies, some of which, e.g., the $H_0$ tension, are quite significant, and that their solutions turned out to be challenging, suggesting the need of new physics. Thus, to better understand the universe we live in, it is important not only to explore in detail the pure geometric generalizations of the $\Lambda$CDM model, but also to explore such generalizations of models that suggest solutions to cosmological tensions.

\begin{acknowledgments}
The authors thank the anonymous referee for their comments and suggestions that helped improve the paper. The authors also thank Shahin Sheikh-Jabbari for valuable comments. \"{O}.A. acknowledges the support by the Turkish Academy of Sciences in the scheme of the Outstanding Young Scientist Award  (T\"{U}BA-GEB\.{I}P). E.D.V. is supported by a Royal Society Dorothy Hodgkin Research Fellowship. S.K. gratefully acknowledges the support from Science and Engineering Research Board (SERB), Govt. of India (File No. CRG/2021/004658). This article is based upon work from COST Action CA21136 Addressing observational tensions in cosmology with systematics and fundamental physics (CosmoVerse) supported by COST (European Cooperation in Science and Technology).
\end{acknowledgments}

\end{document}